\documentclass[sn-mathphys,Numbered]{sn-jnl}

\usepackage{multirow}%
\usepackage{amsmath,amssymb,amsfonts}%
\usepackage{amsthm}%
\usepackage{mathrsfs}%
\usepackage[title]{appendix}%
\usepackage{textcomp}%
\usepackage{manyfoot}%
\usepackage{booktabs}%
\usepackage{algorithm}%
\usepackage{algorithmicx}%
\usepackage{algpseudocode}%
\usepackage{listings}%

\usepackage{float}
\usepackage[dvipsnames]{xcolor}
\usepackage{hyperref}
\hypersetup{
    colorlinks,
    linkcolor=NavyBlue,
    citecolor=BrickRed,
    urlcolor=MidnightBlue
}
\usepackage{graphicx}
\usepackage{nth}
\usepackage[nointegrals]{wasysym}
\usepackage{physics}
\usepackage{float}
\usepackage{nameref}
\usepackage{xspace}

\usepackage[separate-uncertainty=true]{siunitx}
\usepackage{caption}
\newcommand{\vect}[1]{\boldsymbol{#1}}
\newcommand{\diff}{\,\mathrm{d}}
\newcommand{\NA}{\mathrm{NA}}
\newcommand{\exc}{\mathrm{exc}}
\newcommand{\emi}{\mathrm{em}}

\newcommand{\Int}{\int\displaylimits}
\newcommand{\FT}{\mathcal{F}}

\newcommand{\xs}{\vect x_s}
\newcommand{\xd}{\vect x_d}
\newcommand{\x}{\vect x}

\newcommand{\ks}{\vect k_s}
\newcommand{\kd}{\vect k_d}

\newcommand{\sism}{s\textsuperscript{2}ISM\xspace}
\newcommand{\sflism}{s\textsuperscript{2}FLISM\xspace}

\DeclareMathOperator*{\argmin}{arg\,min} 
\DeclareMathOperator*{\argmax}{arg\,max}

\begin{document}

\title{Structured Detection for Simultaneous Super-Resolution and Optical Sectioning in Laser Scanning Microscopy}



\author[1]{\fnm{Alessandro} \sur{Zunino}}
\email{alessandro.zunino@iit.it}
\equalcont{These authors contributed equally to this work.}

\author[1,2]{\fnm{Giacomo} \sur{Garrè}}
\email{giacomo.garre@iit.it}
\equalcont{These authors contributed equally to this work.}

\author[1,3]{\fnm{Eleonora} \sur{Perego}}
\email{eleonora.perego@unil.ch}

\author[1,2]{\fnm{Sabrina} \sur{Zappone}}
\email{sabrina.zappone@iit.it}

\author[1]{\fnm{Mattia} \sur{Donato}}
\email{mattia.donato@iit.it}

\author*[1]{\fnm{Giuseppe} \sur{Vicidomini}}\email{giuseppe.vicidomini@iit.it}

\affil[1]{\orgdiv{Molecular Microscopy and Spectroscopy}, \orgname{Istituto Italiano di Tecnologia}, \orgaddress{\street{Via Enrico Melen, 83}, \city{Genoa}, \postcode{16152}, \country{Italy}}}

\affil[2]{\orgdiv{Dipartimento di Informatica, Bioingegneria, Robotica e Ingegneria dei Sistemi}, \orgname{University of Genoa}, \orgaddress{\street{Via Dodecaneso 35}, \city{Genoa}, \postcode{16146}, \country{Italy}}}

\affil[3]{Current address: \orgdiv{Center for Integrative Genomics (CIG)}, \orgname{University of Lausanne (UNIL)}, \orgaddress{ \city{Lausanne}, \postcode{1015}, \country{Switzerland}}}


\abstract{
Fast and sensitive detector arrays enable image scanning microscopy (ISM), overcoming the trade-off between spatial resolution and signal-to-noise ratio (SNR) typical of confocal microscopy.
However, current ISM approaches cannot provide optical sectioning and fail with thick samples, unless the size of the detector is limited. Thus, another trade-off between optical sectioning and SNR persists.
Here, we propose a method without drawbacks that combines uncompromised super-resolution, high SNR, and optical sectioning.
Furthermore, our approach enables super-sampling of images, relaxing Nyquist's criterion by a factor of two.
Based on the observation that imaging with a detector array inherently embeds axial information about the sample, we designed a straightforward reconstruction algorithm that inverts the physical model of ISM.
We present the comprehensive theoretical framework and validate our method with synthetic and experimental images of biological samples captured using a custom setup equipped with a single-photon avalanche diode (SPAD) array detector.
We demonstrate the feasibility of our approach exciting fluorescence emission both in the linear and non-linear regime.
Moreover, we generalize the algorithm for fluorescence lifetime imaging, fully exploiting the single-photon timing ability of the SPAD array detector.
Our method outperforms conventional approaches to ISM and can be extended to any LSM technique.
}




\maketitle

\section{Main}

Optical microscopy is an invaluable tool for both industry and research, with applications ranging from material science to life science.
In particular, fluorescence imaging stands out as a powerful technique, offering fast, highly specific, and minimally invasive observation of biomolecule distribution within live cells, tissues and model organisms \cite{Lichtman2005}.
In the last decades, super-resolution microscopy effectively pushed these observations beyond the diffraction limit, overcoming one of the major limitations of traditional optical microscopy \cite{Prakash2022}.
The performance of a microscope further increase with optical sectioning, namely the capability to reject out-of-focus light and image a single plane within a three-dimensional sample \cite{Agard1984, Conchello2005}.
The ways for optical sectioning can be categorized into three non-mutually exclusive strategies: (i) avoiding excitation of out-of-focus planes, (ii) removing out-of-focus fluorescence before detection, and (iii) computationally removing out-of-focus light.
Category (i) comprises techniques that do not excite the sample outside the focal region.
A representative example is selective plane illumination microscopy (SPIM), which only excites a thin sample slice \cite{Huisken2004, Zunino2021, Stelzer2021}.
SPIM is a fast and gentle technique but achieving high resolution requires sophisticated optical architectures \cite{Chen2014, Yang2022}.
Another technique that falls in the same category is multi-photon excitation microscopy (MPEM), where the non-linear interaction with the sample happens only in the focal volume \cite{Denk1990, Diaspro2006}.
Thanks to long wavelengths, MPEM is robust to scattering and enables deep penetration into tissues.
However, MPEM requires high optical power, which increases both costs and the risk of photo-damaging to the sample.
Category (ii) involves blocking light before reaching the detector.
This goal can be achieved by physically filtering the out-of-focus light with a pinhole, as in confocal laser scanning microscopy (CLSM) \cite{Sheppard1977, Sheppard1990, Bayguinov2018}. However, complete rejection of out-of-focus light inevitably compromises the detection of in-focus photons and, consequently, the signal-to-noise ratio (SNR).
Another strategy employs destructive interference to remove out-of-focus light, as in 4Pi \cite{Sheppard1991, Hell1992, Hao2022} and I\textsuperscript{5}M microscopy \cite{Gustafsson1999, BEWERSDORF2006}.
These techniques provide axial super-resolution and superior optical sectioning, but they increase experimental complexity and are suitable only for thin samples.
Finally, category (iii) comprises techniques that do not avoid generating or detecting out-of-focus light but rely on encoding the axial information into the data to remove the background in post-processing. The most important representative of this category is structured illumination microscopy (SIM), which can be used to either achieve optical sectioning (OS-SIM) \cite{Neil1997, Neil1998} or super-resolution (SR-SIM) \cite{Gustafsson1999}, depending on the periodicity of the illumination pattern \cite{Chen2023}. The two effects can be combined using a three-dimensional structured illumination (3D-SIM) \cite{Gustafsson2008}, even in a single-plane acquisition \cite{Jost2015, Soubies2018}. Thus, SIM enables high-contrast super-resolution imaging. However, its capability to decode additional information from the images depends on the contrast of the illumination pattern, which may be severely reduced in thick and crowded samples \cite{Manton2022}.

In this scenario, image scanning microscopy (ISM) has the potential to provide exquisite optical sectioning combined with super-resolution while maintaining versatility in terms of sample types, architectural simplicity, and integration with various other spectroscopy techniques, such as multi-color imaging and fluorescence lifetime imaging.
ISM was originally conceived as an improvement over CLSM \cite{Sheppard1988,Bertero1989}.
Indeed, the pinhole plays a dual role in confocal microscopy: the closer the pinhole, the less out-of-focus light is collected and the better the lateral resolution, up to a theoretical limit of twice the diffraction limit \cite{Sheppard1982}.
However, CLSM achieves super-resolution only when the pinhole is so closed to irremediably deteriorate the SNR of the image.
ISM allows for harnessing the benefit of the confocal effect by replacing the pinhole and single-element detector with a detector array in a CLSM architecture.
The initial ISM implementation utilized a conventional camera \cite{Muller2010}, but the frame rate (\SI{\sim 1}{kHz}) significantly limited the imaging temporal resolution.
Later, different optomechanical ISM implementations addressed this temporal resolution limitation \cite{Roth2013, York2013} at the cost of sacrificing the laser-scanning architecture.
Thus, they increased the complexity and renounced to versatility and compatibility with other advanced microscopy and spectroscopy techniques.
More recently, tailored fast and small array detectors, such as the AiryScan \cite{Huff2017} and the asynchronous read-out single-photon avalanche diode (SPAD) array \cite{Antolovic2018,Buttafava2020} allowed for the original ISM implementation without compromise its imaging temporal resolution.
Thanks to the single-photon sensitivity and temporal resolution of SPAD array detectors, they are becoming the tool-of-choice for ISM \cite{Slenders2021}.
The asynchronous read-out approach facilitates ISM implementations based on resonant scanners, enabling video-rate imaging. At the same time, the single-photon timing ability allows for the combination of ISM with time-resolved measurements, such as fluorescence lifetime microscopy \cite{Castello2019} -- a well-established technique for functional imaging.
The essence of ISM lies in its operational principle: each element of the detector array is designed to function as a closed pinhole, while the entire array guarantees good light collection efficiency. Thus, an image scanning microscope can be seen as multiple confocal microscopes observing the same sample in parallel, each from a slightly different point of view. Indeed, the raw dataset is four-dimensional, consisting of a set of confocal-like images as many as the number of detector elements.
The goal of ISM data processing is to fuse the images to utilize all the collected photons while preserving the super-resolution. Such a task is traditionally performed either by adaptive pixel reassignment (APR) \cite{Sheppard2013, Castello2019, Ancora2024} or multi-image deconvolution \cite{Ingaramo2014, Zunino2023a}. Both techniques are capable of pushing the SNR and the lateral resolution beyond the diffraction limit but do not provide any optical sectioning. Thus, the rejection of out-of-focus light is conventionally carried out through a pinhole -- virtual or physical -- as in conventional CLSM.
In order to preserve the sectioning capabilities of ISM, the size of the detector array is typically kept small \cite{Sheppard2020}, constraining the collection efficiency and limiting the SNR of the images. Furthermore, the pinhole strategy cannot fully block out-of-focus light from reaching the detector. However, the axial information required to achieve optical sectioning is inherently encoded into the raw ISM dataset, as reported by a recent proof-of-principle \cite{Tortarolo2022} and a few theoretical publications \cite{Sheppard2023a, Sheppard2023b}.

Here, we propose a novel computational method to construct the ISM image by leveraging all the information encoded in the raw ISM dataset. Our approach aims to achieve the conventional benefits of ISM, namely super-resolution, together with improved optical sectioning, excellent SNR, and compatibility with other advanced microscopy techniques.
%
Since optical sectioning is enabled in post-processing, there is no need to constrain the detector size. Thus, our approach allows uncompromised collection efficiency, paving the way for faster and gentler super-resolution and high-contrast multi-modal imaging. Our idea stems from the observation that ISM can also be seen as a special case of SIM \cite{Sheppard2021}. Indeed, the focused excitation encodes the high-frequency component needed to increase the resolution and exclude the out-of-focus light (Supp. Note \ref{subsec:A1}). We extract the information encoded in a single-plane ISM dataset using a maximum likelihood estimation technique. The result is a couple of images containing the in-focus and out-of-focus portion of the signal, respectively. The out-of-focus image is discarded, while the in-focus image is kept as the final reconstruction with enhanced resolution and SNR.
Some approaches have already been proposed to enhance the axial sectioning capabilities of ISM, such as refocusing after scanning using helical phase engineering (RESCH) \cite{Jesacher2015, Roider2016} and engineered ISM (eISM) \cite{Roider2017}. They robustly encode the axial information in the data through wavefront shaping, enabling full volumetric imaging from a single-plane ISM dataset. However, those benefits come at the cost of increasing the experimental complexity and sacrificing some lateral resolution.
Conversely, we do not rely on light shaping. Thus, we preserve the simplicity and the super-resolution capability of ISM. Since super-resolution and optical sectioning are achieved simultaneously, we named our technique \sism (Super-resolution Sectioning Image Scanning Microscopy).

In this work, we present a comprehensive theoretical framework and validate our method on synthetic and experimental data acquired with a custom ISM setup equipped with a SPAD array detector. We validate our reconstruction strategy with resolution targets to measure the benefits of our method quantitatively, and we apply it to various biological samples, including cells and tissues, to demonstrate the feasibility of our approach. To simplify the application of \sism and alleviate the user from additional experimental tasks, we propose a rigorous strategy to automatically extract the relevant parameters needed to run our algorithm directly from the data. We demonstrate that the abundance of data in the ISM dataset can also be used to achieve both optical and digital super-resolution, relaxing the Nyquist sampling criterion by a factor of two.
Our approach is not limited to conventional ISM, but synergizes with any incoherent microscopy technique based on focused illumination and a detector array.
We demonstrate the broad scope of our work by combining \sism with 2PE \cite{Koho2020} and multi-color imaging.
Finally, we apply \sism also to fluorescence lifetime imaging microscopy (FLIM) \cite{Tortarolo2024}, increasing the robustness of the technique and valorizing the additional temporal information provided by the SPAD array detector.

\section{Results}
\label{sec:principle}

\subsection{Principle of \sism}

\begin{figure*}[t!]
    \centering
    \includegraphics[width=\textwidth]{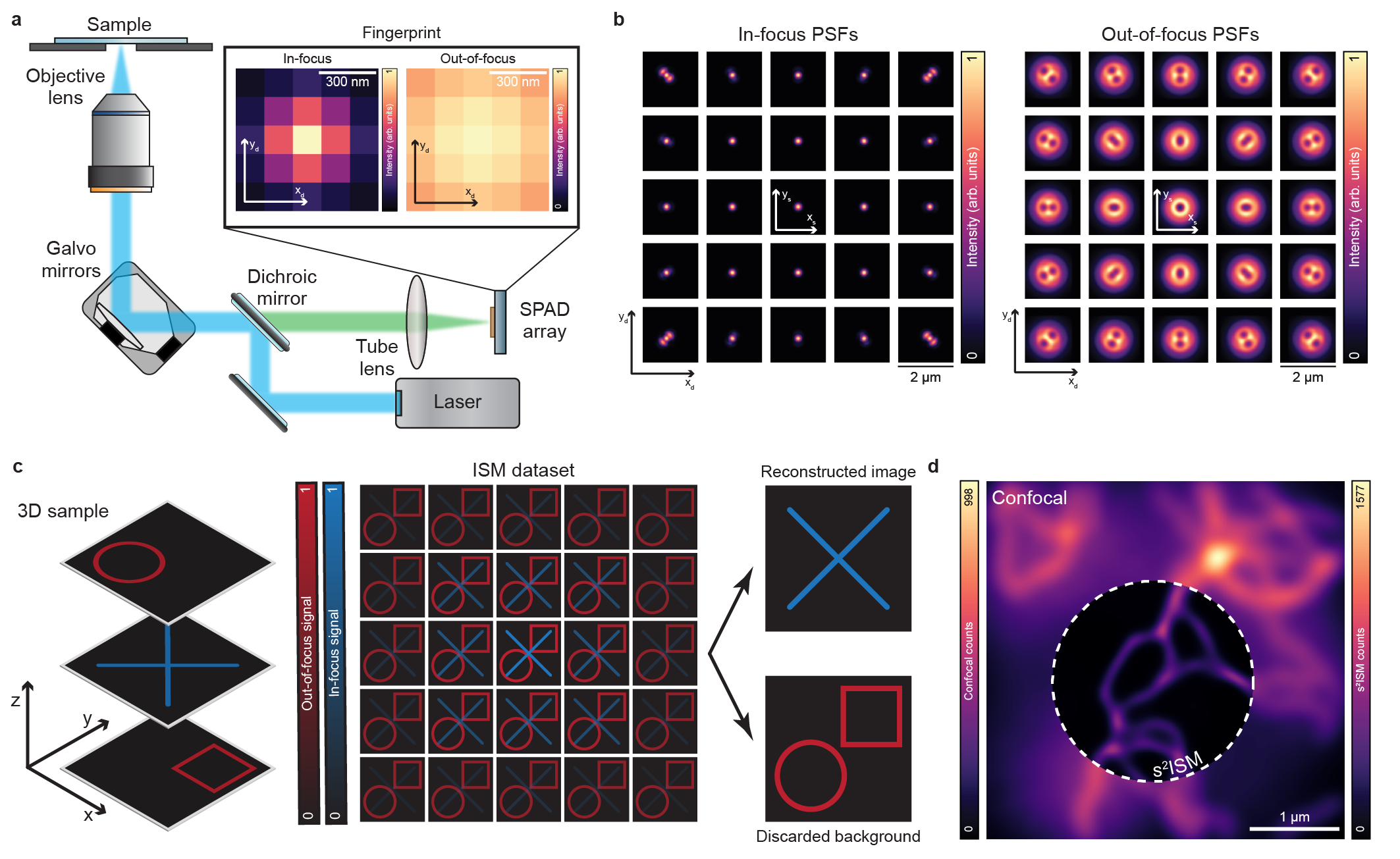}
    \caption{\textbf{Image Scanning Microscopy.} \textbf{a}, Sketch of an image scanning microscope. The inset shows fingerprints relative to the different axial positions of the sample. \textbf{b}, Simulated PSFs of an ideal ISM system at $z=\SI{0}{nm}$ (in-focus) and $z=\SI{720}{nm}$ (out-of-focus) with $\lambda = \SI{640}{nm}$. \textbf{c}, Sketch of the ISM image formation. The images of an in-focus sample appear brighter at the centre of the detector coordinate and dimmer at the periphery. The brightness of the images of out-of-focus samples decays slower along the detector coordinate, encoding the axial information into the ISM dataset. The reconstruction algorithm builds two images from a single-plane dataset, one with the in-focus image and the background discarded. The out-of-focus sections of the sample are projected into a single image, which is subsequently discarded. \textbf{d}, comparison of a confocal image with the ISM image reconstructed by the s$^2$ISM algorithm on synthetic tubulin filaments.}
    \label{fig:1}
\end{figure*}

For this work, we developed a custom ISM microscope -- sketched in Fig. \ref{fig:1}a and more detailed in Supp. Fig. \ref{supfig:setup}.
The instrument is a laser scanning microscope (LSM) equipped with a $5 \times 5$ asynchronous-readout single photon avalanche diode (SPAD) array detector \cite{Buttafava2020}.
A laser beam focused at the scan point $\xs = (x_s, y_s)$ excites the fluorescent molecules from a diffraction-limited volume within the specimen.
The array detector collects the emitted light from each pixel location $\xd = (x_d, y_d)$, building a four-dimensional dataset.
The dimensionality is higher for time-resolved acquisitions. In this case, time acts as an additional scanning coordinate.
For the sake of simplicity, we leave the treatment to spatiotemporal dataset to Supp. Note \ref{subsec:A5} and we now focus only on the spatial dimensions.

The sensitive elements of the detector can be seen as independent pinholes, making the ISM microscope equivalent to multiple confocal microscopes observing the same sample in parallel, each from a different point of view.
Thus, the ISM dataset can be interpreted as a set of confocal-like images, each blurred by a unique point spread function (PSF).
In detail, the PSF recorded by the detector at position $\xd$ is given by
\begin{align}
    h(\vect x_s, z_s | \vect x_d) &= h_{\exc}(-\vect x_s, z_s) \cdot \qty[ h_{\emi}(\vect x_s, z_s) * p(\vect x_s - \vect x_d) ] = \nonumber \\
    & =  h_{\exc}(-\vect x_s, z_s) \cdot  h_{\det}(\xs - \xd, z_s)
    \label{eq:PSF_ISM}
\end{align}
where $*$ is the convolution operator with respect to the coordinate $\xs$, $z_s$ is the axial coordinate of the emitter, $p$ is the function describing the geometry of the detector element's active area, $h_{\exc}$ and $h_{\emi}$ are the excitation and emission PSF of the microscope. We define the detection PSF $h_{\det}$ as the convolution of the emission PSF with the pinhole. The above equation explains the physical reason behind the super-resolution in ISM. Indeed, for sufficiently small sensors, $h_{\det}$ is very similar to $h_{\exc}$. Thus, the product of the two PSFs is sharper than the individual ones. Conversely, if the sensor size is too large, $h_{\det}$ approaches a constant value and the super-resolution effect is washed out -- as for non-confocal laser-scanning microscopy.

The position $\xd$ of the detector element plays a key role in determining the shape and the relative intensity of each PSF (Fig. \ref{fig:1}b). These latter are rescaled by a normalization factor known as the \textit{fingerprint}
\begin{align}
    f(\xd, z_s)  = \int_{\mathbb{R}^2} h(\xs, z_s | \xd) \diff \xs = h_{\exc}(\xd, z_s) \star h_{\det}(\xd, z_s)
    \label{eq:fingerprint}
\end{align}
where $\star$ is the cross-correlation operator. The fingerprint is a bell-shaped distribution peaked at the centre of the detector. The more the emitter is defocused, the broader the distribution is (Supp. Fig.  \ref{supfig:In focus and bkg PSFs}).
In general, samples have a three-dimensional structure. Given the linearity of incoherent image formation, the ISM dataset $i(\vect x_s | \vect x_d)$ is given by the superposition of the fluorescence light emitted from any plane
\begin{equation}
    i(\xs|\xd) = \int_{\mathbb{R}} o(\xs,z) * h(\xs,z|\xd) \diff z
    \label{eq:image_formation}
\end{equation}
where $o$ is the object (i.e., the specimen's 3D distribution of emitters).
Even though the shape of the PSFs depends on the axial coordinate, it cannot be used to identify the depth of the emitters.
Indeed, the lateral structure of the specimen is convolved with the PSFs, and their relative contributions cannot be disentangled unless some prior knowledge is available.
However, Eq. \ref{eq:fingerprint} suggests that the light stemming from different axial positions distributes differently on the detector plane, regardless of the lateral sample structure. Thus, the fingerprint inherently encodes the axial information into the detector plane dimension.
Despite not allowing for a full volumetric reconstruction of the specimen, the fingerprint map is informative enough to enable the discrimination of the defocused light in a single-plane acquisition.

Assuming that the relevant out-of-focus contributions stem from a finite and discrete number of planes, we re-write the forward model of ISM image formation as follows
\begin{equation}
    i(\vect x_s | \vect x_d) = \sum_{k=1}^N f_k(\xd) \qty[ o_k(\xs) * \hat{h}_k(\xs | \xd) ]
    \label{eq:forward}
\end{equation}
where $k$ is the discretized axial position and $\hat{h}$ are the normalized PSFs. Thus, in-focus and out-of-focus emitters are weighted by a different fingerprint function. Namely, the modulation of contrast on the detector plane varies according to the axial position of the sample. The images generated by the central and peripheral elements of the detector array collect light mainly from in-focus and out-of-focus planes, respectively (Fig. \ref{fig:1}c).
Our goal is to leverage the fingerprint map -- uniquely provided by the ISM architecture -- to design a reconstruction procedure that builds a single super-resolution image starting from a single-plane ISM dataset, excluding defocused contributions. To this end, we consider only two axial planes ($N=2$) to account for the in-focus and out-of-focus sections of the sample. Furthermore, we leverage the SPAD array detector's single-photon sensitivity and negligible read-out noise to consider the photon shot noise as the only noise source in our data. Under the aforementioned assumptions, we can write an explicit Poisson likelihood functional (Supp. Note \ref{subsec:A2}), whose maximization leads to the following iterative solution
\begin{equation}
\begin{array}{l}
o_{1}^{(m+1)}(\xs)=o_{1}^{(m)} (\xs) \displaystyle\int h_1(-\xs|\xd)*\displaystyle\frac{i(\xs|\xd)}{\sum_{k=1}^2 o^{(m)}_k(\xs)*h_k(\xs|\xd)} \diff\xd  \\[20pt]
o_{2}^{(m+1)}(\xs)=o_{2}^{(m)}(\xs) \displaystyle\int h_2(-\xs|\xd)*\displaystyle\frac{i(\xs|\xd)}{\sum_{k=1}^2 o^{(m)}_k(\xs)*h_k(\xs|\xd)} \diff\xd
\\
\end{array}
\label{eq:s2ism}
\end{equation}
where $m$ and $k$ are the iteration and axial index, respectively. Namely, one estimated image contains the projection of the out-of-focus light and is discarded. The other contains only the signal originating from the focal plane and is built by fusing and deconvolving the twenty-five images of the ISM dataset. Indeed, the proposed algorithm can be considered a generalization of the multi-image deconvolution method \cite{Zunino2023a}, which can be recovered simply by putting $N=1$ in Eq. \ref{eq:forward} (Supp. Fig. \ref{supfig:MID_vs_3D_ISM}). The result is an image with enhanced resolution optical sectioning compared to its confocal counterpart (Fig. \ref{fig:1}d). Importantly, we achieve both benefits without discarding in-focus light, enhancing the SNR \cite{Roth2016}.

\subsection{Data-Driven Parameters Estimation for \sism}

\begin{figure*}[t!]
    \centering
    \includegraphics[width=\textwidth]{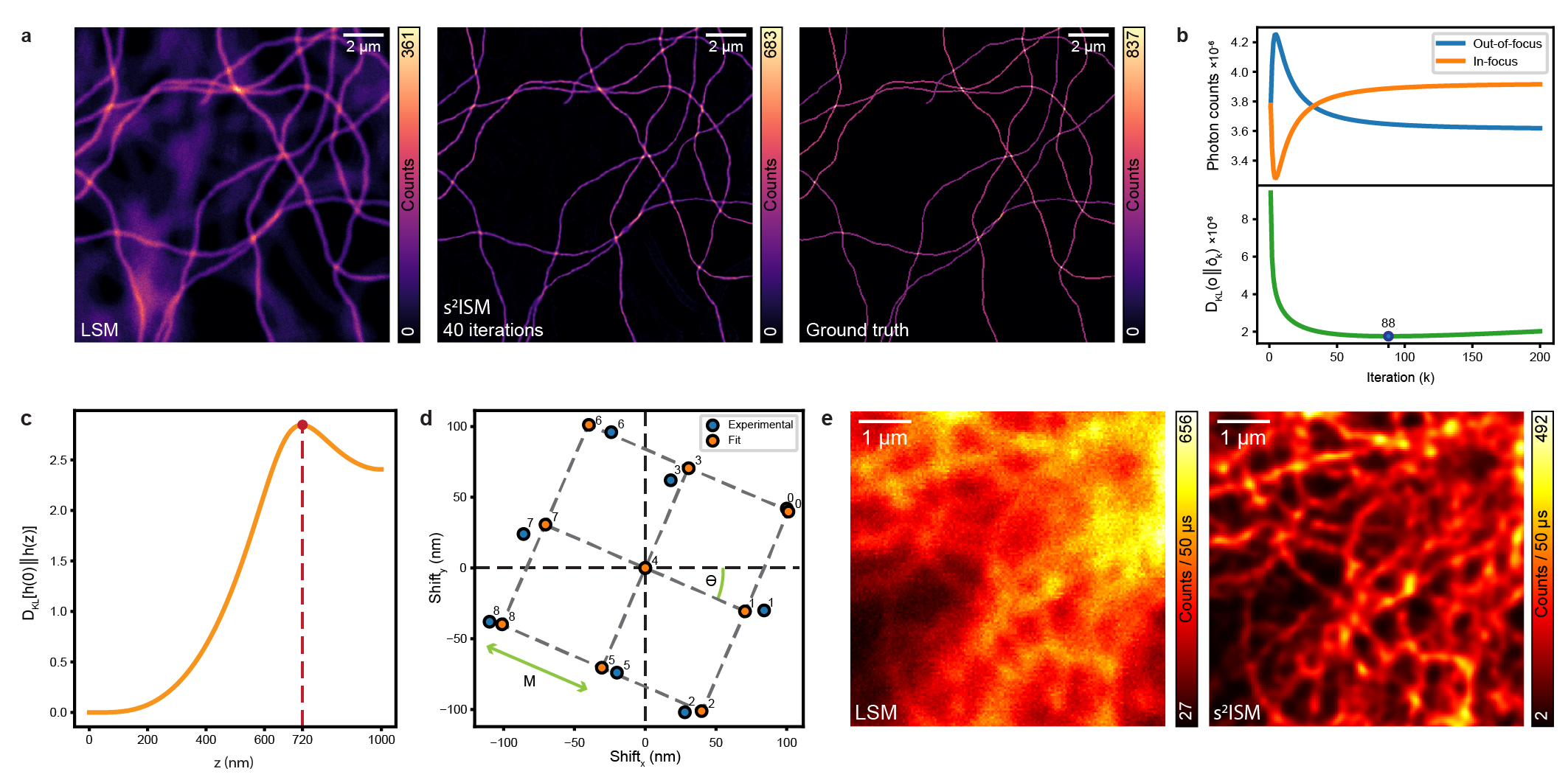}
    \caption{\textbf{Data-driven estimation of the parameters.} \textbf{a}, Simulated experiments at $\lambda = \SI{640}{nm}$ and $z=\SI{720}{nm}$. Comparison of a confocal (1.4 AU) image, \sism reconstruction, and ground-truth of the focal plane. \textbf{b}, Total number of photons per plane (top) and Kullback-Leibler divergence of the reconstruction to the focal ground truth (bottom) at varying iteration numbers. \textbf{c}, Kullback-Leibler divergence of the out-of-focus PSF dataset to the in-focus one. The maximum position defines the plane used for the \sism reconstruction. \textbf{d}, Experimental and fitted shift-vectors from the inner $3\times3$ array detector. The fit returns the magnification, orientation, and rotation parameters. \textbf{d},  Comparison of confocal (1.4 AU) and \sism image of the tubulin network of a HeLa cell (mouse anti-$\alpha$-tubulin combined with anti-mouse Abberior STAR RED).}
    \label{fig:2}
\end{figure*}

In order to apply the proposed algorithm (Eq. \ref{eq:s2ism}), we need to access a set of PSFs for each axial plane involved in the reconstruction. Such a requirement is not challenging for \textit{in silico} experiments, where we can easily validate the proposed approach. Thus, we first explored the capabilities of \sism in a simplified scenario, where the image is corrupted by fluorescence stemming from a single defocused plane at a known axial position. Using vectorial diffraction theory, we simulated the PSFs at the required axial positions and applied the iterative reconstruction. \sism successfully eliminates the out-of-focus light, building an image with enhanced resolution and optical sectioning (Fig. \ref{fig:2}a).
The algorithm's iterative nature raises the question of when to terminate it. To address this, we compared the reconstruction to the ground truth at different iterations, demonstrating that the algorithm exhibits a semiconvergent behaviour, similar to conventional deconvolution (Fig. \ref{fig:2}b).
Therefore, an early stop criterion should be adopted to avoid noise amplification.
Despite the optimum number of iterations being unknown in the absence of ground truth, we can estimate the optimal iteration range by exploiting the photon conservation property of our algorithm. 
Specifically, the total number of photons in the raw dataset equals the sum of the two reconstructed images regardless of the iteration (Supp. Note \ref{subsec:A3}).
Thus, \sism is not able to generate or destroy light but only to reassign the photons in the in-focus or out-of-focus plane.
The dynamics of photon exchange between the two planes mostly happen during the first iterations, later reaching a plateau (Fig. \ref{fig:2}b), indicating that convergence has been reached and the algorithm should stop.

In a realistic scenario, out-of-focus light stems from multiple planes.
While we could address this case by generalizing Eq. \ref{eq:forward} to account for a larger number $N$ of axial contributions, doing so would significantly escalate the computational cost of the algorithm and exacerbate the ill-posed nature of the problem we aim to solve.
Using only two axial planes, we are essentially asking the algorithm if the light collected by the detector array is more likely to have originated from the in-focus or out-of-focus plane.
Such an approach is simple, improves the conditioning of the inverse problem, and effectively enhances the optical sectioning of ISM, regardless of the actual thickness of the sample.
To demonstrate the effectiveness of our choice, we simulated a thick sample and generated the ISM dataset for a single plane.
In this case, multiple defocused planes contribute to the background in the data.
Thus, the axial position of the out-of-focus plane is not clearly defined.
To optimize the conditioning of the algorithm, we choose the out-of-focus plane as the one that maximizes the diversity of PSFs with respect to the focal plane.
We measured the similarity between PSFs at different axial positions either using Person correlation or the Kullback-Leibler divergence (Fig. \ref{fig:2}c) and chose the out-of-focus position as their minimum or maximum, respectively.
Regardless of the metric used, the calculated out-of-focus position is the same (Supp. Fig. \ref{supfig:conditioning}).
With such an approach, we successfully applied \sism to the simulated data, obtaining a reconstruction without out-of-focus background (Supp. Fig. \ref{supfig:3D-ISM_vol_synth_tub}). 
To better visualize the working principle of \sism, we applied it plane-by-plane to a simulated 3D dataset of a point source (Supp. Fig. \ref{supfig:optical_sectioning_wls}).
The results demonstrate that the algorithm effectively discards the light when the emitter is defocused but leaves intact the signal and performs a deconvolution on the planes within the focal volume.
Thus, the two-plane approach is effective in enhancing the optical sectioning of ISM.

However, in an experimental scenario, accessing the PSFs of the microscope is not straightforward and typically burdens the microscopist with the need for a direct measurement using point-like sources.
Furthermore, the goodness of the reconstructions strictly depends on the quality of the PSF dataset, which is likely to be degraded by various sources of noise (e.g., shot noise, vibrations, stray light), while misalignments and other non-idealities may distort the shape of the PSFs. 
Therefore, we propose to leverage the information naturally encoded into the ISM dataset to extract the relevant parameters needed to provide a reliable theoretical estimate of the PSFs.
To this end, we leverage an inherent property of ISM.
Indeed, the PSFs associated to the inner detector elements ($3\times 3$) are approximately identical in shape but shifted in position.
Therefore, the corresponding images are translated by a quantity called shift-vectors, which can be evaluated without requiring a direct PSF measurement.
The shift-vectors are uniquely related to the geometrical structure of the array detector and the magnification of the microscope.
More precisely, we can measure the detector rotation angle and the orientation of the detector elements (Fig. \ref{fig:2}c and Supp. Fig. \ref{supfig:par_retrieving}) using a tailored optimization procedure (Supp. Note \ref{subsec:A6}).
Assuming that the microscope does not suffer from significant aberrations, we use the aforementioned method to estimate the relevant PSF parameters directly from the dataset we aim to reconstruct.
Importantly, shift-vectors depend very weakly on the sample's axial position, and their estimation is not hindered by the presence of an out-of-focus background in the images.
Thus, we leverage the information retrieved from the dataset to simulate the PSFs using a vectorial diffraction model \cite{Caprile2022, Zunino2023b} that accounts for the high numerical aperture of our system.
Imaging of tubulin network demonstrates that our approach effectively removes the defocused light and improves resolution, revealing details previously hidden by the background (Fig. \ref{fig:2}d and Supp. Fig. \ref{supfig:cell_A}).
Even if we assumed that no aberrations are present in the microscope, we report that \sism is robust against small aberrations that do not break the circular symmetry of the PSF -- such as spherical aberration (Supp. Fig. \ref{supfig:exp_vs_synth_goldbeads}).

Finally, we compared our approach to \sism with the same algorithm applied using experimental PSF (Supp. Note \ref{subsec:A7}).
To this end, we acquired a volumetric dataset of images of a gold bead.
We define the focal plane as the one with the largest integral of the modulation transfer function.
The out-of-focus plane maximises the discrepancy from the focal plane (Supp. Fig.  \ref{supfig:synth_exp_PSFs}).
Once the axial planes are correctly defined, we reconstructed an image using \sism with the experimental PSFs  (Supp. Fig. \ref{supfig:3D-ISM_w_exp_and_synt_PSFs}).
Reconstruction with synthetic PSFs and experimental PSFS are qualitatively similar, demonstrating that the two approaches are both viable.
Nonetheless, the image reconstructed with the experimental PSFs shows a slightly inferior resolution, most likely due the inferior quality of the PSFs dataset which is inevitably corrupted by various sources of noise.
Given the ease of simulation of PSFs and the reliable results they ensure, we used synthetic PSFs to perform the reconstructions showed in the rest of the manuscript.

\subsection{Experimental Validation of \sism}

\begin{figure*}[t!]
    \centering
    \includegraphics[width=\textwidth]{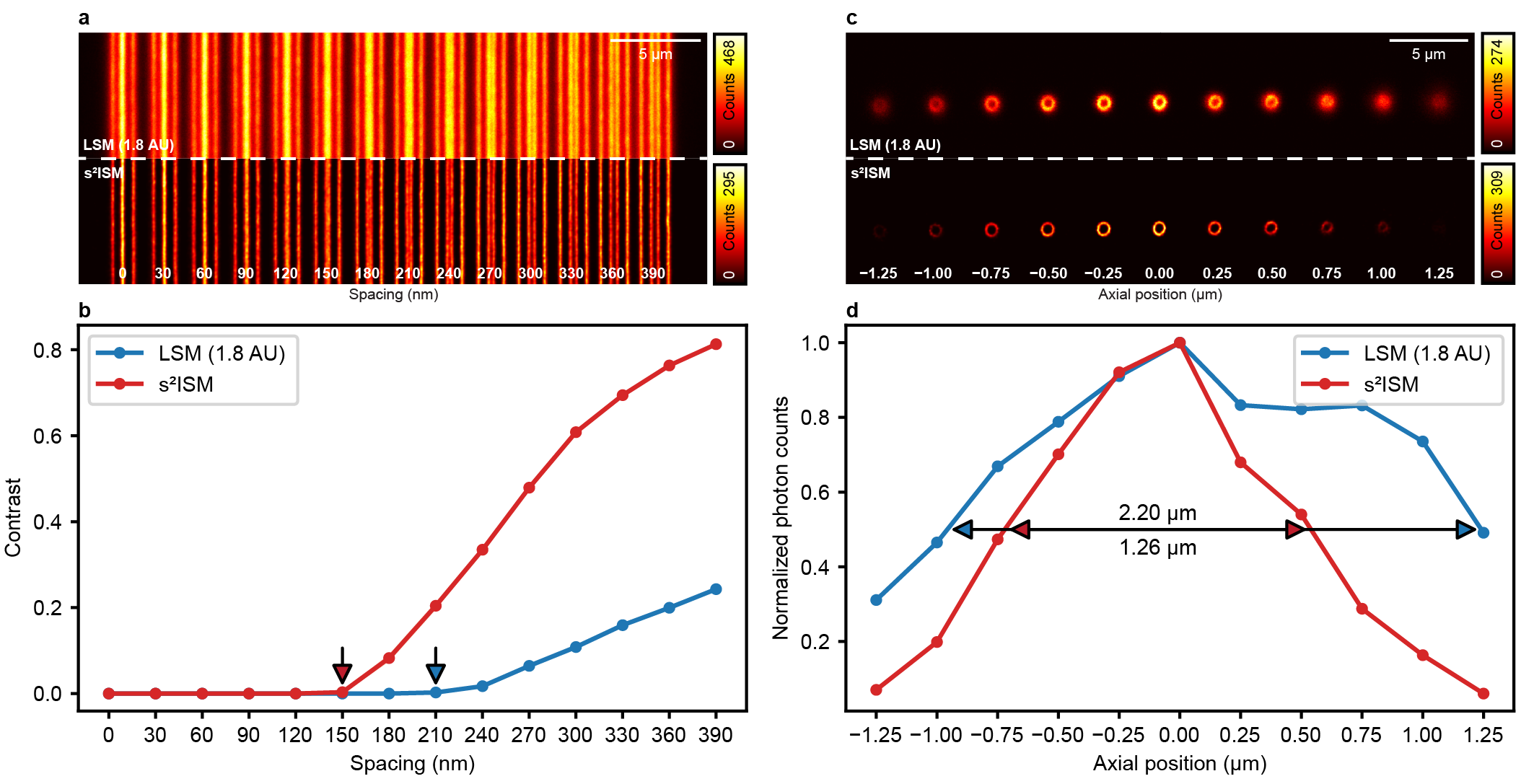}
    \caption{\textbf{Lateral resolution and optical sectioning.} \textbf{a}, Compared images (LSM and \sism) of a resolution target composed of gradually spaced lines. \textbf{b}, Corresponding modulation transfer function, experimentally measured by calculating the contrast of the dip relative to the two adjacent lines. When no dip is discernible, the contrast is set to zero. \textbf{c}, Compared images (LSM and \sism) of a three-dimensional stair of rings evenly spaced on the axial direction ($\Delta z = \SI{250}{nm}$). \textbf{d}, Corresponding normalized optical sectioning function, calculated by summing the photon counts from each ring. The values indicate the FWHM of the curves. We excited both targets using the laser wavelength $\lambda = \SI{488}{nm}$.}
    \label{fig:3}
\end{figure*}

In order to quantify the performance of our ISM reconstruction method, we performed imaging on a fluorescent resolution target \cite{argolight}.
First, we estimated the resolution gain provided by \sism using a pattern of gradually spaced lines. More in detail, we reconstructed the image either by summing over all the detector elements -- obtaining the corresponding open-pinhole confocal image -- and by applying \sism using the procedure described in the previous section -- namely, by estimating the PSFs directly from the data (Fig. \ref{fig:3}a). Then, we measured the visibility of the dip between adjacent lines at each spacing. The results shown in Fig. \ref{fig:3}b demonstrate that our reconstruction method enhances the lateral resolution of the final image, beating the diffraction limit. Indeed, the cut-off spacing for the confocal image is \SI{210}{nm}, which is pushed to \SI{150}{nm} thanks to our algorithm. Furthermore, the contrast improves at any spacing, outperforming existing reconstruction procedures (Supp. Fig. \ref{supfig:Argo_sample_diag_lines} and \ref{supfig:Argo_sample_resolution}). 

Then, we quantified the capability of \sism to reject out-of-focus light by imaging a 3D stair composed of ring-shaped targets uniformly spaced along the axial coordinate (Fig. \ref{fig:3}c and Supp. Fig. \ref{supfig:Argo_sample}). As expected, the more defocused the sample, the more light is discarded -- effectively increasing the contrast of the in-focus content of the image. The optical sectioning improves roughly by 75\% while the resulting image gets sharper thanks to the resolution improvement (Fig. \ref{fig:3}d). Importantly, the removal of the out-of-focus background does not rely on properties of the structure on the image coordinate (such as brightness of sharpness), but only on the modulation on the detector coordinate -- namely, by the fingerprint. In other words, we make no prior assumptions about the structure of the specimen, and we leverage only the unique information provided by the detector array. Indeed, similar results would not be possible with low-pass filtering techniques (Supp. Fig. \ref{supfig:low_freq_sample}). 

Finally, we estimated the SNR enhancement achieved by \sism by comparing the reconstruction of multiple noisy realizations of the same ISM dataset.
This comparison   highlights how SNR dramatically increases with a few iterations of the algorithm (Supp. Fig. \ref{supfig:SNR_enhancement_charac}).
On the other hand, if the reconstruction is applied for too many iterations, noise amplification might happen, degrading the SNR. This result underlines once again the importance of an early stop.

\begin{figure*}[t!]
    \centering
    \includegraphics[width=\textwidth]{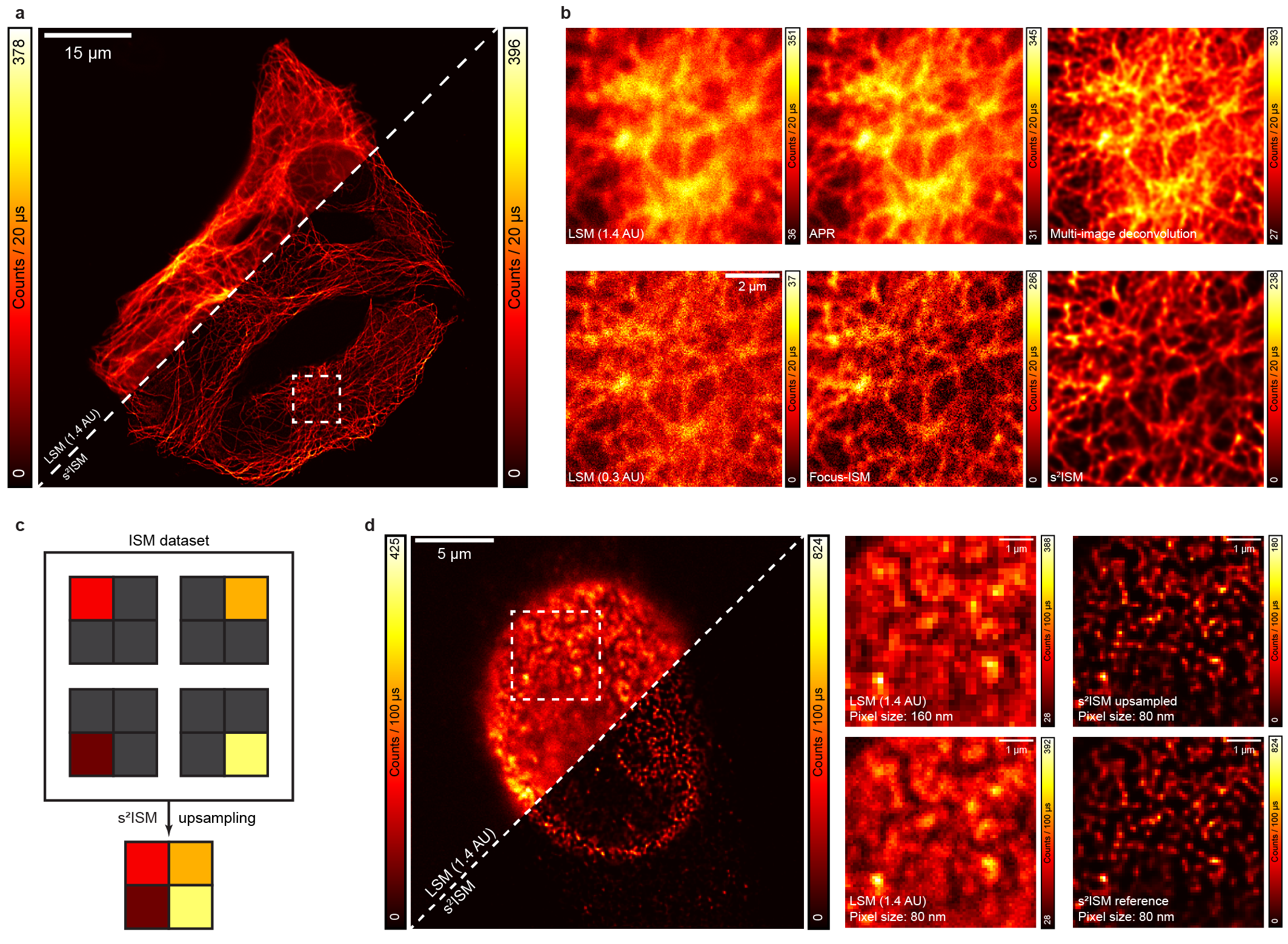}
    \caption{\textbf{Comparison of ISM reconstruction methods.} \textbf{a}, Confocal image (left) of the tubulin network of a HeLa cell (mouse anti-$\alpha$-tubulin combined with anti-mouse Abberior STAR RED) compared to the \sism reconstruction (right). \textbf{b}, Detail of the image in \textbf{a} (white dashed box) reconstructed using different algorithms. From left to right, lateral resolution and SNR are improved. From top to bottom, optical sectioning is improved. Both multi-image deconvolution and \sism algorithms are stopped at 20 iterations.
    \textbf{c}, Sketch of the upsampling working principle. The images of the ISM dataset are shifted, and the mutual redundancy can be used to fill the gaps and reconstruct an image on a finer grid than that generated by the acquisition process. \textbf{d}, Results of \sism reconstruction with and without upsampling at the same target pixel size of \SI{80}{nm}. The sample is an immunostained HeLa cell for nuclear pore complexes on the nucleus surface (rabbit anti-Nup-153 combined with anti-rabbit Abberior STAR 635P).
    }
    \label{fig:4+5}
\end{figure*}

To further validate our approach to ISM image reconstruction, we compared the performances of \sism with conventional reconstruction methods (Fig. \ref{fig:4+5}a,b and Supp. Fig. \ref{supfig:cell_B}). The simplest technique is the summation of the images of the ISM dataset, which leads to an image equivalent to that generated by a confocal microscope with a pinhole size the same as that of the detector array -- in our case, 1.4 Airy units (AU). Similarly, extracting only the image built by the central element of the detector array corresponds to generating a confocal image with a closed pinhole (0.3 AU). It is well known that closing the pinhole improves lateral resolution and optical sectioning but irremediably compromises the SNR. Conversely, by opening the pinhole, the photon collection efficiency improves, but the resolution is limited by diffraction, and optical sectioning deteriorates according to the size of the detector. 
The aforementioned reconstruction methods completely disregard the spatial information the SPAD array detector provided, and we use them only as a reference towards CLSM. The results should not be considered ISM images.
A smarter reconstruction method is adaptive pixel reassignment (APR), an algorithm that first estimates the shifts between the images of the raw datasets, later registers them and finally performs the summation \cite{Castello2019}. As a result, the image preserves the super-resolution enabled by the confocal effect while exploiting all the photons collected and achieving an excellent SNR. However, APR is designed assuming all the images of the ISM dataset are identical but shifted and rescaled. Such an assumption is only approximately correct and leads to sub-optimal reconstruction, even if always superior to CLSM.
A more rigorous approach to ISM reconstruction is multi-image deconvolution \cite{Zunino2023a}, which takes into account the unique structure of the PSFs of the imaging system. Although slightly more computationally expensive, multi-image deconvolution can produce better results than APR. However, APR and multi-image deconvolution cannot reject out-of-focus light. Thus, their feasibility is limited in the context of thin samples or using detector arrays with a limited size -- compromising SNR.
A solution to reject defocused light is focus-ISM \cite{Tortarolo2022}, which builds upon APR. After reassignment, each micro-image -- namely the distribution of light on the detector space -- is fitted to a two-component Gaussian mixture. The broadest distribution is interpreted as background and discarded, generating an optically-sectioned image pixel-by-pixel. While effective, focus-ISM stands on an approximated model of ISM image formation, and its local nature makes it very sensitive to noise. Indeed, each pixel is analyzed individually, and the algorithm cannot leverage the full information contained in the ISM dataset. Consequently, the reconstruction is sub-optimal and typically noisy. Furthermore, it is incompatible with multi-image deconvolution and cannot benefit from it.
Finally, \sism is the most comprehensive reconstruction algorithm designed for ISM, enabling all the benefits provided by the previously described methods without sacrificing any image feature. The result is an image with enhanced lateral resolution, optical sectioning, and improved SNR.
We carried out the comparison also in Fourier space by calculating the radial spectra of the images from Fig \ref{fig:4+5}b (Supp. Note \ref{subsec:A9} and Supp. Fig. \ref{supfig:radial_spectra}). The results confirm that \sism outperforms the conventional reconstruction methods. Indeed, the high-frequency contrast uniquely correlates with superior lateral resolution and suppression of low-frequency background. Furthermore, the spectrum of the reconstructed image indicates a lower noise level, confirming the improvement of the SNR.

Finally, we compared \sism with conventional 3D deconvolution, reconstructing a full volumetric dataset using a volumetric PSF. In this scenario, the axial information is explicitly contained in the images of the 3D stack, and each slice of the volume can be reconstructed with enhanced sectioning. However, such an approach requires the experimental acquisition of multiple images at different axial positions of the sample. While the acquired information is larger and enables a complete 3D reconstruction of the specimen, such a task is time-consuming and increases the light dose shined onto the sample. Instead, \sism requires only a planar dataset to achieve optical sectioning, the only difference being that out-of-focus light is discarded instead of reassigned to a different axial plane. Thus, \sism reconstructions are typically less bright than that obtained through 3D deconvolution. Nonetheless, our method correctly reconstructs the structure of the specimen (Supp. Fig. \ref{supfig:3D-ISM_vs_MID-3D}), enabling fast and gentle measurements -- especially suited for live cell imaging and time lapses.

As recently demonstrated \cite{Zunino2023a}, the ISM dataset contains enough redundancy to allow for a two-fold lateral upsampling. Indeed, the raw images of the ISM dataset are inherently shifted by a quantity -- the shift-vector -- which depends on the $\xd$ coordinate. 
If the pixel size $\Delta \xs$ is chosen to equal the pitch of the array detector $\Delta \xd$  projected in the sample plane, then the images $i(\xs|\xd)$ are mutually shifted by half the pixel size.
Since the coordinates between two pixels of an image are sampled by another image of the ISM dataset, fusing the images allows for doubling the number of scanned points in the final reconstruction (Fig. \ref{fig:4+5}c).
An important implication is that ISM enables super-resolution also in the digital sense.
Indeed, it is possible to recover an image with a pixel size respecting the Nyquist sampling criterion, even though this latter was not respected by the raw images.
In other words -- if the upsampling condition $\Delta \xs = \Delta \xd$ is respected -- ISM enables surpassing both Nyquist's and Abbe's limits. 

The \sism reconstruction method can also leverage the extra information encoded into the ISM dataset to achieve image upsampling without sacrificing the benefits demonstrated so far (Supp. Note \ref{subsec:A4}).
To demonstrate it, we acquired a dataset of nuclear pore complexes on the nuclear membrane of a HeLa cell (Fig. \ref{fig:4+5}d and Supp. Fig. \ref{supfig:nup}).
We used two different values of pixel size, \SI{160}{nm} and \SI{80}{nm}, the first not respecting the Nyquist criterion but respecting the upsampling criterion.
Then, we reconstructed an image using the \sism method on both datasets, reaching a target pixel size of \SI{80}{nm}. Despite having roughly one-fourth of the photon counts of the reference data, the upsampled image is correctly reconstructed, as demonstrated by the similarity with the reference image.
More quantitatively, the structural similarity index measure (SSIM) calculated on the upsampled and reference images reaches local values as high as 0.98, with a median value of 0.87 (Supp. Note \ref{subsec:A8} and Supp. Fig. \ref{supfig:SSIM}).
Thus, the \sism method enables high-fidelity reconstruction of undersampled raw data, paving the way for gentler and faster imaging.

\subsection{Versatility of \sism}

\begin{figure*}[t!]
    \centering
    \includegraphics[width=\textwidth]{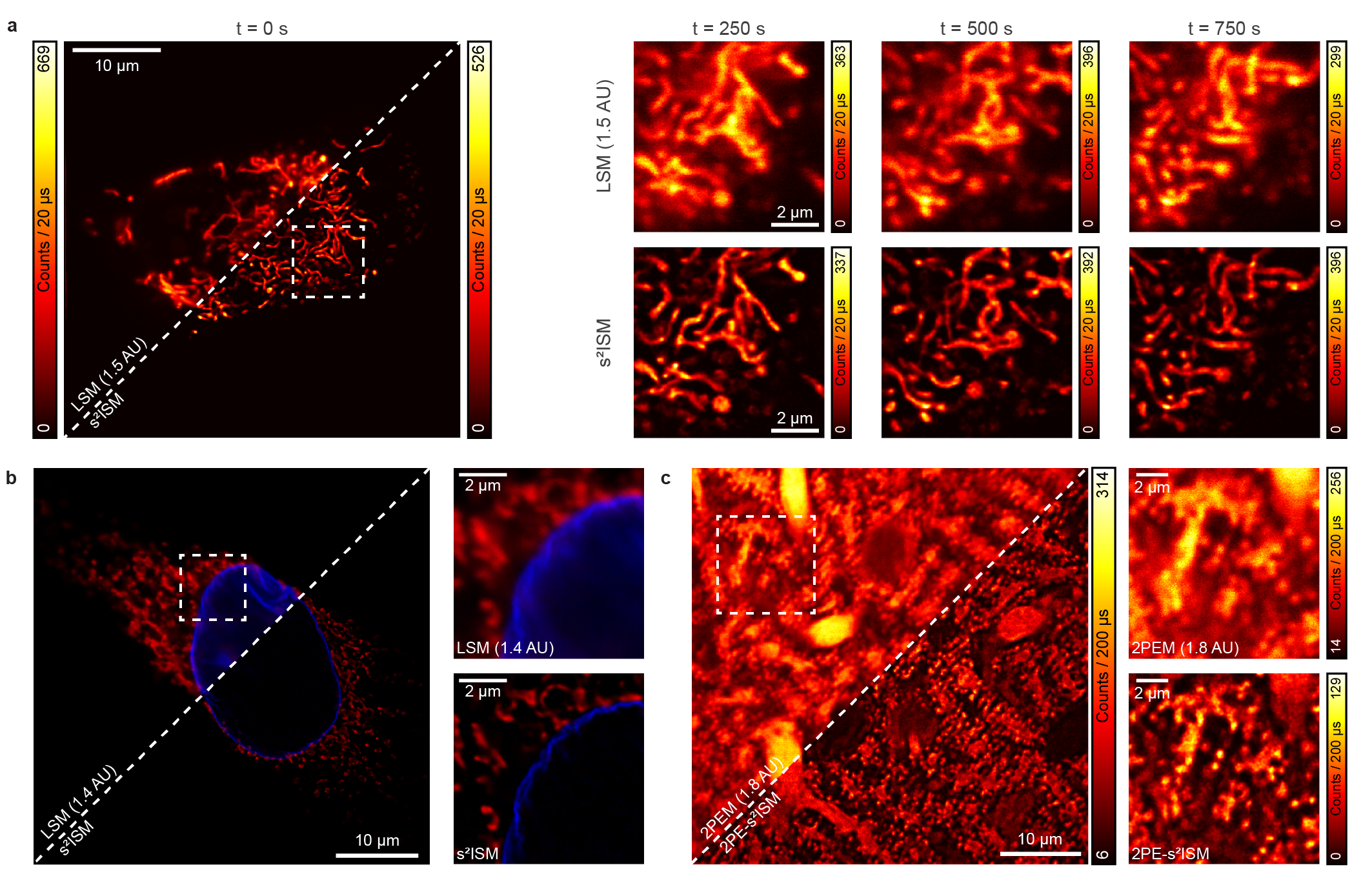}
    \caption{\textbf{Generalization of \sism.} \textbf{a}, time series of a live HeLa cell with stained mitochondria (Mito Tracker Orange). \textbf{b}, Multicolor imaging of a fixed HeLa cell with mitochondria and nuclear membrane immunostained with two different fluorophores (mouse anti-ATP Synthase combined with anti-Mouse Alexa647 and rabbit anti-lamin B1 combined with anti-rabbit Alexa488, respectively). \textbf{c}, two-photon excitation imaging of a cerebellum rat slice at a depth of roughly \SI{10}{\micro m}.}
    \label{fig:6}
\end{figure*}

As described in the previous sections, \sism requires only a single plane dataset, improves the SNR, and enables faster acquisition. Such benefits are especially useful in the context of live cell imaging over a long period of time. We demonstrate the feasibility of \sism on a time series of live mitochondria (Fig. \ref{fig:6}a) images. We performed the reconstruction frame-by-frame, obtaining a sequence of images with high resolution and optical sectioning. Our results demonstrate how high-quality continuous imaging of live cells can be performed at moderate pixel dwell times (Supp. movie).

Similarly, we extend \sism to multi-colour imaging by applying the reconstruction algorithm to each channel. More in detail, we modelled the image formation of each channel using the corresponding excitation and emission wavelengths. The results demonstrate that \sism can easily be applied to multi-channel datasets with a simple sequential reconstruction to obtain high-resolution and high optical sectioning multi-colour images (Fig. \ref{fig:6}b and Supp Fig. \ref{supfig:Multicolor}).

The benefits of \sism are also particularly useful while investigating thick samples whose images are likely to be corrupted by out-of-focus background, given the inherent 3D structure of the specimen. We demonstrate the advantage of \sism by acquiring a volumetric dataset of a cell (Supp. Fig. \ref{supfig:stack_cell}). Then, we moved to an even more challenging sample, namely a cleared rat cerebellum slice, which we imaged up to a depth of \SI{10}{\micro m} (Supp. Fig. \ref{supfig:Purkinje_stack}). Thanks to the improved optical sectioning, each slice of the 3D stack is free from defocused blur. At the same time, the size of the detector array ensures a good collection efficiency and a correspondingly good SNR.
To further push the capability to explore thick samples at depth, we demonstrate the feasibility of the combination of two-photon excitation (2PE) with \sism. Indeed, our reconstruction method is general and -- in principle -- can be used with any laser scanning microscope equipped with a detector array. More in detail, 2PE exploits near-infrared light -- which is less affected by scattering, compared to visible light -- and the non-linear excitation process further improves the optical sectioning of the microscope. We imaged the same rat cerebellum slice described previously, demonstrating an improvement in the image quality in the more general case of multi-photon excitation (Fig. \ref{fig:6}c and Supp. Fig. \ref{supfig:Purkinje}).

\subsection{\sism for Fluorescence Lifetime Imaging}

\begin{figure*}[t!]
    \centering
    \includegraphics[width=\textwidth]{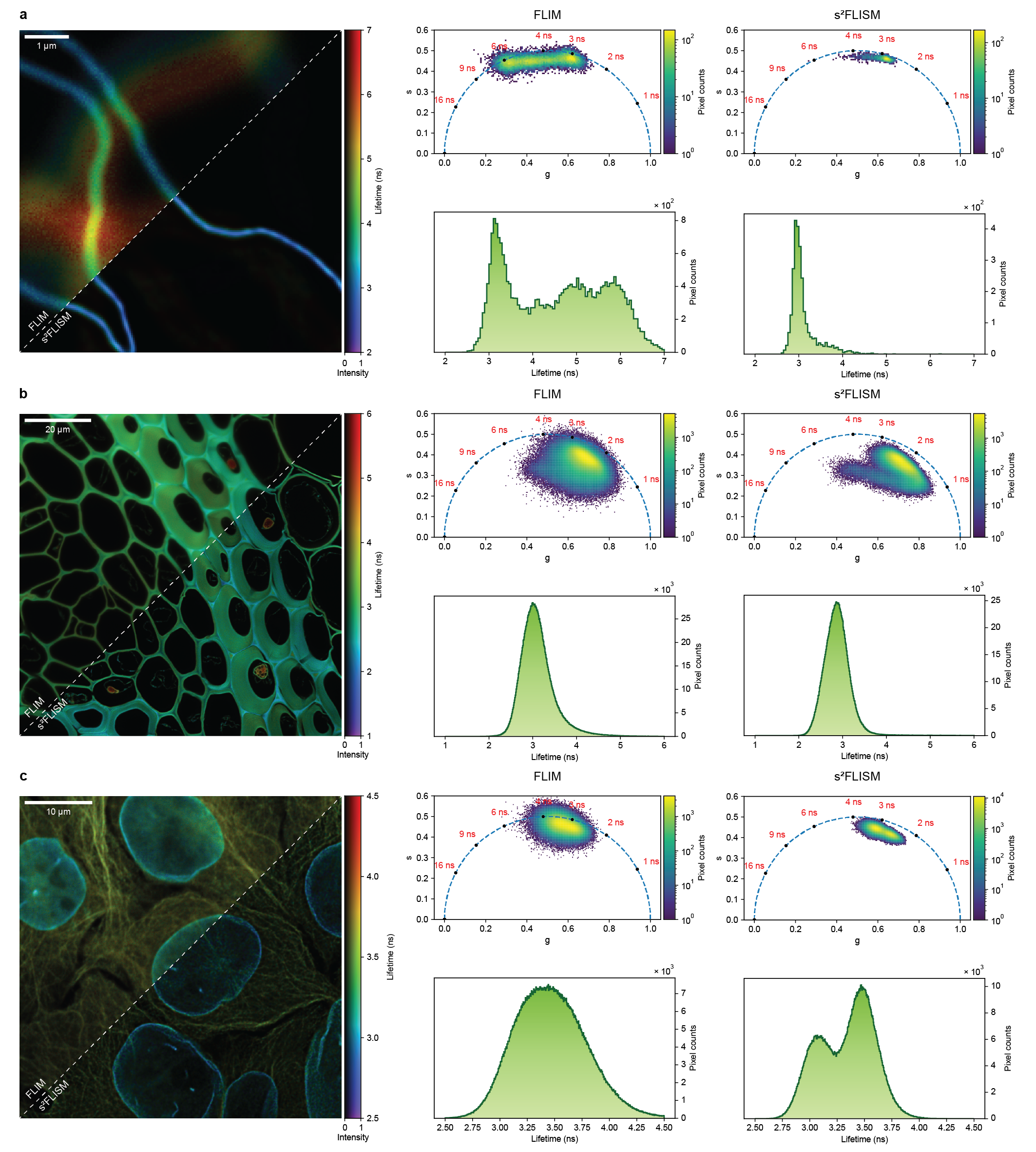}
    \caption{\textbf{Fluoresce lifetime imaging with \sism.} \textbf{a}, simulation of tubulin filaments with different lifetime values (NA = 1.4, $\lambda = \SI{640}{nm}$). The in-focus filaments have a lifetime value of $\tau = \SI{3}{ns}$, while the out-of-focus ($z = \SI{720}{nm}$) filaments have a lifetime value of either $\tau = \SI{3}{ns}$ or $\tau = \SI{6}{ns}$. \textbf{b}, experimental image of a rhizome of Convallaria Majalis stained with acridine orange, excited with $\lambda = \SI{488}{nm}$. \textbf{c}, experimental image of HeLa cells with tubulin stained with STAR RED ($\tau = \SI{3.4}{ns}$) and lamin-A on the nuclear membrane stained with STAR 635 ($\tau = \SI{2.8}{ns}$). Both fluorophores are excited with the same source at $\lambda = \SI{640}{nm}$. The intensity of each image is normalized to its maximum. The phasor plots and the histograms are thresholded at 5\% of the maximum intensity of the corresponding image. Lifetime values are calculated from the magnitude of the phasors \cite{Digman2008}.}
    \label{fig:7}
\end{figure*}

Finally, we extend the capabilities of \sism also into the time domain, demonstrating how the additional spatial information provided by the array detection can help in better estimating the lifetime of fluorescence emission. Indeed, if structures located at different axial planes of the specimen are characterized by different lifetime values, the out-of-focus light hinders the correct estimation of the lifetime of the in-focus structure.
We depict such a scenario by simulating filaments with a lifetime value of $\tau=\SI{3}{ns}$ (in-focus) and a combination of $\tau=\SI{3}{ns}$ and $\tau=\SI{6}{ns}$ (out-of-focus) (Fig. \ref{fig:7}a). Using the phasor approach \cite{Digman2008}, we calculated a lifetime value for each pixel. It is noticeable in the confocal FLIM image that where the out-of-focus light superimposes with the in-focus structure, the lifetime value cannot be correctly estimated. However, the array detector encodes the axial information, as described previously. Indeed, the fractional components of the emitters in a thick sample depend on the $\xd$ coordinate, following the fingerprint distribution. Such information can be exploited to remove the defocused decays from the temporal dimension.
Thus, we generalized the \sism algorithm to also take into account the temporal dimension in our data. Briefly, we extended Eq. \ref{eq:forward} by adding a temporal dimension in the data -- which contains the fluorescence dynamics -- and in the PSFs -- which contains the impulse response functions (IRF) of each sensitive element of the SPAD array. Thus, the convolution is now understood to be with respect to the scanning and temporal coordinates (see Supp. Note \ref{subsec:A5}). A key point of our generalization is that we do not impose an exponential model, paving the way for broader applications where the fluorescence decay follows more complex dynamics -- such as time-resolved STED microscopy \cite{Lanzano2015, Tortarolo2019}.
We validated the generalization of \sism to fluorescence lifetime image scanning microscopy (FLISM) on the simulations shown in Fig. \ref{fig:7}a. In the reconstructed image, the out-of-focus filaments have a highly suppressed intensity, enabling a more robust estimation of the lifetime of the in-focus fluorescence. Indeed, the phasor plot and the lifetime histogram of the raw FLIM image depict multiple components, while only a single component remains after the \sism reconstruction.
Importantly, removing out-of-focus light corresponding to a different lifetime value using phasor decomposition \cite{Tortarolo2024} or related techniques would have required prior knowledge of the lifetime distribution of the sample, which is not needed with our method. Furthermore, \sism is also capable of removing background fluorescence with an identical lifetime to that stemming from the focal plane, a task that would be infeasible with the sole measurement of the decay without array detection.
We experimentally validated our approach by collecting the complete fluorescence decay for each scan point and each detector element using a multi-channel digital frequency domain (DFD) acquisition scheme \cite{Tortarolo2024}. We acquired the dataset of a rhizome of Convallaria Majalis stained with a single fluorophore, whose lifetime is highly sensitive to the environment. We compared the conventional FLIM and \sflism reconstruction. As expected, this latter shows all the previously demonstrated benefits -- such as improved resolution and optical sectioning -- and also enables a more robust phasor analysis and estimation of the
lifetime. Indeed, the phasor plot and lifetime histogram of Fig. \ref{fig:7}b have a narrower spread, indicating a more accurate analysis of the fluorescence dynamics due to both the improved SNR and the reduced axial cross-talk.
Such benefits are of particular interest when different structures are labeled with fluorophores with different lifetime. In such a scenario, the correct lifetime estimation is of paramount importance for image segmentation or related tasks. To demonstrate the benefits of \sflism in this context, we imaged a sample of HeLa cells with tubulin filaments and nuclei stained with two fluorophore with the same absorption spectrum, but different lifetime values. Therefore, a time-resolved measurement enables multi-species imaging with a single excitation laser and a single detector, as long as the different lifetime values can be correctly discriminated. As shown in Fig. \ref{fig:7}c, in the conventional FLIM image it is possible to distinguish two different populations of emitters. However, the phasor and lifetime histograms are broad and individual distributions cannot be resolved. With the \sflism reconstruction, the two populations are clearly identified both in the phasor plot and in the lifetime histogram. Therefore, segmenting the image becomes an easier task thanks to \sflism (Supp. Fig. \ref{supfig:phasor_segmentation}).

\section{Discussion}
\label{sec:discussion}

The ISM dataset is rich with information that can be exploited to improve the optical quality of the image and to access additional knowledge on the specimen.
This information significantly increases when the ISM implementation is based on an asynchronous readout SPAD array detector, where the detector's single-photon sensitivity and photon-timing resolution bring photon-resolved microscopy to reality \cite{Rossetta2022}: The fluorescent photons emitted by the sample are registered one-by-one, along with a series of detailed spatial and temporal signatures typically discarded in conventional microscopy. 
Nonetheless, decoding the information encoded in the raw data is a delicate task that requires tailored tools.
Existing image reconstruction algorithms are effective in achieving specific goals, such as improving the lateral resolution, but fail to get all the theoretically predicted benefits.
Thus, previous approaches required the application of multiple algorithms sequentially, which is a sub-optimal strategy and not always feasible.
Instead, \sism is the first algorithm capable of achieving all the improvements of array detection in a comprehensive algorithm without compromises or drawbacks.
Indeed, \sism utilizes the ISM dataset to generate a single super-resolution image surpassing both Abbe's and Nyquist's limits and with enhanced optical sectioning.

As demonstrated in this work, \sism is a general concept which can be applied to any laser scanning microscope equipped with a detector array.
Indeed, we proved its feasibility in the context of 2PE fluorescence microscopy for tissue imaging.
Nonetheless, other non-linear effects can be combined with \sism, such as fluorescence saturation in saturated excitation ISM \cite{Temma2022} to further improve spatial resolution.
Similarly, single-molecule ISM \cite{Radmacher2023} will also benefit from \sism, improving localization uncertainty thanks to the combined improvement in resolution and reduced background.
Such benefits synergistically translate to any additional dimension contained in the dataset.
Thanks to their sub-nanosecond temporal resolution, SPAD array detector naturally extend the data to the temporal dimension.
We leveraged this latter to demonstrate how \sism greatly improves the robustness of fluorescence lifetime measurements
Nonetheless, equivalent benefits are to be expected for hyperspectral imaging \cite{Strasser2019}.
The temporal dimension could also be exploited to measure other fluorescence dynamics.
Indeed, the current version of \sism does not require a specific fluorescence dynamics model and can be applied over a wide range of scenarios, such as intensity fluctuations induced by the transition of the marker to dark states \cite{Sroda2020} or anti-bunching from single-photon emitters \cite{Tenne2019}.
%
Another intriguing extension of \sism lies in stimulated emission depletion (STED) microscopy  \cite{Vicidomini2018}, particularly in time-resolved STED microscopy \cite{Vicidomini2011, Vicidomini2013}.
In the latter case, the fluorescence lifetime signature could be used to further improve the spatial resolution thanks to the separation-by-lifetime tuning concept \cite{Lanzano2015, Tortarolo2022}.
%
%
By properly modelling the multi-spot excitation of camera-based implementations of ISM, the \sism algorithm could be generalized to such architectures, enabling optical sectioning without the need for physical or virtual pinholing \cite{York2012}.
With some caution, the concept of \sism could also be extended to coherent variations of ISM.
One approach is to modify the forward model to account for the coherent image formation process, but it is unclear if such a modality would grant the same benefits as in the incoherent case (e.g., fluorescence, spontaneous Raman and Brillouin scattering, photothermal imaging).
An alternative approach is to adjust the microscope to enable interferometric detection \cite{Raanan2022, Zhitnitsky2024}.
In this case, \sism can be used simply by replacing intensities with fields.

Despite the promising potentialities of \sism, one should carefully evaluate when to stop the iterations to avoid noise-amplification -- a well-known side effect of maximizing likelihood functionals.
This limitation could be relaxed by adding explicit regularization terms, which recent works have already demonstrated to drastically improve the quality of reconstructions \cite{Zhao2022}.
However, explicit regularizers require prior information on the specimen's structure, which might not always be available.
A more general approach would be the combination of \sism with a local stop criterion \cite{gradient_consensus} or with denoisers, such as Noise2Noise \cite{Lehtinen2018, Qu2024} and related methods \cite{Qiao2024}.
Indeed, the excellent temporal resolution of the SPAD array detector easily enables the measurement of multiple realizations of the same dataset in a single experiment, paving the way for a smooth integration of statistical denoisers with our reconstruction technique.
Finally, we designed \sism assuming that the microscope suffers negligible optical aberrations. However, our algorithm can also be used on each isoplanatic patch as long as the correct wavefront can be measured and used to calculate the corresponding PSFs.
Indeed, the combination of wavefront sensing techniques \cite{Ancora2021} with \sism could dramatically extend its range of usability, pushing the imaging depth.

In conclusion, \sism increases the capabilities of any scanning microscope equipped with a detector array by extracting the complete information encoded into the raw data.
All the benefits of our reconstruction method -- super-resolution, enhanced optical sectioning, and faster sampling -- are available without changing the microscope architecture or compromising any of the features of the ISM microscope.
The potential of \sism further increases when the detector is a SPAD array, fully exploring the perspectives opened by the photon-resolved microscopy paradigm.
Therefore, we are positive that \sism will be widely adopted by the community of microscopy users as it is and by developers as a starting point for new advanced reconstruction methods.

\section{Methods}
\label{sec:methods}

\subsection{Microscope Architecture}

For this work, we built a custom ISM setup (Suppl. Fig. \ref{supfig:setup}). The excitation beams are provided by three triggerable pulsed (\SI{80}{ps} pulse-width) diode lasers emitting at \SI{640}{nm}, \SI{561}{nm}, and \SI{488}{nm} (LDH-D-C-640, LDH-D-C-560, and LDH-D-C-488 -- Picoquant).
We control the coarse power of the visible laser using their respective drivers and control software. We performed the fine control of the power using acoustic optical modulators (AOM, MT80-A1-VIS, AAopto-electronic).
All laser beams are coupled into a different polarising-maintain fibre (PMF) to transport the beams to the microscope. In all cases, we used a half-wave plate (HWP) to adjust the beam polarization parallel to the fast axis of the PMF.
The beam for two-photon excitation is provided by a tunable ultrafast laser (Chameleon Vision -- Coherent), emitting at \SI{900}{nm} (\SI{140}{fs} pulse-width). The power is controlled with HWP and a polarizing beam-splitter (PBS), which redirects a fraction of the light onto a beam dump, depending on the rotation angle of the HWP. The beam is magnified by a factor of 3 using a telescope.
A set of dichroic mirrors (491 short-pass, 590 short-pass, 750 short-pass) allows the combination of all laser beams. The excitation and fluorescence light are separated by a different dichroic mirror (multi-reflection band 488-560-640-775 or 720 short-pass), depending on the excitation modality (one or two photons, respectively).
Two galvanometer scanning mirrors (6215HM40B, CT Cambridge Technology), a scan lens and a tube lens -- of a commercial confocal microscope (C2, Nikon) -- deflect and direct all the beam towards the objective lens (CFI Plan Apo VC 60$\times$, 1.4 NA, Oil, Nikon) to perform the raster scan on the specimen. 
The objective lens is mounted over a nanopositioner (FOC.100, Piezoconcept), enabling z-scanning.
The fluorescence light is collected by the same objective lens, de-scanned, and sent towards the detection path. This latter consists of a set of lenses to form a telescopic system that conjugates the sample plane onto the detector plane with an overall magnification of 450$\times$.
Spectral filters are installed in the detection path to discard residual excitation light. Depending on the experiments, fluorescence light is selected by using a dedicated set of filters
(red set: ZET633TopNotch and ET685/70M,
green set: ZET561NF and ET575LP,
blue set: ZET488NF and ET525/50M,
two-photon set: 720SP and ET525/50M).
The detector is a $7 \times 7$ SPAD array (PRISM-light kit, TTL version, Genoa Instruments) with a pixel pitch of \SI{75}{\micro m}, but only the inner $5 \times 5$ array is read due to limitations of the read-out system used in this work. Every photon detected by any element of the SPAD array generates a TTL signal that is delivered through a dedicated channel to a multifunction FPGA-based I/O device (NI USB-7856R from National Instruments), which acts both as a data-acquisition system and a control unit. The BrightEyes-MCS software \cite{mcs} controls the entire microscope, including the galvanometric mirrors, the FOC, and the AOMs. The software also provides real-time image visualisation during the scan and saves the raw data in a hierarchical data format (HDF5) file.
The saved file contains metadata as a dictionary and data as a six-dimensional array (repetition, axial position, vertical position, horizontal position, time, detector channel).

\subsection{Digital Frequency Domain}

We used the multi-channel digital frequency domain (DFD) method \cite{Tortarolo2024} to measure the fluorescence decay at each scan point and for each detector array element.
The DFD strategy enables the acquisition of periodic signals with a timing precision superior to direct sampling through a heterodyne measurement.
Laser pulses are emitted at frequency $f_{exc}$, and the fluorescence is sampled at frequency $f_s$. Those frequencies are slightly detuned, $kf_{exc} = (k-1) f_{s}$.
Therefore, the sampling accumulates a delay with every cycle, resulting in a sliding window that spans over $k-1$ excitation periods until the two signals are back in phase.
Each period is more finely sampled in $n$ shorter windows of duration $T_w$ by a frequency $f_w = nf_s$.
Defining the counters $w \in [0, n)$ and $\varphi \in [0, k)$, we reconstruct the time index $\gamma$ as follows
\begin{equation}
      \gamma = \qty( mw - \varphi )  \bmod k
      \label{eq:dfd}
\end{equation}
where and $m = \frac{k - 1}{n}$, and $k \in \mathbb{N}$, $n \in \mathbb{N}$ such that $m \in \mathbb{N}$.
Finally, the photon arrival time is given by $t =  \gamma T_{exc} / k$.

In our implementation on an FPGA board (NI USB-7856R, National Instruments), we used a base clock of $f_0 = \SI{40}{MHz}$. From this latter, we derived the frequencies 
$f_{\exc} = \frac{28}{27} f_0 = \SI{41.48}{MHz}$, 
$f_s = \frac{21}{20} f_0 = \SI{42}{MHz}$, 
$f_w = \frac{21}{2} f_0 = \SI{420}{MHz}$.
The DFD parameters are $n = 10$, $k = 81$, and $m = 8$. Therefore, we obtain a timing precision of $\Delta t = T_{\exc}/k = \SI{298}{ps}$ and a temporal resolution of $T_w = \SI{2.38}{ns}$.

The DFD system builds the fluorescence decay histogram using Eq. \ref{eq:dfd} for each detector coordinate $\xd$. Additionally, the system returns an extra channel that samples the laser trigger signal. This latter is used as a reference to align different measurements to a common reference frame, whose origin is the instant of emission of the laser pulse.

\subsection{Numerical simulations}

We simulated the phantoms and the PSFs of the ISM microscope using the open-source Python package BrightEyes-ISM \cite{Zunino2023b}, based on the vectorial diffraction model provided by the package pyFocus \cite{Caprile2022}.
We modelled the SPAD array detector as a $5 \times 5$ array of pinholes with a pitch of \SI{75}{\micro m} and an individual size of \SI{50}{\micro m}.
For all simulations, we set the magnification to $M = 450$, the numerical aperture to $\NA = 1.4$, and the refractive index to match the one of the immersion oil $n = 1.51$.
We generate the synthetic datasets using an excitation and emission wavelength of \SI{640}{\nano m} and \SI{660}{\nano m}, respectively.
Finally, we applied Poisson noise to the generated images.
In all cases, we assumed that we illuminated the back aperture of the objective lens with a uniform plane wave -- namely, we did not apply any aberrations.
We modelled the IRFs of the time-resolved synthetic data using a rectangular window smoothed by a Gaussian kernel ($w = \SI{2}{ns}$ and $\sigma = \SI{0.3}{ns}$). We simulated the fluorescence decay as a single exponential. We used 81 time bins separated by $\Delta t = \SI{298}{ps}$ to match our DFD acquisition system.
To better distinguish the simulations from the experimental data, we present them using the magma and hot colormap, respectively.


\subsection{Image processing}

\subsubsection{Sum}

For each ISM dataset, we generated the corresponding confocal image by summing all the raw images
\begin{equation}
    i_s(\xs)= \sum_{x_d} i(\xs|\xd) 
\end{equation}
The result is equivalent to a confocal image acquired with a pinhole as large as the detector array.

\subsubsection{Adaptive Pixel Reassignment}
We calculated the shift-vectors of an ISM dataset as
\begin{equation}
    \boldsymbol{\mu}(\xd)= \argmax_{\xs} \{ \mathcal{R}(\xs|\xd) \}
\end{equation}
where $\mathcal{R}$ is the phase correlation of the raw images to the central one
\begin{equation}
    \mathcal{R}(\xs|\xd)= \mathcal{F}^{-1} \left\{ \frac{\mathcal{F}\{i(\xs|\xd)\} \overline{\mathcal{F}\{i(\xs|\bold{0})\}}}{|\mathcal{F}\{i(\xs|\xd)\}\overline{\mathcal{F}\{i(\xs|\bold{0})\}}}  \right\}
\end{equation}
The APR reconstruction \cite{Castello2019} is calculated as the sum of the aligned images
\begin{equation}
    i_{APR}(\xs) = \sum_{x_d}i(\xs + \boldsymbol{\mu}(\xd)|\xd)
\end{equation}

\subsubsection{Focus ISM}
The focus-ISM algorithm \cite{Tortarolo2022} exploits the APR approach to register the images of the ISM dataset. Instead of summing the result, the algorithm fits each reassigned and normalized micro-image to the following two-components Gaussian mixture model
\begin{equation}
    i(\xd) = \alpha \cdot g(\xd|\bold{0},\sigma_{sig})+ (1 - \alpha) \cdot g(\xd|\bold{0},\sigma_{bkg})
\end{equation}
where $g(\xd|\bold{0},\sigma)$ is a centred Gaussian function and $\sigma_{sig}$ is kept fixed following calibration procedure.
Finally, the map of weights $\alpha(\xs)$ is applied to the APR image to remove the background.

\subsubsection{\sism}

The first step is to simulate the PSFs of the microscope. To this end, we assume that aberrations are absent or negligible. Then, we need to estimate the orientation and rotation of the detector array, the magnification of the system, and the position of the out of-focus-plane.
We retrieve the first three parameters from the shift-vectors of the ISM dataset, using a minimization procedure explained in detail in the Supplementary Note \ref{subsec:A6}. The out-of-focus position $z_2$ is found as the one that maximizes the difference among the in-focus ($z_1 = 0$) and defocused PSFs. Normalizing each set of PSFs to the flux at the corresponding axial plane we remove the total intensity as a possible source of discrepancy. Thus, we maximize the difference of the spatial structure on the $\xd$ and $\xs$ coordinates
\begin{equation}
\label{eq:bkg_as_kl_minimized}
    z_2 = \argmax_z  D\qty[ h(\xs, 0|\xd) \parallel h(\xs,z|\xd) ]
\end{equation}
where $D(\cdot \parallel \cdot)$ is a discrepancy measure. Since the PSFs are symmetric along the z-axis, we explore only the positive axial range. We sample up to the depth of field in 40 steps.
In this work we used the Kullback-Leibler divergence or the negative Pearson correlation, which led to the same result (Supp. Fig. \ref{supfig:conditioning}). Therefore, they can be used interchangeably.
Other parameters required to simulate the PSFs are easily found from the design of the experiments (excitation wavelength, numerical aperture, etc.).
Once all the required parameters are known, we numerically simulated the PSFs using the BrightEyes-ISM Python package \cite{Zunino2023b}. We set the size of the simulation box adaptively to contain the full PSFs at every axial plane.
We estimated the temporal IRFs $h(t, \xd)$ as the average of multiple ($\sim 10^4$) experimental recordings of the scattering of a gold bead. Finally, we combined the PSFs with the IRFs as follows
\begin{equation}
    h_k(t, \xs | \xd) = h_k(\xs | \xd) \cdot h(t | \xd)
\end{equation}
and normalized them such that
\begin{equation}
    \sum_{t} \sum_{\xd} \sum_{\xs} h_k(t, \xs, \xd) = 1
\end{equation}
where $k \in \qty{1,2}$ is the depth index.

The last step consists of the image reconstruction, which is carried out by applying the following iterative rule
\begin{equation}
    o_{k}^{(m+1)}(t,\xs) = o_{k}^{(m)}(t,\xs)
    \sum_{\xd} 
    h_k(-t, -\xs|\xd) * 
    \frac{i(t,\xs|\xd)}
    {\sum_{k} o^{(m)}_k(t, \xs)*h_k(t,\xs|\xd)}
    \label{eq:fullsism}
\end{equation}
where $m$ is the iteration index and $*$ is the convolution operator with respect to the coordinates $t$ and $\xs$. The out-of-focus reconstruction $o_2$ is discarded and the in-focus image $o_1$ is the final result. We stop the algorithm at an arbitrary number of iterations. To avoid noise amplification, we iterate at most 20 times the reconstruction of experimental images.
Using a single axial plane and no temporal dimension, the algorithm corresponds to multi-image deconvolution \cite{Zunino2023a}.

The computation time is in the order of 1 second per iteration for a $2000\times2000\times25$ dataset on a computer equipped with a
8-core CPU (\SI{3.6}{GHz}), 32 GB of RAM, and a GPU with 8 GB of dedicated RAM.

\subsubsection{Phasor calibration and analysis}

The signal $f(t)$ acquired by the DFD system for each scan and detector coordinate is

\begin{equation}
    f(t) = d(t) * h(t)
\end{equation}
where $d(t)$ is the fluorescence decay and $h(t)$ is the impulse response function (IRF).
The corresponding phasor $F$ is given by the complex conjugate of Fourier Transform, evaluated at $\omega = 2 \pi f_{\exc}$. Using the convolution theorem and writing the result in exponential form, we have
\begin{equation}
    F = m_Fe^{i\phi_F} = \qty(m_D \cdot m_H)\exp\qty[ i(\phi_D + \phi_H) ]
\end{equation}
where $m$ and $\phi$ are the magnitude and phase of the corresponding phasor, respectively.
The phasor of the IRF $m_H e^{i\phi_H}$ can be estimated indirectly from a sample with a known decay or directly by measuring an almost-instantaneous response, such as from a quenched fluorophore or a reflective sample. We preferred the latter strategy, being less error-prone. We used the back-scattering signal from a gold bead, which provides a very good signal-to-noise ratio. However, we had to measure the signal without spectral filters, having the same wavelength as the excitation laser. Thus, we cleaned the signal from multiple reflections by multiplying the IRFs with a rectangular window centred at the centroid of the IRF and with a length of \SI{4}{ns}. Finally, we used the DFD acquisition system's extra channel (laser trigger) to set a common reference frame for all the measurements. In practice, we subtract the phase $\varphi$ of the reference channel to the phase of the corresponding phasor. Finally, the complete calibration procedure is:

\begin{align}
\phi_D(\xs, \xd) &= \phi_F(\xs, \xd) - \phi_H(\xd) + \varphi_H - \varphi_F \\
m_{D}(\xs, \xd) &= m_F(\xs,\xd) /  m_H(\xd),
\end{align}
The above procedure is a Wiener deconvolution performed in frequency space. The \sism algorithm inherently compensates for the effect of the IRF by performing a temporal deconvolution.
Thus, when calculating the phasor from the decays reconstructed by the \sism method, there is no need to consider again the effect IRF and the calibration is performed only by correcting the phase with the reference channel.
Furthermore, the result of \sism is a single image, and the dependency from $\xd$ is lost.
To calculate the phasor of the corresponding CLSM image, we also removed the dependency from $\xd$ by temporally aligning and summing the decays and the IRFs. Then, we analyzed the result as in a single-channel scenario.

Once calculated the magnitude and phase for each pixel, we can estimate the lifetime map using the following equations \cite{Digman2008}, derived using the assumption that the fluorescence dynamics consists of a single exponential decay:
 \begin{equation}
     \tau_{\phi} = \frac{1}{ 2 \pi f_{\exc}} \tan\qty(\phi_D) \quad \quad \quad \tau_m = \frac{1}{ 2 \pi f_{\exc}}\sqrt{\frac{1}{m_D^2}-1}
 \end{equation}
If the above assumption holds true, the two estimates should match. In practice, the lifetime estimated using the phasor's magnitude is more robust and less sensitive to miscalibrations. Therefore, we used this latter to calculate the lifetime values for this work.

\subsection{Samples}

\subsubsection{Argolight calibration slide}

We used the Argo-SIM v1 slide (Argolight). In detail, we imaged two patterns \cite{argolight}. The first is the resolution target of gradually spaced lines rotated by \SI{45}{\degree}. The second is a 3D crossing stair axially spaced by \SI{250}{nm}.


\subsubsection{Cell culture}

We cultured HeLa cells in Dulbecco’s Modified Eagle Medium (DMEM, Gibco, ThermoFisher Scientific) supplemented with 10\% fetal bovine serum (Sigma-Aldrich) and 1\% penicillin/streptomycin (Sigma-Aldrich) at 37°C in 5\% CO$_2$.
The day before the staining, we seeded HeLa cells on coverslips in a 12-well plate (Corning Inc., Corning, NY) for immunostaining or a $\mu$-Slide eight-well plate (Ibidi, Grafelfing, Germany) for live-cells imaging.

\paragraph*{Fixed cells}

HeLa cells were fixed with either ice methanol, when cytoskeletal proteins were imaged, for 20 minutes at -20°C, or with a solution of 3.7\% paraformaldehyde (Sigma-Aldrich) in phosphate-buffered saline (PBS, Gibco™, ThermoFisher Scientific) buffer for 15 min at room temperature. 
Cells were washed three times with PBS buffer and treated with blocking buffer (5\% bovine serum albumin (BSA, Sigma-Aldrich) supplemented with 0.2\% Triton X-100 in PBS buffer) for 1 hour at room temperature.
Cells were incubated with primary antibodies diluted in the blocking buffer for 1 hour at room temperature.
The primary antibodies used in this study were: monoclonal mouse anti-$\alpha$-tubulin antibody (1:1000, Sigma-Aldrich), rabbit polyclonal anti-lamin B1 antibody (Abcam, ab16048, 1:500), rabbit polyclonal Nup-153 antibody (Abcam, ab84872, 1:500) and mouse monoclonal anti-ATP Synthase $\beta$ antibody (Sigma, A9728, 1:250).
After incubation with the antibody, cells were washed three times with blocking buffer and incubated with a secondary antibody diluted into blocking buffer for 1 hour at room temperature. The secondary antibodies used in this study were: Anti-Mouse IgG-Abberior STAR Red (Abberior, 1:1000), Anti-Mouse IgG-Abberior STAR 635P (Abberior, 1:1000), anti-Rabbit IgG Alexa-488 (ThermoFisher Scientific, 1:1000), Anti-Mouse IgG-Alexa647 (ThermoFisher Scientific, 1:500).
We rinsed HeLa cells three times in PBS for 15 min. Finally, we mounted the coverslips onto microscope slides (Avantor, VWR International) with ProLong Diamond Antifade Mountant (Invitrogen, ThermoFisher Scientific).

\paragraph*{Live cells}

For the mitochondrial staining in living cells, seeded HeLa cells were incubated with MitoTracker\textsuperscript{\texttrademark} Orange (ThermoFisher Scientific) at a concentration of 100 nM in DMEM supplemented with 10\% fetal bovine serum and 1\% penicillin/streptomycin for 10 minutes at 37°C in 5\% CO$_2$.After the incubation, cells were washed three times with PBS and placed in live cells imaging solution (LICS, ThermoFisher Scientific) immediately before the measurement.  
 
\section{Statements}

\subsection{Acknowledgments}
The authors thank Ryu Nakamura (Nikon Instruments) for providing the cerebellum slice sample, Agnieszka Pierzyńska-Mach for providing the sample for lifetime-based multispecies imaging, Federico Benvenuto (University of Genoa), Max Tillman (PicoQuant), Francesco Fersini, and Eli Slenders (Istituto Italiano di Tecnologia, Genoa, Italy) for providing useful insights. Finally, we thank the Genoa Instruments team for their support.

\subsection{Funding}
This project has received funding from: the European Research Council, \textit{BrightEyes}, ERC-CoG No. 818699 (G.V.); the European Union - Next Generation EU, PNRR MUR - M4C2 – Action 1.4 - Call "Potenziamento strutture di ricerca e creazione di "campioni nazionali di R\&S" (CUP J33C22001130001), \textit{National Center for Gene Therapy and Drugsbased on RNA Technology} No. CN00000041 (M.D. and G.V.); the Fondazione San Paolo, \textit{Augmented fluorescence correlation spectroscopy with a novel SPAD array detector to observe complex biological processes in living cells}, Trapezio No. 71100 (E.P.).

\subsection{Disclosures}
G.V. has a personal financial interest (co-founder) in Genoa Instruments, Italy.

\subsection{Authors' contributions}
A.Z. conceived the idea and designed the study. A.Z. developed the theoretical model. A.Z. and G.G. developed the code for the data analysis. M.D. developed the microscope control software and firmware. A.Z. built the image scanning microscope and performed the experiments. S.Z. and E.P. prepared the biological samples. A.Z., G.G., and G.V. wrote the manuscript. G.V. supervised the project. All authors discussed the results and commented on the manuscript.

\subsection{Code and Data Availability}
The Python code used to perform image processing is available at the following GitHub repository:
\url{https://github.com/VicidominiLab/s2ISM}.

The experimental data generated for this study are available at the following Zenodo database:
\url{https://doi.org/10.5281/zenodo.11284051}.

\bibliography{references}
\addtocounter{section}{1}


\clearpage
\begin{appendices}

\section{Supplementary notes}
\label{sec:A}

\subsection{Lateral resolution and optical sectioning in ISM}
\label{subsec:A1}

Image scanning microscopy (ISM) can be regarded as a special case of sequential structured illumination microscopy. Indeed, for each scan point $\xs$, we are exciting the sample only on the diffraction-limited region described by the excitation PSF $h_{\exc}$. The resulting image recorded by the detector array is
\begin{equation}
    i(\xd | \xs) \eval_{z_d = 0} = \qty[ o(\xd - \xs) \cdot h_{\exc}(\xd) ] * h_{\emi}(\xd) \eval_{z_d = 0}
\end{equation}
where $\xd = (x_d, y_d, z_d)$, $\xs = (x_s, y_s, z_s)$, and the symbol $*$ represents the 3D convolution with respect the the $\xd$ coordinate. The corresponding Fourier transform is
\begin{equation}
    I(\kd | \xs) = \qty{ \qty[ e^{i\kd\cdot\xs}O(\kd) ] * H_{\exc}(\kd) } \cdot H_{\emi}(\kd)
    \label{eq:fourier_image}
\end{equation}
where $\kd = (k_x, k_y, k_z)$, and the capital letters represent the Fourier transform of the corresponding quantity in real space.
The excitation 3D optical transfer function (OTF) contains multiple spatial frequencies, up to following lateral and axial cut-off frequencies
\begin{equation}
    k^\text{max}_{xy} = \frac{2n\sin \alpha}{\lambda} \qquad k^\text{max}_{z} = \frac{2n\sin ^2 \alpha/2}{\lambda}
\end{equation}
where $\alpha$ is the semi-angular aperture of the objective lens.
Thus, the excitation OTF can be rewritten using the following identity
\begin{equation}
    H_{\exc}(\kd) = \int_{\Omega} H_{\exc}(\vect k) \delta(\kd - \vect k) \diff \vect k
    \label{eq:H_exc}
\end{equation}
where $\Omega$ is the support of the 3D OTF, namely $\Omega = \qty{ \kd \in \mathbb R^3 : \qty|H_{\exc}(\kd)| > 0 }$.
Plugging eq. \ref{eq:H_exc} into eq. \ref{eq:fourier_image}, we obtain
\begin{align}
    I(\kd | \xs) &= H_{\emi}(\kd) \cdot 
    \int_{\Omega} H_{\exc}(\vect k) 
    \qty[ e^{i\kd\cdot\xs}O(\kd) * \delta(\kd - \vect k)] \diff \vect k \\
    &= H_{\emi}(\kd) \cdot 
    \int_{\Omega} H_{\exc}(\vect k) 
    \qty[ e^{i(\kd - \vect k)\cdot\xs}O(\kd - \vect k) ] \diff \vect k
\end{align}
Namely, the emitted fluorescence from the sample using a focused illumination spot has the effect of scanning the object's spatial spectrum in the frequency space.
More in detail, the frequency content of the illumination enables access to the object's lateral frequencies up to twice the wide-field cut-off, allowing for super-resolution.
%
At the same time, the illumination frequencies enable filling the \textit{missing cone} of the microscope's OTF, granting higher optical sectioning compared to wide-field microscopy. The object's accessible frequencies are those allowed by the ideal confocal 3D OTF, calculated as the auto-convolution of the wide-field 3D OTF. Unlike confocal microscopy, ISM does not discard light to extend the OTF, enabling high resolution and optical sectioning without compromising the signal-to-noise ratio.
The shifted spectra are available as a weighted sum in the Fourier transform of the raw images. Thus, reconstructing the sample's image with enhanced resolution and optical sectioning requires a computational approach.

\subsection{Algorithm derivation}
\label{subsec:A2}

Assuming an incoherent image formation process, the complete forward model for an image scanning microscope is
\begin{equation}
    i(\xs|\xd) = o(\xs) * h(\xs|\xd)
\end{equation}
where $\xs = (x_s, y_s, z_s)$ and $\xd = (x_d, y_d)$ are the coordinates of the sample plane and detector plane, respectively. The operator $*$ is the 3D convolution with respect to the coordinates $\xs$, $o(\xs)$ is the 3D distribution of fluorescence emitters, $h(\xs|\xd)$ is the set of 3D point spread functions (PSFs) for each detector element, and $i(\xs|\xd)$ is the set of 3D images composing the ISM dataset.
Evalutating the above equation at $z_s = 0$ we obtain the forward model for single-plane imaging
\begin{equation}
    i(\xs|\xd) = \int o(\xs,z) * h(\xs,z|\xd) \diff z
\end{equation}
where we redefined $\xs$ as $(x_s, y_s)$ and $*$ as the 2D convolution operator.
The PSFs -- and, correspondingly, the fingerprint -- evolve along the axial direction on a scale of the order of the depth-of-field (DOF). However, emitters that are more out-of-focus than a few DOFs are so weak they provide a negligible contribution.
Therefore, we discretize the integral along the axial coordinate, assuming that the detected light stems from a finite number $N$ of planes
\begin{equation}
    i(\xs|\xd) = \sum_{k=1}^N o_k(\xs) * h_k(\xs|\xd)  
\end{equation}
where $k$ indexes the axial position.

Our goal is to estimate the distribution of emitters located at the focal plane. To this end, we infer the full vector of planar distributions $\bold{o}=(o_1, \cdots, o_N)$. According to Bayes' theorem, the posterior probability is
\begin{equation}
    P[\bold{o}(\xs)|i(\xs|\xd)]= \frac{P[i(\xs|\xd)|\bold{o}(\xs)] P[\bold{o}(\xs)]}{P[i(\xs|\xd)]} 
\end{equation}
An unbiased estimator of $\bold{o}(\xs)$ is found by maximizing the posterior probability. Since we have no prior information on the specimen, the aforementioned task is equivalent to maximizing the likelihood probability.
In this work, we used a single-photon avalanche diode (SPAD) array detector with no read-out noise, low dark count rates, and negligible cross-talk. Then, we assume that the signal is corrupted only by shot noise. Therefore, we can see the photon counts for each pixel as random variables following a Poisson distribution. Neglecting cross-talk among different detector elements, we assume independent realizations of noise. Thus, the likelihood is the product of individual probabilities for each scan and detector point

\begin{equation}
\label{eq:prior-prob}
    P[i(\xs|\xd) | \bold{o}(\xs)]=\prod_{x_d}\prod_{x_s}\frac{\sum_k[o_k(\xs)*h_k(\xs|\xd)]^{i(\xs|\xd)} e^{-\sum_k[o_k(\xs)*h_k(\xs|\xd)]}}{i(\xs|\xd)!}
\end{equation}
The corresponding negative log-likelihood functional is
\begin{align}
\mathcal{L}[\bold{o}(\xs)] = -\log \{ P[i(\xs|\xd) | \bold{o}(\xs)] \} = \int l[\bold{o}(\xs)|\xd]\diff\xd
\end{align}
where we defined
\begin{align}
l[\bold{o}(\xs)|\xd]=\int \sum_k o_k(\xs)*h_k(\xs|\xd) - i(\xs|\xd) \cdot \log \qty{ \sum_k o_k(\xs)*h_k(\xs|\xd) } \diff\xs
\end{align}
and discarded the constant terms.

The solution to our inverse problem is given by the vector $\hat{ \vect o}$ that minimizes the log-likelihood
\begin{equation}
    \hat{ \vect o}(\xs) = \argmin_{\vect o(\xs)} \mathcal{L}[\bold{o}(\xs)]
\end{equation}
The minimum is found by setting to zero the functional derivative of the log-likelihood. Using the fact that the adjoint operator of the convolution is the convolution with the mirror-reflected kernel, we find

\begin{equation}
    \frac{\delta l}{\delta o_j} = 
    h_j(-\xs,\xd)*\left[1- \frac{i(\xs|\xd)}{\sum_{k}[o_k(\xs)*h_k(\xs|\xd)]}  \right]
\end{equation}
We assume the PSFs to be normalized as follows
\begin{equation}
    \iint h_j(\xs|\xd)\diff\xs \diff\xd=1  \qquad \forall j \in \qty{1, \ldots, N}
\end{equation}
While other normalization choices could be made, the one above preserves the fingerprint information and allows for interpreting the PSFs as probability distributions.
Therefore, we obtain
\begin{equation}
      \frac{\delta \mathcal{L}}{\delta o_j} = 1-\int h_j(-\xs|\xd)*\frac{i(\xs|\xd)}{\sum_{k}[o_k(\xs)*h_k(\xs|\xd)]}\diff\xd 
\end{equation}
We minimize the log-likelihood with an iterative gradient descent method
\begin{equation}
    o_{j}^{(m+1)}=o_{j}^{(m)}-\gamma_k^{(m)} o_{j}^{(m)} \frac{\delta \mathcal{L}}{\delta o_j}
\end{equation}
where $m$ is the iteration index and $\gamma_j^{(m)}$ is the step of the descent. Choosing $\gamma_j^{(m)} = 1 \; \forall \; j,m$, we obtain a multiplicative iteration rule
\begin{equation}
    o_{j}^{(m+1)}(\xs) = o_{j}^{(m)}(\xs)
    \int h_j(-\xs|\xd)*\frac{i(\xs|\xd)}{\sum_{k}o^{(m)}_k(\xs)*h_k(\xs|\xd)}\diff\xd
    \label{eq:sism}
\end{equation}
where all the planes are updated at the same iteration. As a consequence of the multiplicative structure of the above equation, the solution is constrained to have non-negative values if initialized with a positive starting-point.

The Hessian matrix of the log-likelihood is the following
\begin{equation}
    \frac{\delta^2 \mathcal{L}}{\delta o_i \delta o_j} = 
    h_i(-\xs,\xd)* \frac{i(\xs|\xd)}{[\sum_{k} o_k(\xs)*h_k(\xs|\xd)]^2} *h_j(\xs|\xd)
\end{equation}
Since we work with light intensities (photon counts), all the quantities in the above equation are non-negative. Therefore, the Hessian matrix is positive and the minimized functional convex. As a result, the algorithm is guaranteed to converge to a unique solution.

Finally, the \sism algorithm is completed by choosing $N=2$ with the two set of PSFs evaluated at the focal plane and an out-of-focus plane. We chose the starting point $\vect o^{(0)}$ as constant and strictly positive, whose integral equals that of the raw dataset to be reconstructed.

\subsection{Flux conservation}
\label{subsec:A3}

The photon counts of individual images $o_k^{(m)}$ may change with each iteration. However, the overall quantity of photons is unchanged after each iteration. The proof is the following:
\begin{equation}
\begin{split}
    &\sum_k \int o_k^{(m+1)}(\xs)\diff\xs=\\
    =&
    \sum_k \iint o_k^{(m)}(\xs) \left[h_k(-\xs|\xd)* \frac{i(\xs|\xd)}{\sum_j[o^{(m)}_j(\xs)*h_j(\xs|\xd)]}
    \right] \diff\xs \diff\xd  =\\
    =&
    \sum_k \iiint o_k^{(m)}(\xs) h_k(\x-\xs|\xd) \frac{i(\x|\xd)}{\sum_j[o^{(m)}_j(\x)*h_j(\x |\xd)]}\diff\xs \diff\xd \diff \x  =\\
    =&
     \sum_k \iint o_k^{(m)}(\x) *h_k(\x|\xd) \frac{i(\x|\xd)}{\sum_j[o^{(m)}_j(\x)*h_j(\x|\xd)]} \diff\xd \diff \x =\\
     =&
     \iint \sum_k  [o_k^{(m)}(\x) *h_k(\x|\xd) ]\frac{i(\x|\xd)}{\sum_j[o^{(m)}_j(\x)*h_j(\x|\xd)]} \diff\xd \diff\x= \\
     =& \iint i(\xs|\xd) \diff\xs \diff\xd
\end{split}
\end{equation}
Indeed, \sism works by reassigning photons to the correct axial plane while preserving the total photon counts.

\subsection{Generalization to upsampling}
\label{subsec:A4}

If the pixel size $\Delta \xs$ of an ISM acquisition is identical to the detector pitch $\Delta \xd$, the ISM dataset contains enough redundancy to enable the reconstruction of an image with twice the pixels per axis.
Here, we demonstrate that the \sism algorithm can achieve this goal.

First, we define the coordinates of the discretized scanning space as
\begin{equation}
    \qty( \frac{x_s}{\Delta x_s}, \frac{y_s}{\Delta y_s} ) = (n_x, n_y) = \vect n \in \mathbb N^2
\end{equation}
We model the downsampling process using a space-variant excitation PSF
\begin{equation}
\tilde{h}_{\exc} =
\begin{cases}
h_{\exc}(\xs)  \quad & \text{if}\; \xs \in \chi\\
0 \quad \quad  & \text{otherwise}
\end{cases}
\end{equation}
where
\begin{equation}
\chi := \qty{ \xs \; | \; (n_x, n_y) \in (2\mathbb{N}+1)^2 }
\end{equation}
is the subset of pixels with odd indices.
Therefore, depending on the position, we can factorize the likelihood probability of Eq. \ref{eq:prior-prob} in two terms

\begin{equation}
P[i(\xs|\xd) | o(\xs)] = \prod_{x_d}
\prod_{x_s \in \chi} P_1[i(\xs|\xd) | o(\xs)]
\prod_{x_s \not\in \chi} P_2[i(\xs|\xd) | o(\xs)]
\end{equation}
We observe that the PSFs and the images are null when no excitation occurs. Therefore, we have that
\begin{align}
P_2[i(\xs|\xd) | o(\xs)] &= 
\frac{\sum_k[o_k(\xs)*h_k(\xs|\xd)]^{i(\xs|\xd)} e^{-\sum_k[o_k(\xs)*h_k(\xs|\xd)]}}{i(\xs|\xd)!} \eval_{\xs \not \in \chi} =  \nonumber\\
& = \frac{0^0\cdot e^0}{0!}
\end{align}
The expression above contains the indeterminate form $0^0$, which is generally undefined. Nonetheless, in the case of fluorescence imaging, the relation between excitation and signal is given by the power law
\begin{equation}
    y = \alpha x^n
\end{equation}
where $y$ is the fluorescence flux, $x$ is the excitation flux, $n \geq 1$ is the order of the excitation process, and $\alpha >0$ is an efficiency factor. Therefore, our context allows for the definition of $0^0$ as the result of the following limit
\begin{equation}
    \lim_{x \rightarrow 0} x^y = 
    \lim_{x \rightarrow 0} x^{\alpha x^n} = 
    \exp\qty(\alpha \lim\limits_{x \rightarrow 0} {x^n\log x}) = 1
\end{equation}
which is readily calculated using L'Hôpital's rule.

Therefore, we obtain that
\begin{equation}
\prod_{x_s \not\in \chi} P_2[i(\xs|\xd) | o(\xs)] = 1
\end{equation}
which means that the total likelihood is unaffected by the lack of signal in the pixels with an even index.
As a result, we can reconstruct an upsampled image by maximizing the likelihood of Eq. \ref{eq:prior-prob} without modifications.
Consequently, the iterative reconstruction rule remains the same as Eq. \ref{eq:prior-prob}, even if the raw dataset contains signal only on the pixels with an odd index
\begin{equation}
\tilde{\imath}(\xs | \xd) =
\begin{cases}
i(\xs | \xd)  \quad & \text{if}\; \xs \in \chi\\
0 \quad \quad  & \text{otherwise}
\end{cases}
\end{equation}
To demonstrate that the reconstruction fills the voids in the dataset by exploiting the redundancy in the ISM dataset, we develop a simplified toy-model.
Without loss of generality, we consider 1D scanning and detector coordinates.
Furthermore, we approximate the ISM PSFs as identical but shifted.
\begin{equation}
    h(x_s|x_d) \approx h(x_s-\mu(x_d)) 
\end{equation}
In the ideal case of Gaussian PSFs, point-like detector elements, and no Stokes-shift, we have an explicit equation for the shift-vector
\begin{equation}
     h(x_s|x_d) \approx h\vect (x_s - \frac{x_d}{2} \vect ) = h\qty(  n \Delta x_s - \frac{ m \Delta x_d}{2})
     \label{eq:discrete_psf}
\end{equation}
where $x_s = n\Delta x_s$ and $x_d = m\Delta x_d$ are the discretized scanning and detector coordinate, respectively.
We assume that the dataset was acquired respecting the upsampling condition $\Delta x_s = \Delta x_d$ and that the reconstruction is taking place on a grid twofold finer than the original. Thus, we have that on the reconstruction grid $2 \Delta x_s = \Delta x_d$. Substituting the latter identity in Eq. \ref{eq:discrete_psf} and dropping the pixel size, we have
\begin{equation}
    h(n | m) \approx  h(n - m)
    = h(n) * \delta(n - m)
\end{equation}
The iterative reconstruction rule of \sism becomes
\begin{align}
    o_{j}^{(m+1)}( n)  \nonumber &= o_{j}^{(m)}( n)
    \sum_m h_j(- n| m)*\frac{\tilde{\imath}( n|m )}{\sum_{k}o^{(m)}_k( n)*h_k( n| m )}
\end{align}
For the sake of simplicity, we consider only the detector elements with indexes $m\in \qty{0,1}$.
Expanding the terms of the summation and exploiting the associative property of the convolution operation, we have
\begin{align}
    o_{j}^{(m+1)}( n)
    & = o_{j}^{(m)}( n) \cdot h_j(- n) * \delta(n)*\frac{\tilde{\imath}( n| 0 )}{\sum_{k}o^{(m)}_k( n)*h_k( n|0 )} + \nonumber \\
    & + o_{j}^{(m)}( n) \cdot h_j(- n) * \delta( n - 1)*\frac{\tilde{\imath}( n| 1 )}{\sum_{k}o^{(m)}_k( n)*h_k( n| 1 )} = \nonumber \\
    & = o_{j}^{(m)}( n) \cdot h_j(- n) *\frac{\tilde{\imath}( n| 0 )}{\sum_{k}o^{(m)}_k( n)*h_k( n| 0 )} + \nonumber \\
    & + o_{j}^{(m)}( n) \cdot h_j(- n) *\frac{\tilde{\imath}( n-1| 1 )}{\sum_{k}o^{(m)}_k( n)*h_k( n| 1 )} 
\end{align}
Thus, the reconstructions at pixel $n$ are given by the contributions coming from the dataset $\tilde{\imath}$ evaluated at pixels with opposite parity, namely $n$ and $n-1$.
Consequently, even if the original dataset did not contain signal in the even pixels, the algorithm fills them exploiting the properties of ISM.
Thus, \sism can generate an upsampled image, relaxing Nyquist's criterion by a factor of two.
The above argument easily generalizes to the 2D case with any number of detector elements, even if some residual dependency of the detector position remains in the shape of the PSFs.

\subsection{Generalization to spatiotemporal data}
\label{subsec:A5}

In conventional fluorescence microscopy, the spatial and temporal impulse response functions are unrelated.
Therefore, the spatiotemporal response function of the ISM setup is given by the product of the spatial PSFs with the temporal IRF
\begin{equation}
    h_k(t,\xs|\xd) = h_k(\xs|\xd) \cdot h(t|\xd)
\end{equation}
The generalized discrete forward model is
\begin{equation}
    i(t,\xs|\xd) = \sum_k o_k(t,\xs) * h_k(t,\xs|\xd)  
\end{equation}
where $*$ is the convolution operator with respect to $(t, x_s, y_s)$.
Since time is effectively an additional scanning coordinate, all the results derived in section \ref{subsec:A2} still hold.

The complete \sism algorithm is
\begin{equation}
    o_{k}^{(m+1)}(t,\xs) = o_{k}^{(m)}(t,\xs)
    \int
    h_k(-t, -\xs|\xd) * 
    \frac{i(t,\xs|\xd)}
    {\sum_{k} o^{(m)}_k(t, \xs)*h_k(t,\xs|\xd)}
    \diff \xd
    \label{eq:s2ism+time}
\end{equation}
where
\begin{equation}
    \iiint h_k(t, \xs, \xd) \diff t \diff \xd \diff \xs = 1
\end{equation}

\subsection{Estimation of PSF parameters from the dataset}
\label{subsec:A6}

In order to run the \sism algorithm presented in Eq. \ref{eq:s2ism+time}, we need a set of PSFs for each axial plane.
If they are not experimentally available, they can be numerically simulated using the following data-driven procedure.

\subsubsection{Rotation and orientation}


First, calculate the phase correlation among the images of the experimental ISM dataset
\begin{equation}
    \mathcal{R}(\xs|\xd)= \mathcal{F}^{-1} \left\{ \frac{\mathcal{F}\{i(\xs|\xd)\} \overline{\mathcal{F}\{i(\xs|\bold{0})\}}}{|\mathcal{F}\{i(\xs|\xd)\}\overline{\mathcal{F}\{i(\xs|\bold{0})  \}} |}  \right\} 
\end{equation}
where $\mathcal{F}$ is the Fourier transform operator, and the overline stands for complex conjugate. Then, we find the shift vectors as the position of the maximum of each correlogram
\begin{equation}
    \boldsymbol\mu^{\text{(exp)}}(\xd)= \argmax_{\xs} \mathcal{R}(\xs|\xd) 
\end{equation}
Since the images with the highest SNR are those close to the centre of the detector array, we select the shift-vectors from the inner $3\times3$ array. Using the discretized detector coordinates, we have
\begin{equation}
    \qty(\frac{x_d}{\Delta x_d}, \frac{y_d}{\Delta y_d}) = \qty(m_x, m_y) = \vect m \in \qty{-1, 0, 1}^2
\end{equation}
We flatten the detector dimension into a single index $j \in [1, 9]$ and indicate the position coordinates using the index $i\in [1,2]$. Thus, the detector coordinates $m_{ij}$ and experimental shift-vectors $\mu^{\text{(exp)}}_{ij}$ are represented as $2 \times 9$ matrices.

Finally, we assume that the transformation from the detector coordinates to the experimental shift-vectors is given by three operations: mirroring, rotation, and dilation.
These transformations are described by the matrix
\begin{equation}
    \vect T(\rho, \theta, \alpha) = \vect A(\alpha) \vect R(\theta) \vect M(\rho)
\end{equation}
The mirroring matrix $\vect M(\rho)$ accounts for the orientation of the detector
\begin{equation}
\vect M(\rho)=
\begin{pmatrix}
1 & 0\\
0 & \rho
\end{pmatrix}
\quad \text{with } \rho \in \{-1,1\};
\end{equation}
The rotation matrix $\vect{R}(\theta)$ accounts for the rotation of the detector
\begin{equation}
\vect{R}({\theta})=
\begin{pmatrix}
\cos\theta & \sin\theta\\
-\sin\theta & \cos\theta
\end{pmatrix}
\quad \text{with } \theta \in [ -\pi, \pi];
\end{equation}
The dilatation matrix $\vect{A}({\alpha})$ accounts for the magnification of the microscope
\begin{equation}
\vect{A}(\alpha) =
\begin{pmatrix}
\alpha & 0\\
0 & \alpha
\end{pmatrix} 
\quad \text{with } \alpha \in \mathbb{R}^+;
\end{equation}
We find the parameters $\hat{\rho}$, $\hat{\theta}$, and $\hat{\alpha}$ that describe the microscope used for the acquisition of the dataset by numerically solving the following minimization problem
\begin{equation}
    \hat{\rho}, \hat{\theta}, \hat{\alpha} = 
    \argmin_{\rho, \theta, \alpha} \norm{
    \vect\mu^{\text{(exp)}} -
    \vect T(\gamma, \theta, \alpha) \vect m }^2_F
\label{eq:shift_loss}
\end{equation}
where we used the Frobenius norm.
While the loss function is not convex in general, the constraints on the minimization values uniquely define a single solution.
We show a graphical depiction of the presented method in Supp. Fig. \ref{supfig:par_retrieving}a-c.

\subsubsection{Magnification}

\begin{figure*}[!htb]
    \centering
    \includegraphics[width=\textwidth]{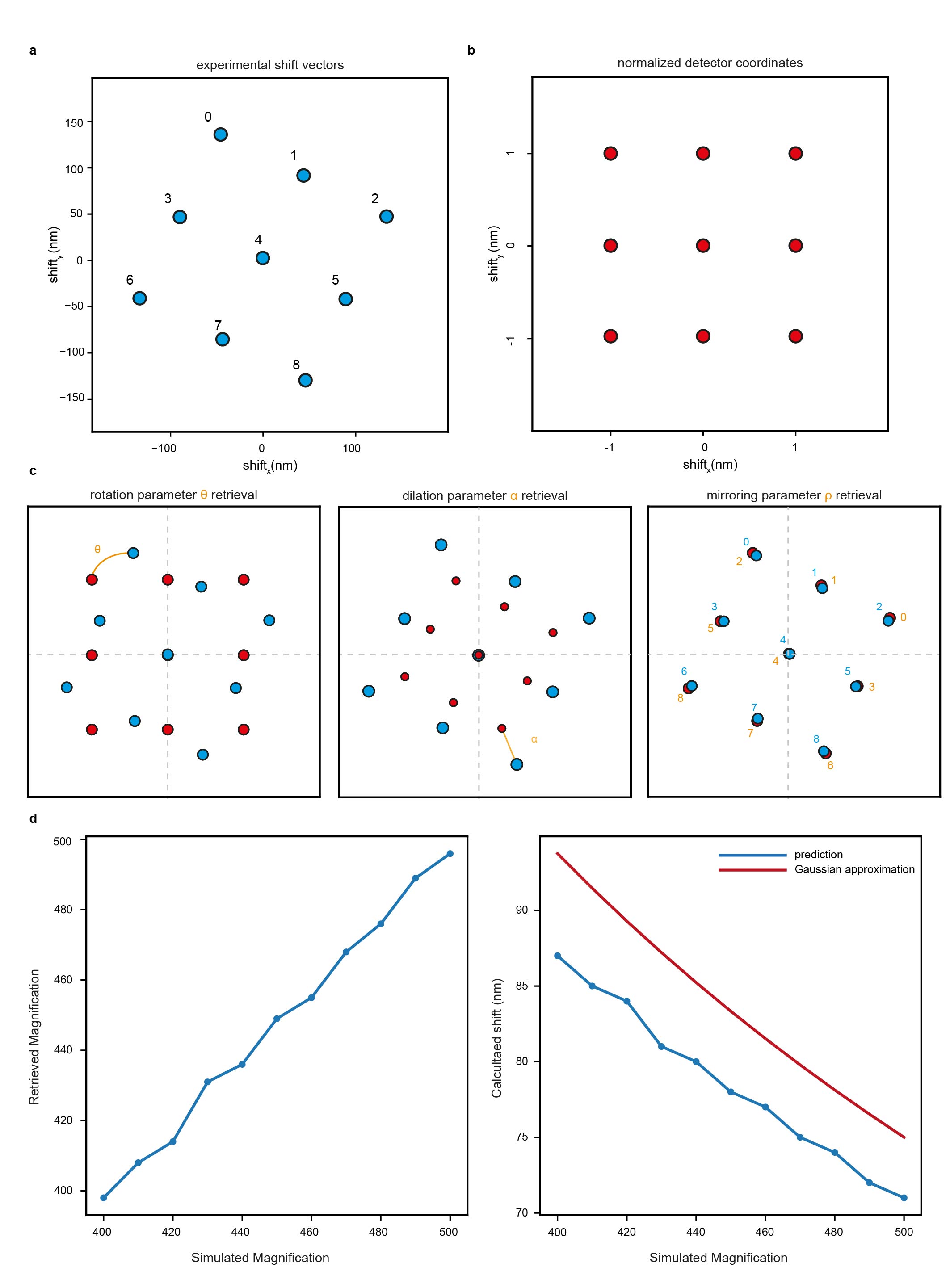}
    \caption{\textbf{Workflow of parameter estimation.} 
    We fit the experimental shift-vectors from the $3 \times 3$ inner array, \textbf{a}, to the normalized detector coordinates, \textbf{b}, transformed by a rotation, dilation, and mirroring operations, \textbf{c}.
    Finally, we convert the dilation parameter into a magnification value minimizing the difference with the shift-vectors value calculated using the scalar approximation of the PSF, \textbf{d}.
    Comparing the values calculated using the vectorial model, we show that the error is negligible (left). Then, we invert the model to retrieve the magnification (right). The simple law derived under the Gaussian approximation of PSF is not accurate enough to be used.
    }
    \label{supfig:par_retrieving}
\end{figure*}

The dilation parameter $\hat{\alpha}$ is the absolute value of the shift-vectors for a detector element located at one pixel pitch from the centre of the detector.
Approximating the PSFs as Gaussian functions, we can calculate the shift-vectors as $\mu(M) = \frac{\Lambda}{2M}$, where $\Lambda$ is the pixel pitch of the detector and $M$ the magnification of the microscope.
However, the Gaussian approximation is too crude and would lead to a poor estimation of the magnification.
Therefore, we used the more accurate scalar model, which requires also the following variables: the pixel size of the detector, the wavelength of the excitation and fluorescence light, and the numerical aperture of the objective lens.
These latter are typically known from the experimental setup and can be used to convert $\hat{\alpha}$ into a magnification value $M$.
To speed up the calculation, we performed a 1D calculation of the following scalar model of the in-focus PSF
\begin{equation}
    h(x | \lambda) = \qty| \frac{J_1\qty( k\NA x )}{k\NA x} | ^2
\end{equation}
where $k = 2\pi / \lambda$ and $J_1$ is the first order Bessel function of the first kind.
We define the pinhole function as
\begin{equation}
    p(x | M) = 
    \begin{cases}
        1 & \text{if } |x| \leq \frac{\Delta}{2M} \\
        0 & \text{otherwise}
    \end{cases}
\end{equation}
where $\Delta$ is the pixel size of the detector.
We calculate the theoretical shift as
\begin{equation}
    \mu(M) = \argmax_x h(x| \lambda_{\exc})
    \qty[ h(x| \lambda_{\emi}) * p\qty(x - \Lambda/M \vert M) ]
\end{equation}
Finally, we estimate the magnification by numerically solving the following minimization problem
\begin{equation}
    \hat{M} = \argmin_M \norm{\hat{\alpha} - \mu(M)}_2^2
\end{equation}
As reported in Supp. Fig. \ref{supfig:par_retrieving}d, the above procedure correctly estimates the magnification value.
We could also have used a vectorial model to calculate the PSFs. Despite being more rigorous, it is more computationally expensive and leads to a negligible improvement in accuracy in estimating the magnification.

\subsubsection{Axial position}

\begin{figure*}[h!]
    \centering
    \includegraphics[width=\textwidth]{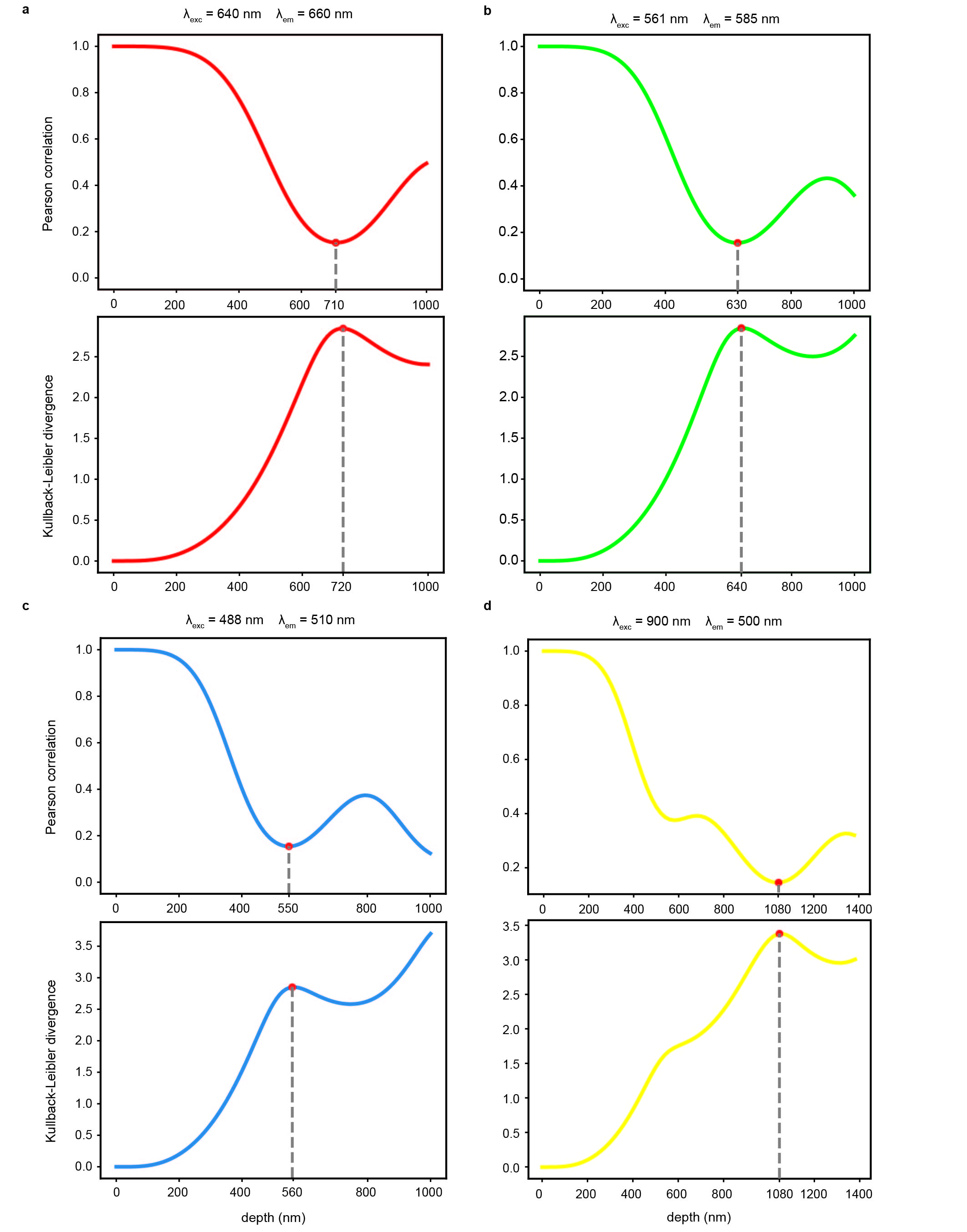}
    \caption{\textbf{Calculation of the out-of-focus position.}
    Optimal out-of-focus depth estimated for red $\bold{a}$, green $\bold{b}$, blue $\bold{c}$ and two-photon $\bold{d}$ excitation. 
    In each case, we report the calculation of Pearson correlation and Kullback-Leibler divergence between in-focus and variably defocused PSFs.
    }
    \label{supfig:conditioning}
\end{figure*}

In this work, we run \sism using two axial planes.
The first one is calculated in the focal plane ($z_1=0$).
We define the out-of-focus plane as the one that maximizes the discrepancy between the defocused and in-focus dataset of simulated PSFs
\begin{equation}
    z_2 = \argmax_z  D\qty[ h(\xs, 0|\xd) \parallel h(\xs,z|\xd) ]
\end{equation}
where  the PSFs are normalized by the flux at the corresponding axial planes
\begin{equation}
    \int_{\mathbb{R}^4} h(\xs,z|\xd) \diff \xd \diff \xs = 1 \qquad \forall z \in \mathbb{R}
\end{equation}
As a discrepancy functional, we used either the Kullback-Leibler divergence or the negative Pearson correlation.
Since the non-aberrated PSFs are axially symmetric, we run the maximization algorithm only for positive $z$ values up to a distance equal to the depth of field of the microscope, calculated using the excitation wavelength.
We explore the axial range in 40 steps.
As shown in Supp. Fig. \ref{supfig:conditioning}, both metrics yield the same result.

\subsubsection{Field of view}

Once all the required parameters are available, we simulate the two PSF datasets using vectorial diffraction theory.
The field of view of the simulation should be large enough to contain the entire structure of the PSFs.
However, too many empty pixels would needlessly increase the computation time.
Therefore, we find the optimal field of view by leveraging the generalized divergence law of Gaussian beams
\begin{equation}
w(z)=w_0 \sqrt{1+\left(\frac{M^2 \cdot z}{z_R}\right)^2}
\end{equation}
where $w(z)$ is the beam waist and at a distance $z$ from the focus, $z_R$ is the Rayleigh range, and $M^2$ is a factor to correct for the PSF's non-Gaussian shape.
We empirically found the optimal value of this latter to be $M^2 = 3$.

We define the in-focus beam size as the radius of the Airy disk
\begin{equation}
    w_0 = 0.61 \frac{\lambda_{\exc}}{\NA}
\end{equation}
and the Rayleigh range as
\begin{equation}
    z_R= \frac{\pi w_0^2 n}{\lambda_{\exc}}.
\end{equation}
where $n$ is the refractive index of the immersion medium.

We also need to consider the shift induced by the off-axis detector elements.
Therefore, we evaluate the size of the field of view $\mathrm{FOV}$ as
\begin{equation}
    \mathrm{FOV} = 2 \qty[ w(z_2) + N \cdot \Delta x_d ]
\end{equation}
where $z_2$ is the position of the out-of-focus plane, and $N$ is the number of elements per axis of the detector array. Thus, $N \cdot \Delta x_d$ is the lateral size of the detector.

\subsection{Reconstruction with experimental PSFs}
\label{subsec:A7}

If the ISM PSFs are experimentally available, they can be used to replace the simulated PSFs in the \sism workflow.
In this case, the axial positions of the axial planes required to perform the reconstruction are unknown \textit{a priori}.
Therefore, we need to acquire the full volumetric PSF and estimate the position of the planes in post-processing.
We estimate the focal plane to be the one with the sharpest PSF.
Equivalently, it is the plane where the normalized 2D modulation transfer function (MTF) has the largest integral.
Since the size of the pinhole does not alter the position of the focal plane, we work with the open-pinhole PSF $h(\xs,z)=\sum_{\xd}h(\xs,z|\xd)$. The normalized MTF integral is
\begin{equation}
    \gamma(z)= \int_{\mathbb R ^2} \qty| \mathcal{F} \qty{\frac{ h(\xs, z) }{ \int h(\xs, z) \diff \xs} }(\ks) | \diff \ks
\end{equation}
where the Fourier Transform $\mathcal{F}$ is calculated with respect to the lateral coordinates $\xs$.
The focal plane is estimated as
\begin{equation*}
    z_1 = \argmax_{z} \gamma(z)
\end{equation*}
The out-of-focus position $z_2$ is calculated as in the simulation case
\begin{equation}
    z_2 = \argmax_z  D\qty[ h(\xs, z_1|\xd) \parallel h(\xs,z|\xd) ]
\end{equation}
However, in the experimental scenario, the PSFs might not be axially symmetric due to the presence of some small aberrations. Therefore, we arbitrarily chose one side of the $z$-axis.
Another valid approach would be to perform the reconstruction using two out-of-focus planes, one for each side.
We show the result of a reconstruction using experimental PSFs in Supp. Fig. \ref{supfig:synth_exp_PSFs}.

\subsection{Structural similarity index measure}
\label{subsec:A8}

\begin{figure*}[h!]
    \centering
    \includegraphics[width=\textwidth]{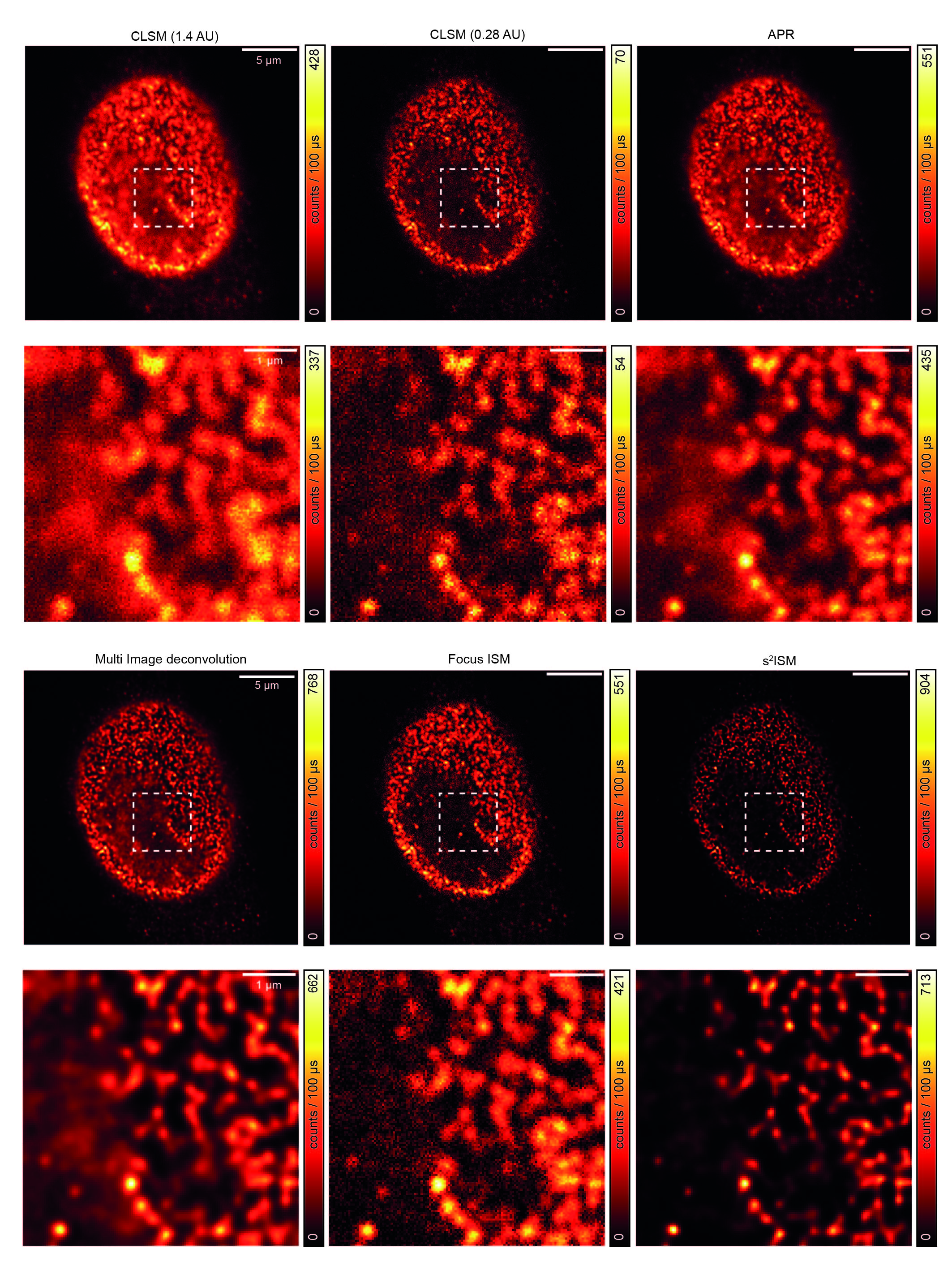}
    \caption{\textbf{Extended data: nuclear pore complexes in a HeLa cell.}
    Extended images from Fig. \ref{fig:4+5}d. Field-of-view: \qtyproduct{25 x 25}{\micro m}, image size: \qtyproduct{625 x 625}{} pixels, \sism and Multi-Image deconvolution iterations: 20.
    }
    \label{supfig:nup}
\end{figure*}

\begin{figure*}[h!]
    \centering
    \includegraphics[width=\textwidth]{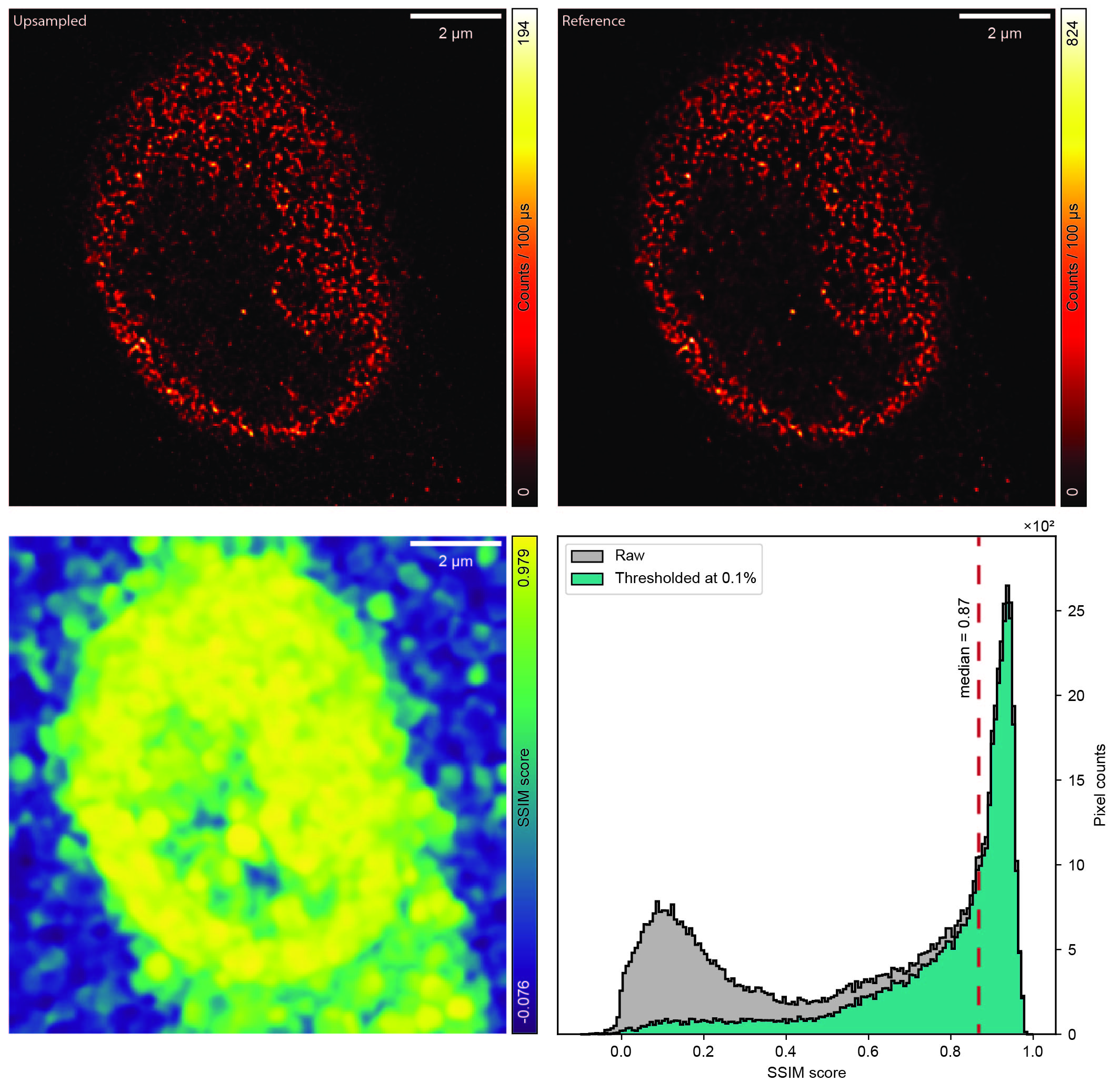}
    \caption{\textbf{Upsampling fidelity.}
    We calculated the structural similarity index measure (SSIM) score between the upsampled and reference images from Fig. \ref{fig:4+5}d. The SSIM score is locally high in the regions of the sample rich with signal. Unsurprisingly, the SSIM score decreases in the regions with little or no signal. Therefore, we removed the pixels with less than 0.1\% of the maximum intensity of the reference images to calculate the thresholded histogram of the SSIM score. The median score calculated on the thresholded histogram is 0.87.}
    \label{supfig:SSIM}
\end{figure*}

In \ref{subsec:A4}, we generalized \sism to enable upsampling.
To measure the quality of the reconstruction, we acquired a dataset using the condition $2 \Delta \xs = \Delta \xd$ and later downsampled the acquisition by a factor of two. Then, we reconstructed the original and downsampled ISM dataset, with and without upsampling. The results are shown in Fig. \ref{fig:4+5}d and extended in Supp. Fig. \ref{supfig:nup}.
We analyzed the reliability of the reconstructions by calculating the structural similarity index measure (SSIM) between the two reconstructions. Naming the images $x$ and $y$, the SSIM index is defined as
\begin{equation}
    \mathrm{SSIM} = \frac{(2 \mu_x \mu_y +c_1)(2 \sigma_{xy} + c_2)}{(\mu_x^2 + \mu_y^2 + c_1)(\sigma_x ^2 + \sigma_y ^2 +c_2)}
\end{equation}
where $\mu_x$, $\mu_y$ are the mean values, $\sigma_x$, $\sigma_y$ the respective standard deviations, and $\sigma_{xy}$ the covariance.
The above quantities are calculated in a subregion of the images defined by a Gaussian window with a standard deviation of 10 pixels.
We set the regularizers constants to zero, $c_1 = c_2 = 0$.
The SSIM metric returns a value in the $[-1,1]$ range, where 1 indicates optimal similarity.
By sliding the window on the full image, we calculated an SSIM map. The result is shown in Supp. Fig. \ref{supfig:SSIM}.

\subsection{Radial spectrum}
\label{subsec:A9}

We used the radial spectrum to compare the reconstructions from different algorithms.
The spatial spectrum of an image $i(\xs)$ as its Fourier transform
\begin{equation}
    \FT\qty{ i(\xs) } = I(\vect k_s) = O(\vect k_s) \cdot H(\vect k_s)
\end{equation}
where the capital letters indicate the Fourier transform of the corresponding quantities and $\ks = (k_s \cos\phi, k_s \sin\phi)$ is the spatial frequency vector written in polar coordinates.
The radial spectrum is given by the absolute value of the average on the angular dimension
\begin{equation}
    S(k) = \frac{1}{2\pi} \Int_0^{2\pi} \qty| I(k, \varphi) | \diff \varphi 
\end{equation}
We report the results of the calculation in Supp. Fig. \ref{supfig:radial_spectra}.

\subsection{SNR estimation}
\label{subsec:A10}

Thanks to the high speed and sensitivity of the SPAD array detector, we can finely tune the duration of photon counting windows on each scan point.
We exploited the detector's speed of to subdivide each pixel dwell time $T$ into $n_T$ temporal bins.
To estimate the signal-to-noise-ratio of the final images, we summed the temporal bins into $l$ realizations of the same ISM dataset.
Then, we independently reconstructed each realization with the \sism algorithm and saved the result at each iteration.
Finally, we calculated pixel-wise mean and standard deviation among the images, obtaining an SNR map for each algorithm update.
We show the result of this analysis in Supp. Fig. \ref{supfig:SNR_enhancement_charac}, where we also report the histogram of the SNR values. To exclude pixels with no signal from the histogram, we bin the values in the range $[5, 60]$ and we calculate the median SNR for each iteration.
The results indicate that the best increase in SNR occurs at a small number of iterations.
However, note that this analysis cannot be used to choose the best \sism iteration since each realization has a lower starting SNR (smaller pixel dwell time) that the dataset integrated over the time bins.

\clearpage

\section{Supplementary figures}
\label{sec:supfigs}

\begin{figure*}[!htb]
    \centering
    \includegraphics[width=\textwidth]{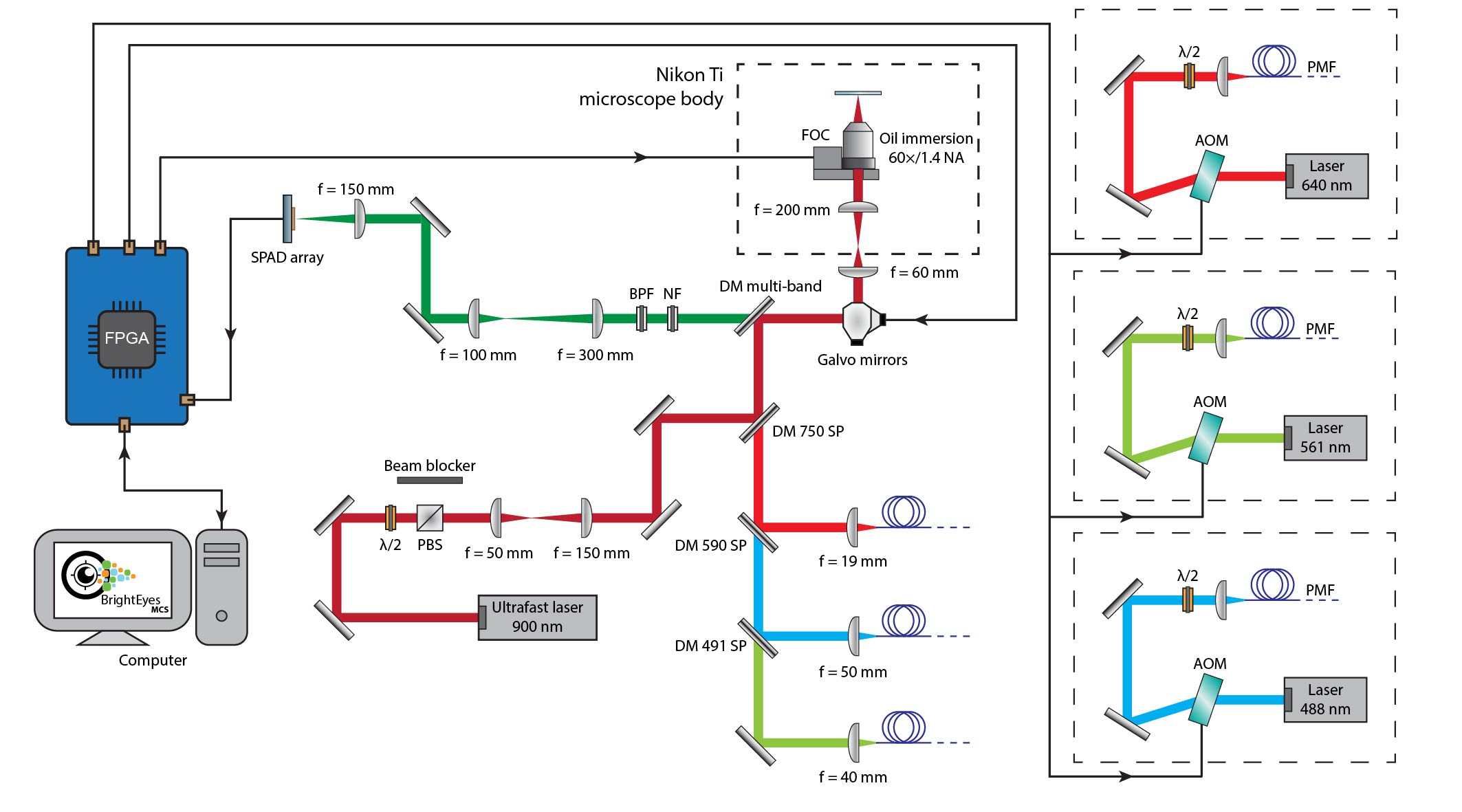}
    \caption{\textbf{Scheme of the image scanning microscope.} AOM: acousto-optic modulator. PMF: single-mode polarization maintaining fibre. DM: dichroic mirror. BPF: band-pass filter. NF: notch filter. PBS: polarizing beam splitter. $\lambda/2$: half-wave plate.}
    \label{supfig:setup}
\end{figure*}

\clearpage

\begin{figure*}[!htb]
    \centering
    \includegraphics[width=\textwidth]{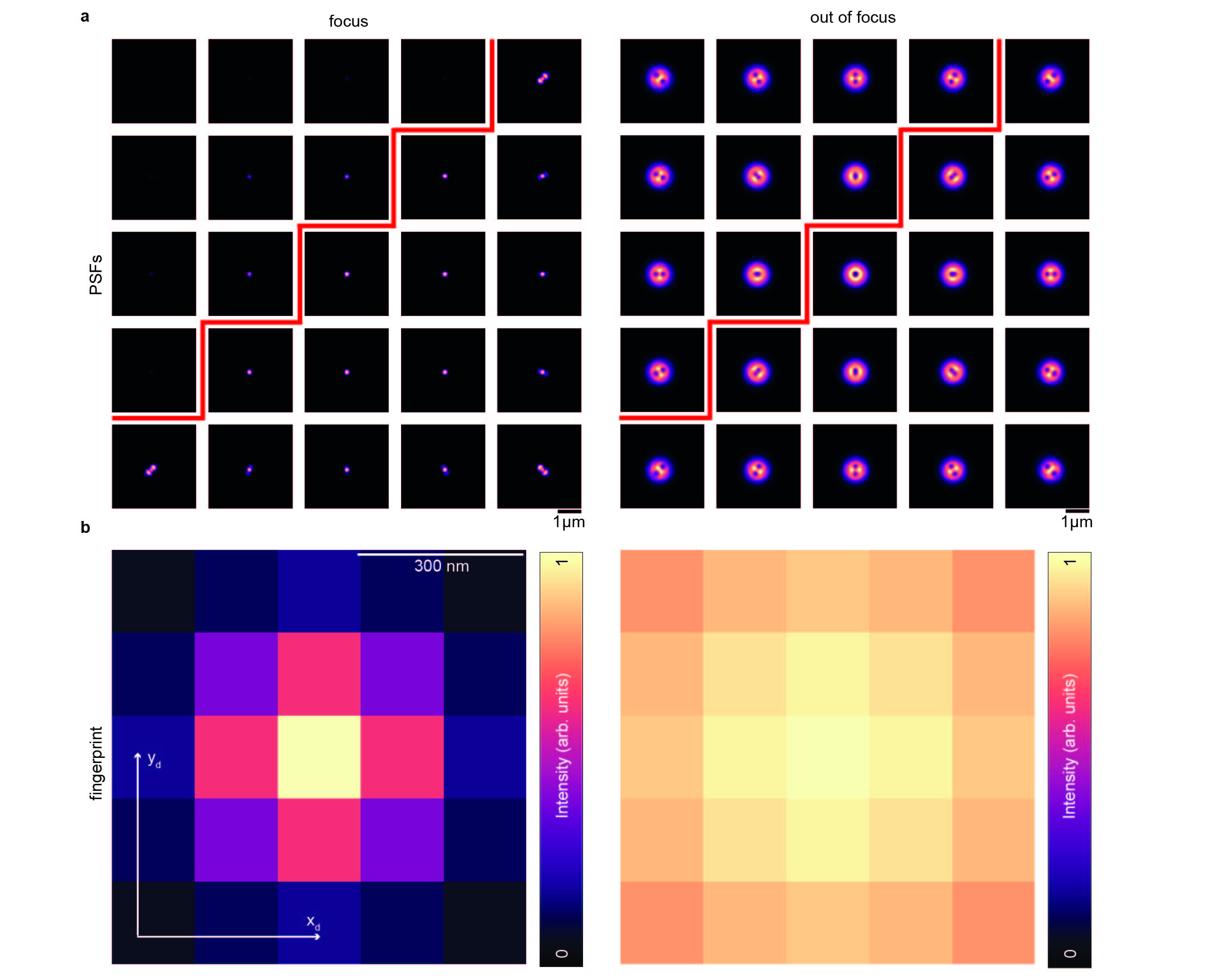}
    \caption{\textbf{In focus and out-of-focus PSFs and fingerprints.} 
    $\bold{a}$ the in focus PSFs and the out-of-focus PSFs. Top left corner : each PSF is normalize with respect to the central element of the SPAD array. Bottom right corner : each PSF normalized to itself. In $\bold{b}$ the respective fingerprints.}
    \label{supfig:In focus and bkg PSFs}
\end{figure*}

\clearpage

\begin{figure*}[!htb]
    \centering
    \includegraphics[width=\textwidth]{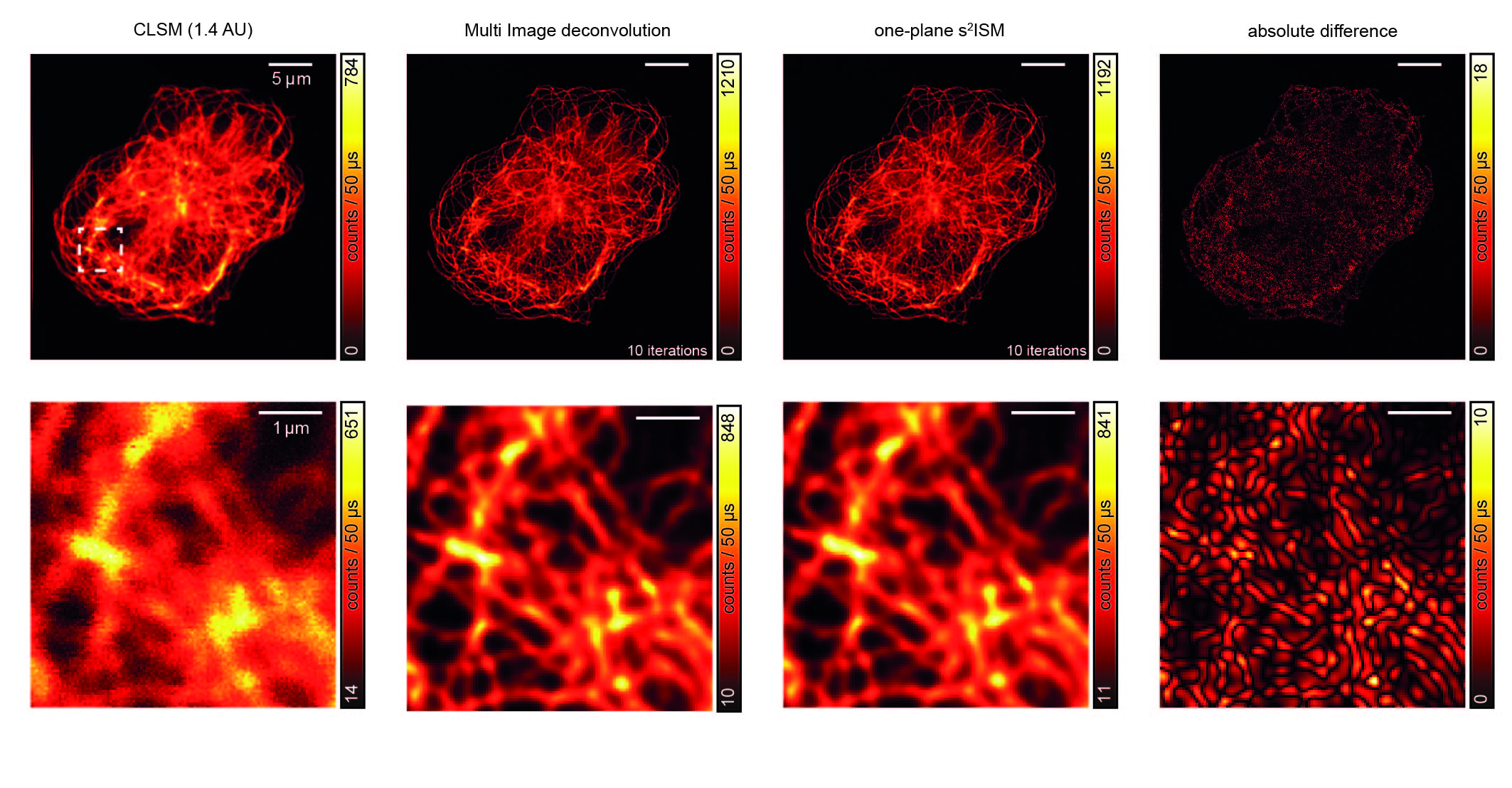}
    \caption{\textbf{Comparison between Multi-Image deconvolution and single-plane \sism}
    The \sism algorithm reconstructing a single plane is equivalent to multi-image deconvolution.
    We quantified the absolute difference of the reconstructions, which is due to numerical errors.
    Pixel size:  \SI{40}{\nano m}. Format of the images: \qtyproduct{875x875}{} pixels. Iterations: 10 for both algorithms.}
    \label{supfig:MID_vs_3D_ISM}
\end{figure*}

\clearpage

\clearpage

\begin{figure*}[!htb]
    \centering
    \includegraphics[width=\textwidth]{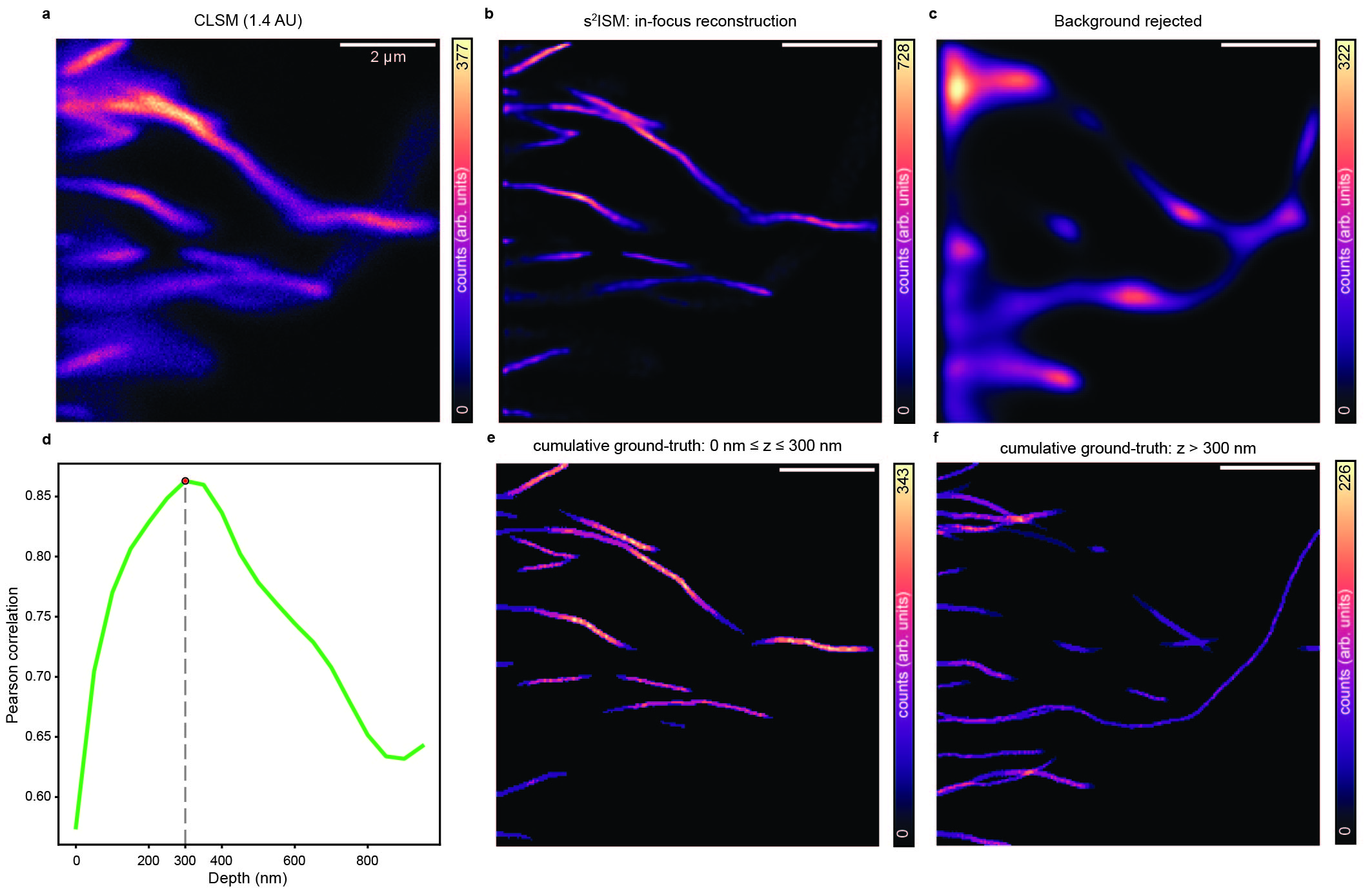}
    \caption{\textbf{\sism on a synthetic thick sample.}
    We generated a volumetric phantom of \qtyproduct{20x201x201}{} voxels $(zxy)$.
    We simulate the corresponding PSF dataset of \qtyproduct{20x201x201x25}{} voxels $(zxyc)$ with a lateral and axial size of \SI{40}{\nano m} and \SI{50}{\nano m}, respectively.
    The raw ISM dataset with size \qtyproduct{201x201x25}{} voxels $(xyc)$ is obtained by convolving plane-by-plane and channel-by-channel the ground truth with the PSF and then summing over the axial dimension.
    \textbf{a}, confocal image obtained summing over the channel dimension.
    Images of the in-focus \textbf{b} and out-of-focus \textbf{c} reconstructed by \sism.
    \textbf{d} correlation of the in-focus reconstruction with the partial axial integral of the ground-truth up to  $z$. The position of the correlation maximum indicates the effective size of the optical section.
    Ground-truth axially integrated up to \textbf{e} and above \textbf{f} the correlation maximum at $z=\SI{300}{nm}$.
    }
    \label{supfig:3D-ISM_vol_synth_tub}
\end{figure*}

\clearpage

\begin{figure*}[!htb]
    \centering
    \includegraphics[width=\textwidth]{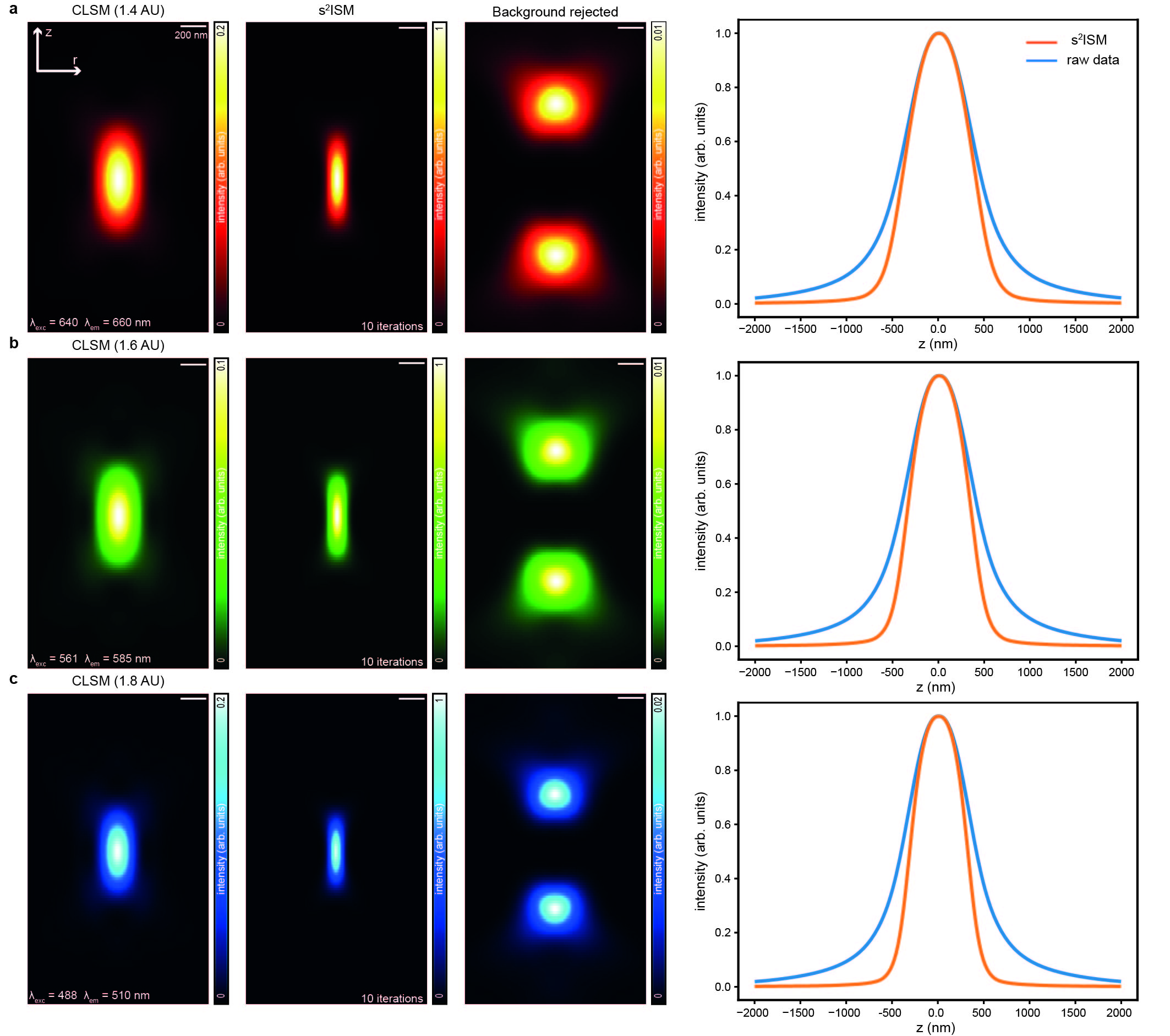}
    \caption{\textbf{\sism reconstruction of synthetic PSFs.}
    3D PSFs for red \textbf{a}, green \textbf{b}, and blue \textbf{c} light.
    We depict the open-pinhole confocal (left) and \sism (middle) images.
    We calculate the integral of the PSF along the lateral dimension, to quantify optical sectioning (right).
    For each color, the PSFs are simulated in a volume of \qtyproduct{4.04x4.53x4.53}{\micro m} with \qtyproduct{202x453x453}{} voxels ($zxy$).
    We reconstruct the stacks with 10 \sism iterations for each axial plane.
    }
    \label{supfig:optical_sectioning_wls}
\end{figure*}

\clearpage

\clearpage

\begin{figure*}[h!]
    \centering
    \includegraphics[width=\textwidth]{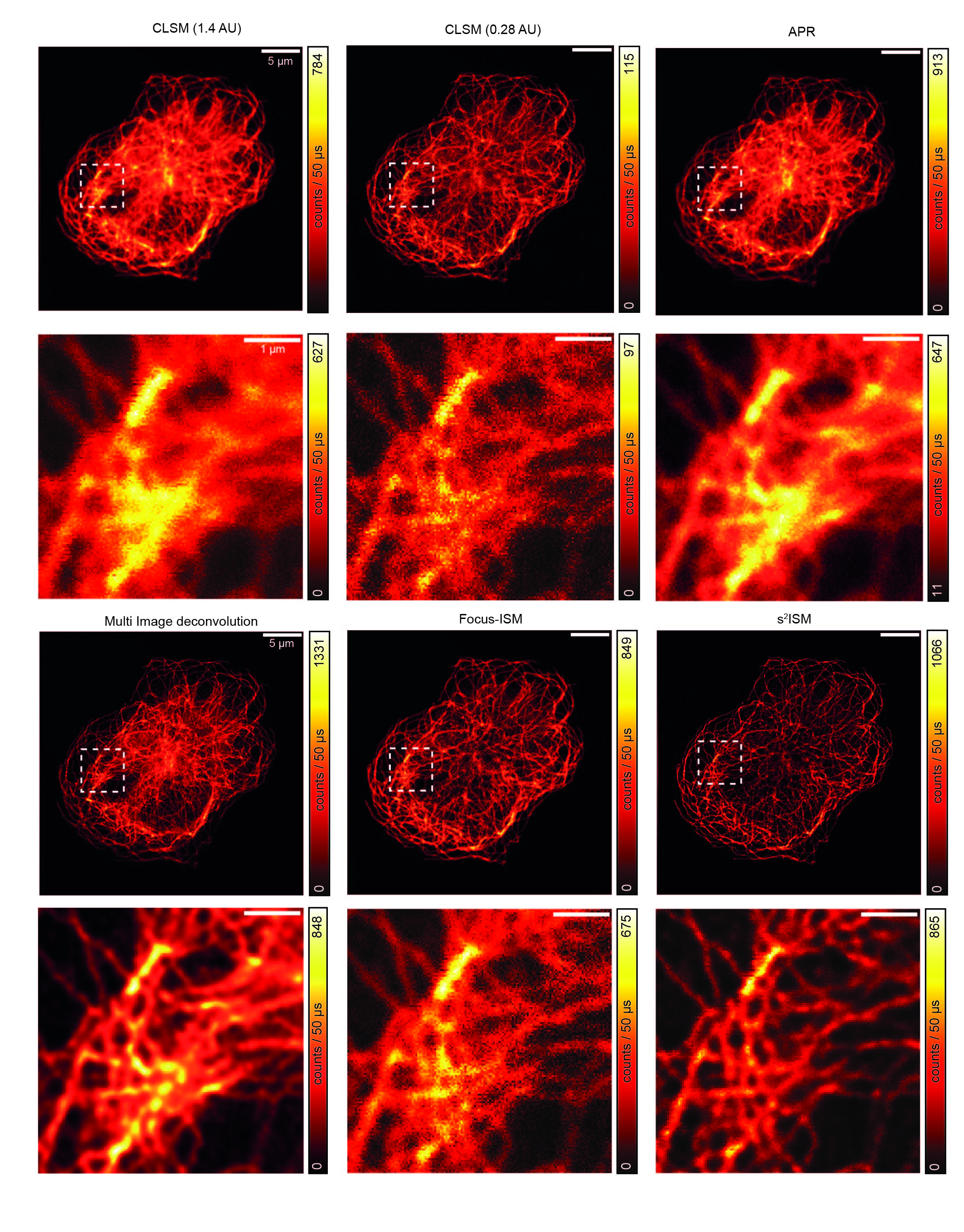}
    \caption{\textbf{Extended data: tubulin network of a HeLa cell.}
    Extended images from Fig. \ref{fig:2}. Field-of-view: \qtyproduct{35 x 35}{\micro m}, image size: \qtyproduct{875 x 875}{} pixels,  \sism and Multi-Image deconvolution iterations: 20.}
    \label{supfig:cell_A}
\end{figure*}

\clearpage

\begin{figure*}[!htb]
    \centering
    \includegraphics[width=\textwidth]{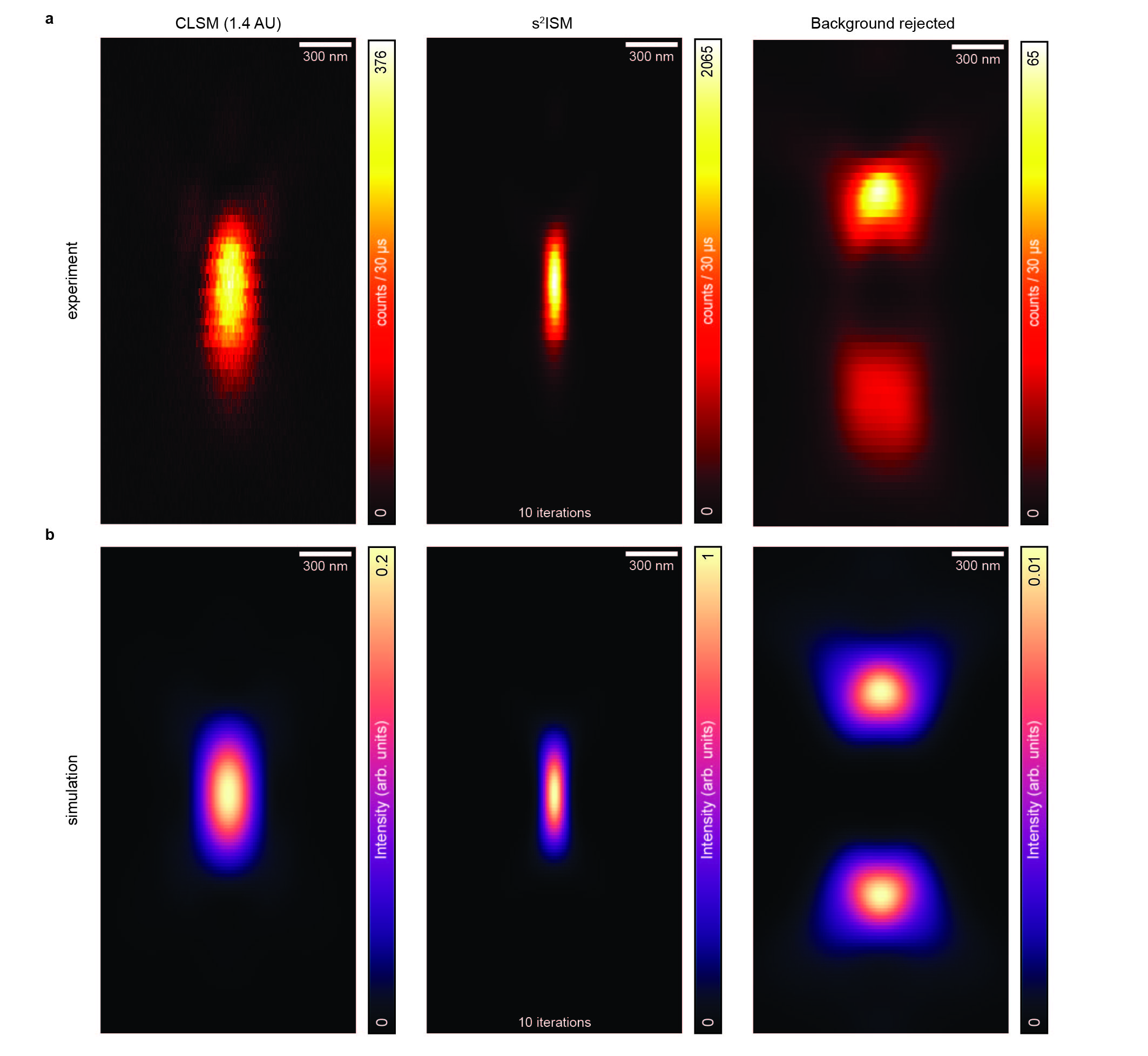}
    \caption{\textbf{\sism applied to experimental and simulated PSFs.}
    \textbf{a}, experimental PSF acquired by imaging a gold bead (\SI{80}{\nano m}) acquired in scattering with at \SI{640}{\nano m} excitation wavelength in pulsed mode (repetition rate = \SI{80}{MHz}). Parameters: $\NA = 1.4$, $n = 1.5$. Voxel size: \qtyproduct{50x10x10}{nm} $(zxy)$. 
    \textbf{b}, PSF simulated to match the experimental parameters reported for \textbf{a}, assuming no aberrations. Despite some spherical aberration is present in the experimental data, \sism is capable to improve resolution and optical sectioning anyway.
    }
    \label{supfig:exp_vs_synth_goldbeads}
\end{figure*}

\clearpage

\begin{figure*}[!htb]
    \centering
    \includegraphics[width=\textwidth]{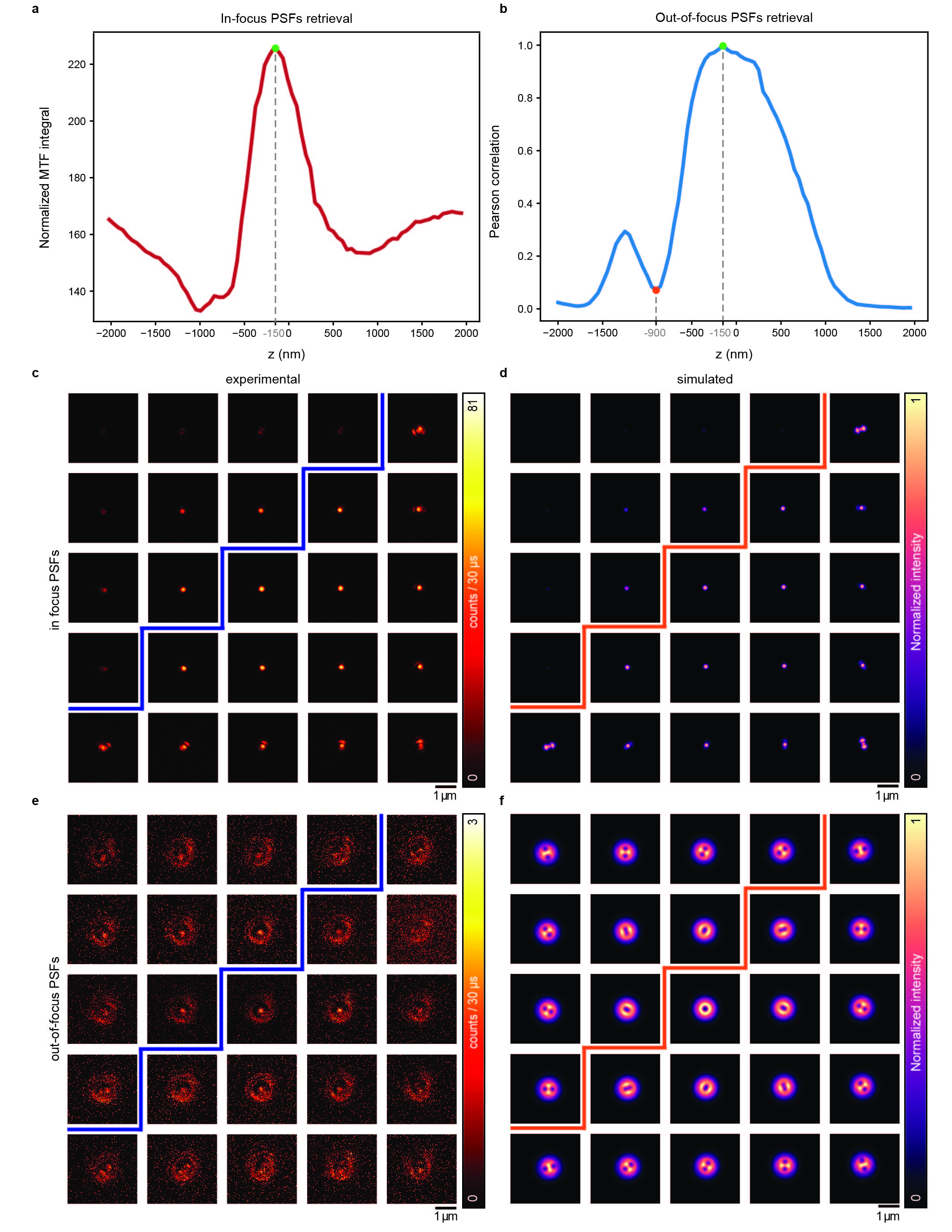}
    \caption{\textbf{Comparison between simulated and experimental PSFs.}
    We calculate the axial planes required to run the \sism algorithm from the 3D ISM dataset of the experimental PSF shown in Supp. Fig. \ref{supfig:exp_vs_synth_goldbeads}.
    \textbf{a}, we find the focal plane as the maximum position of the normalized MTF integral.
    \textbf{b}, we find the out-of-focus plane by minimizing the correlation of the in-focus PSFs with the defocused PSFs.
    We compare the measured \textbf{c} and simulated \textbf{d} in-focus PSFs, calculated using the procedure reported in Supp. Fig. \ref{supfig:par_retrieving}.
    Similarly, we show the measured \textbf{c} and simulated \textbf{d} in-focus PSFs. Note that the defocus position is not the same, but is given by the procedure depicted in \textbf{b} and Supp. Fig. \ref{supfig:conditioning}. For each dataset, the top-left corner images are normalized to the full dataset and the bottom-right images are normalized to themselves.
    }
    \label{supfig:synth_exp_PSFs}
\end{figure*}

\clearpage

\begin{figure*}[!htb]
    \centering
    \includegraphics[width=\textwidth]{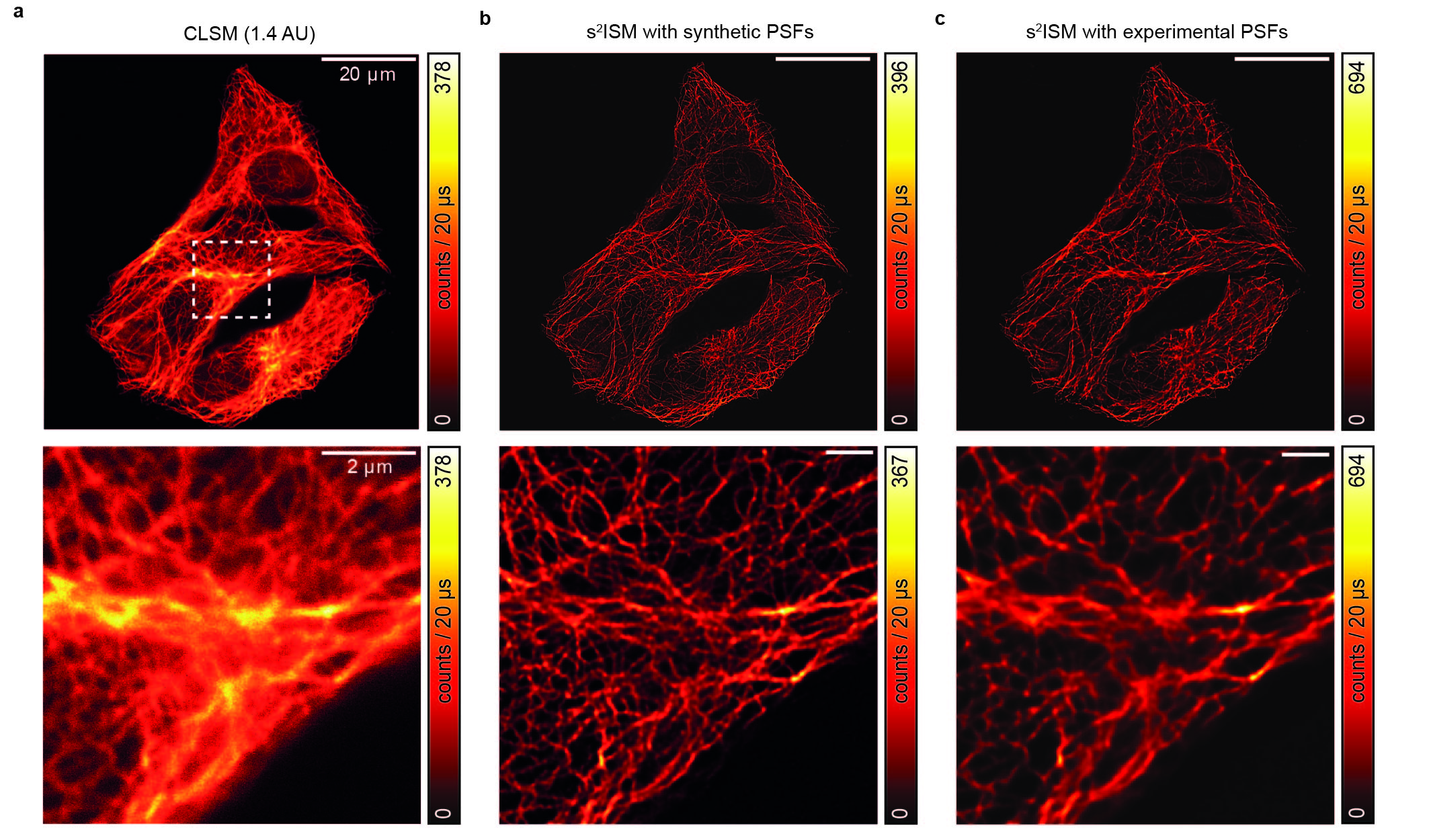}
    \caption{\textbf{\sism with experimental and synthetic PSFs.}
    From the ISM dataset of the tubulin network of a group of HeLa cells, we performed the following reconstructions.
    \textbf{a}, open-pinhole confocal image.
    \textbf{b}, \sism image obtained using synthetic PSFs.
    \textbf{c}, \sism image obtained using esperimental PSFs.
    The used PSFs are shown in Supp. Fig. \ref{supfig:synth_exp_PSFs}.
    Both reconstructions are stopped at 10 iterations.
    }
    \label{supfig:3D-ISM_w_exp_and_synt_PSFs}
\end{figure*}

\clearpage

\begin{figure*}[h!]
    \centering
    \includegraphics[width=\textwidth]{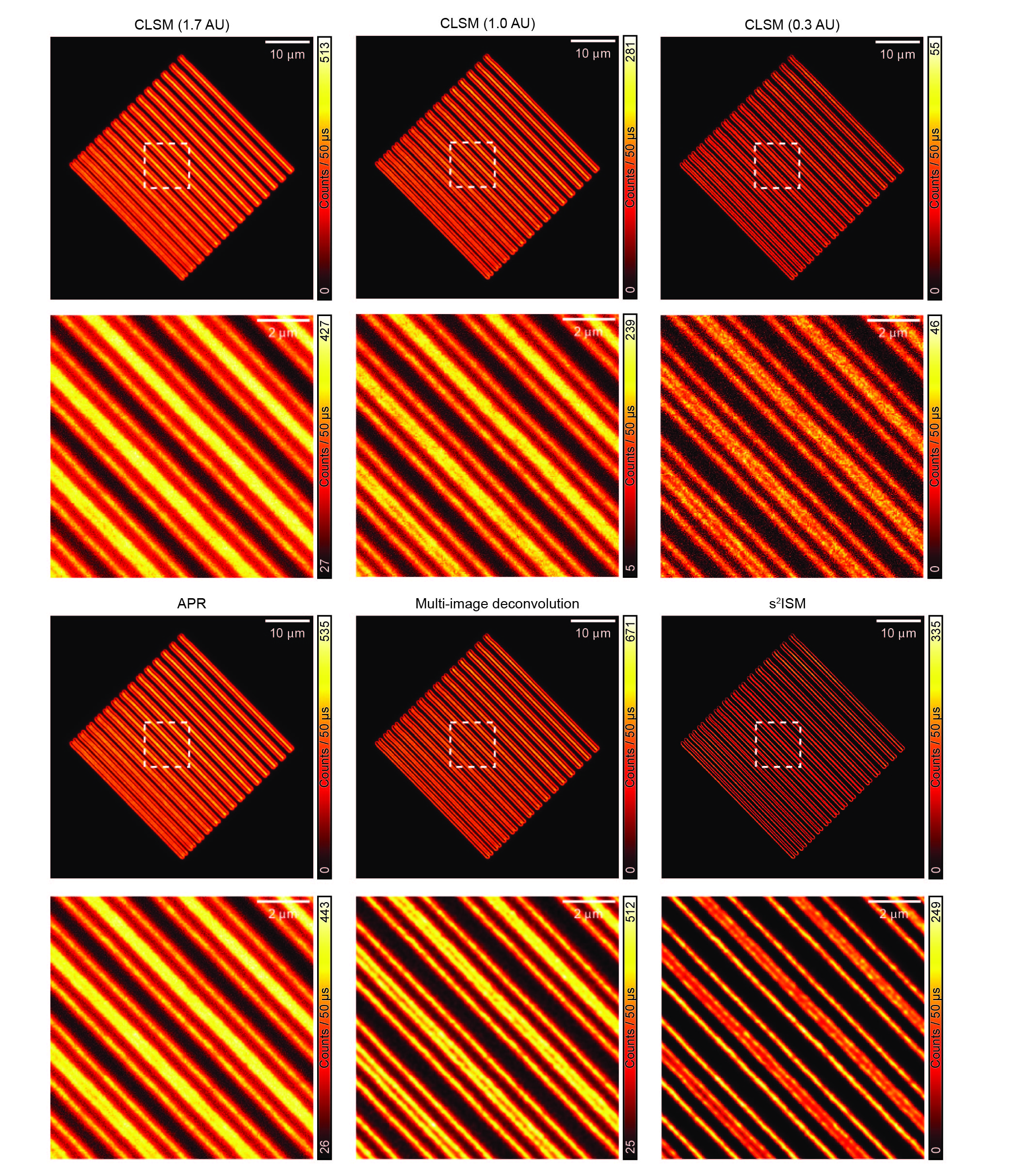}
    \caption{\textbf{Extended data: gradually spaced lines.} 
    Extended images from Fig. \ref{fig:3}a. Field-of-view: \qtyproduct{60 x 60}{\micro m}, image size: \qtyproduct{1500 x 1500}{} pixels, \sism and Multi-Image deconvolution iterations: 20.}
    \label{supfig:Argo_sample_diag_lines}
\end{figure*}

\clearpage

\begin{figure*}[h!]
    \centering
    \includegraphics[width=\textwidth]{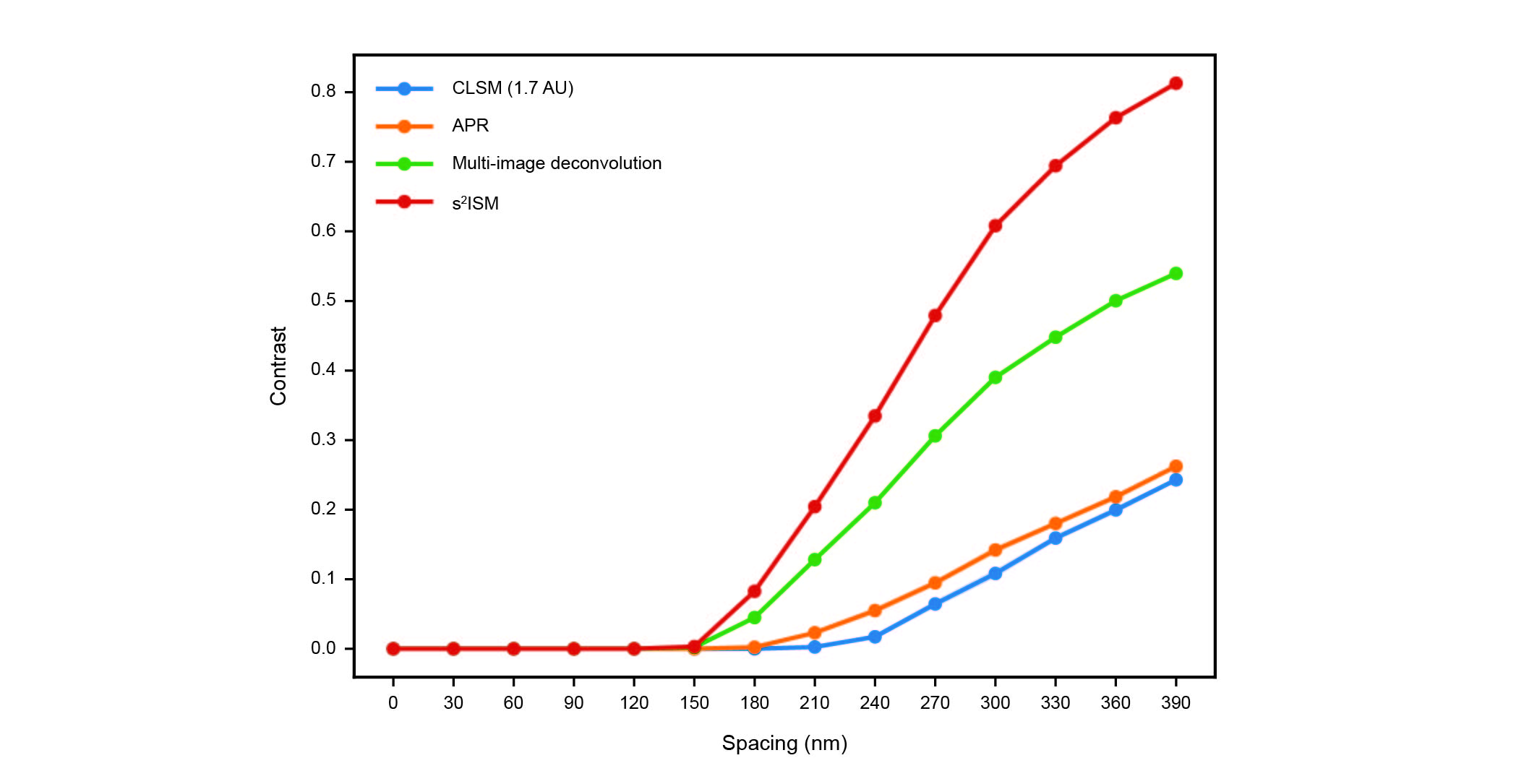}
    \caption{\textbf{Resolution enhancement of ISM reconstruction algorithms.}
    Analysis of the contrast of the gradually spaced lines shown in Fig. \ref{fig:3}a, calculated for each conventional ISM reconstruction algorithm and \sism.
    }
    \label{supfig:Argo_sample_resolution}
\end{figure*}

\clearpage

\begin{figure*}[h!]
    \centering
    \includegraphics[width=\textwidth]{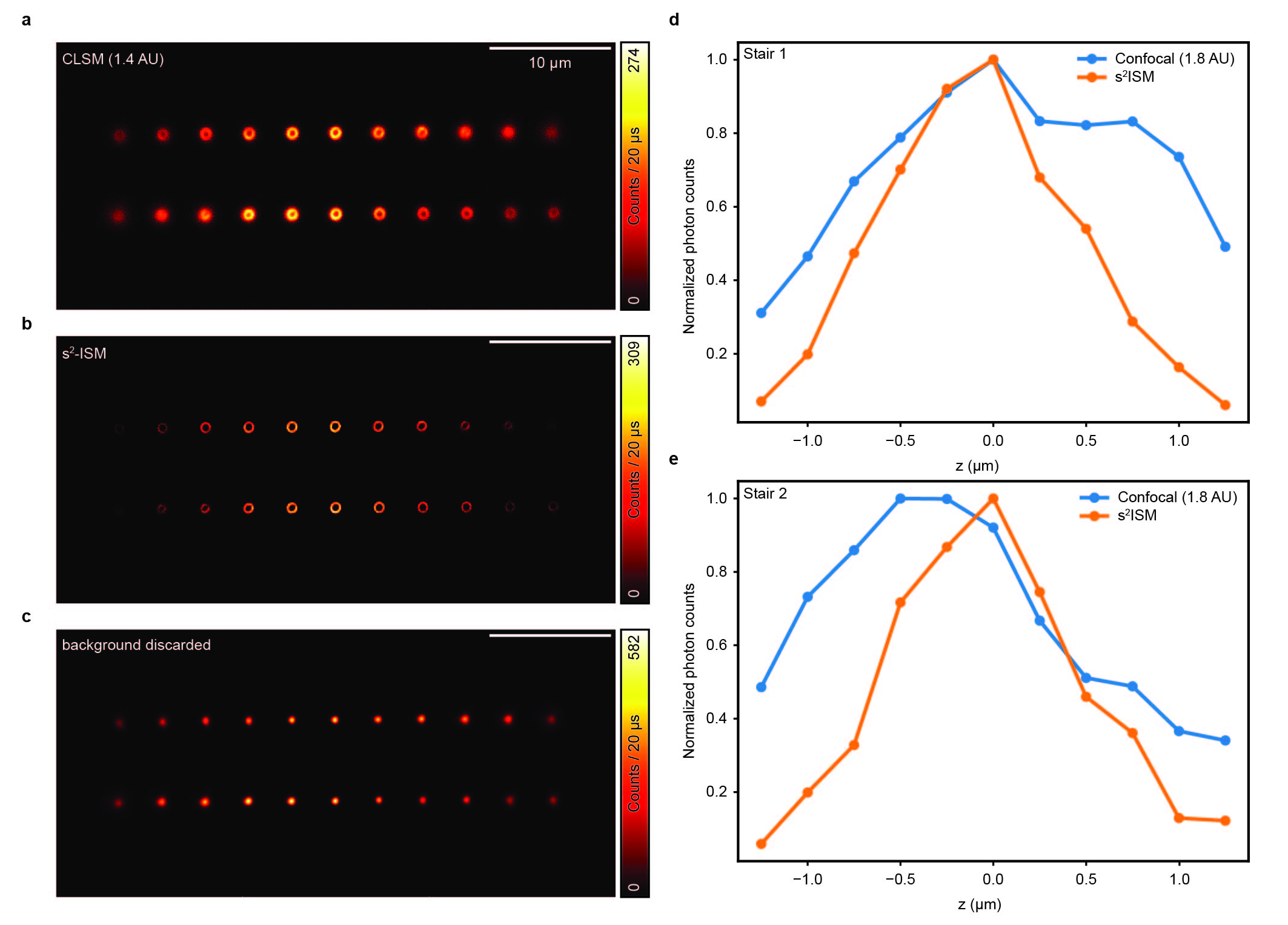}
    \caption{\textbf{Extended data: axially spaced stairs.}
    Extended images from Fig. \ref{fig:3}b.
    \textbf{a}, open-pinhole confocal image.
    \sism reconstructions of the in-focus, \textbf{b}, and out-of-focus plane, \textbf{c}.
    Normalized contrast of the steps at various axial positions calculated for the upper, \textbf{d}, and lower stair, \textbf{e}.
    }
    \label{supfig:Argo_sample}
\end{figure*}

\clearpage

\begin{figure*}[!htb]
    \centering
    \includegraphics[width=\textwidth]{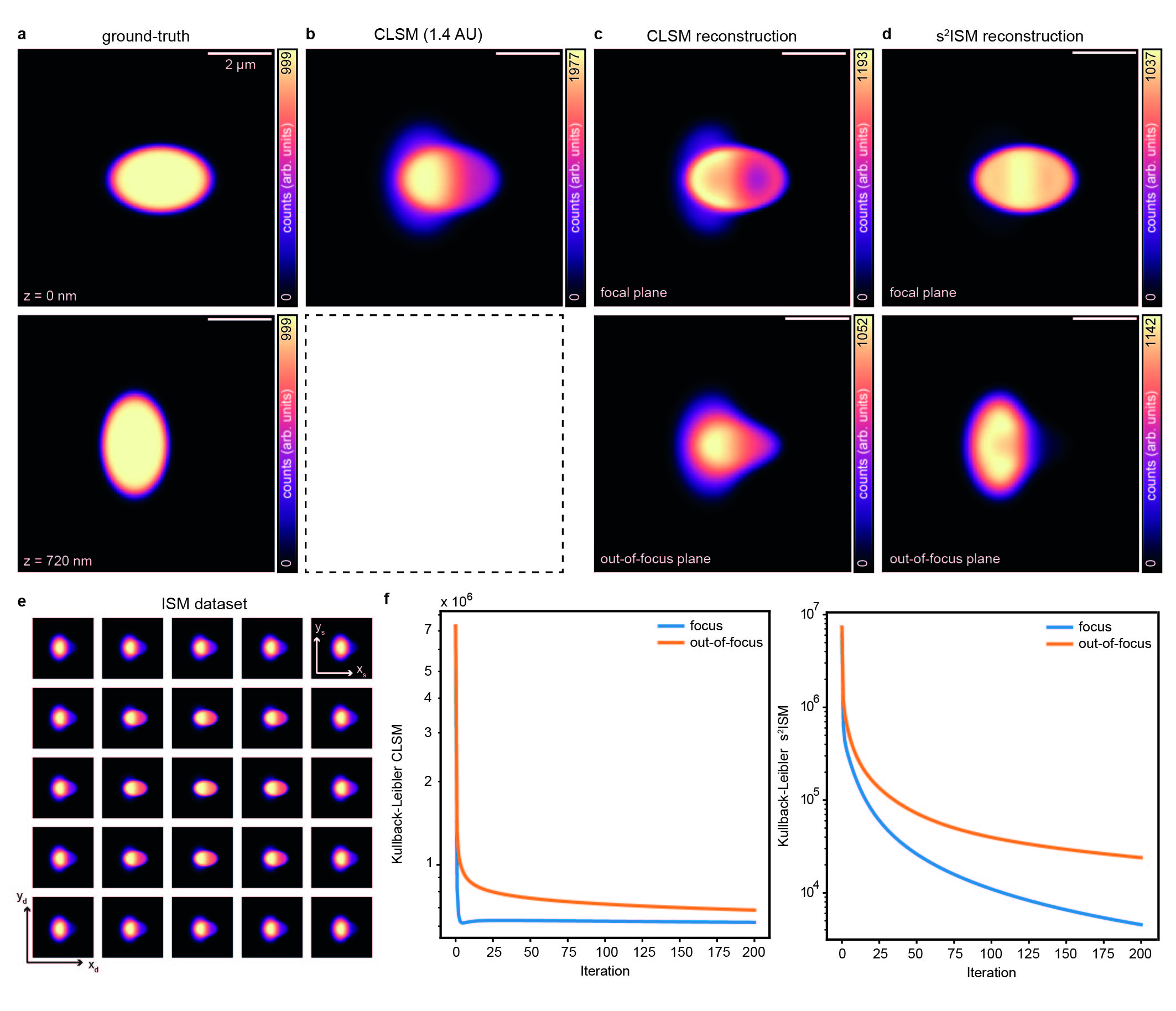}
    \caption{\textbf{Low-frequency sample analysis.}
    \textbf{a}, phantom samples -- one per plane -- generated to be mainly dominated by low-frequency content (large and smooth, compared to the PSF of the system).
    \textbf{b}, corresponding open-pinhole confocal image.
    \textbf{c}, \sism applied to the confocal image, discarding the detector dimension.
    \textbf{d}, \sism applied to the ISM dataset, depicted in \textbf{e}.
    \textbf{f}, Kullback-Leibler divergence as function of the iteration for the reconstructions on the confocal and ISM data.
    The confocal image is not informative enough to enable the discrimination of the axial position of the objects and the reconstruction fails.
    Instead, the detector dimension in the ISM dataset enables the removal of the defocused component without altering the in-focus content. In other words, \sism truly detects the axial position of the emitters without acting as a high-pass filter.
    }
    \label{supfig:low_freq_sample}
\end{figure*}

\clearpage

\begin{figure*}[h!]
    \centering
    \includegraphics[width=\textwidth]{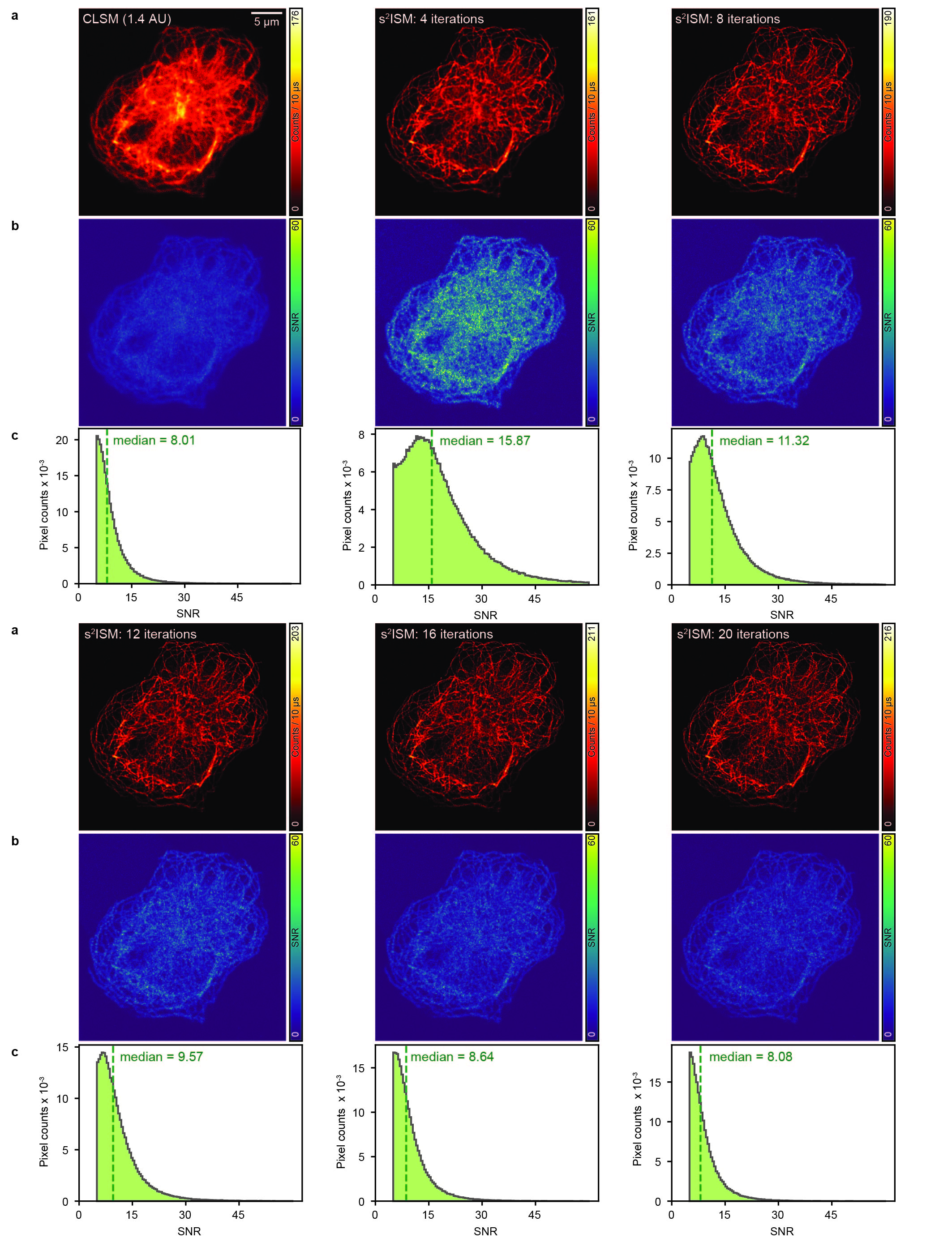}
    \caption{\textbf{Signal-to-noise ratio enhancement.}
    Analysis performed on the data from Supp. Fig. \ref{supfig:cell_A}, sliced into five realizations of the same dataset. We reconstructed each realization independently using the \sism algorithm.
    \textbf{a}, average reconstruction.
    \textbf{b}, signal-to-noise ratio (SNR) map, calculated as the pixel-wise ratio of average and standard deviation of the reconstructions.
    \textbf{c}, histogram of the SNR map, calculated in the range $[5, 60]$.
    }
    \label{supfig:SNR_enhancement_charac}
\end{figure*}

\clearpage

\begin{figure*}[h!]
    \centering
    \includegraphics[width=\textwidth]{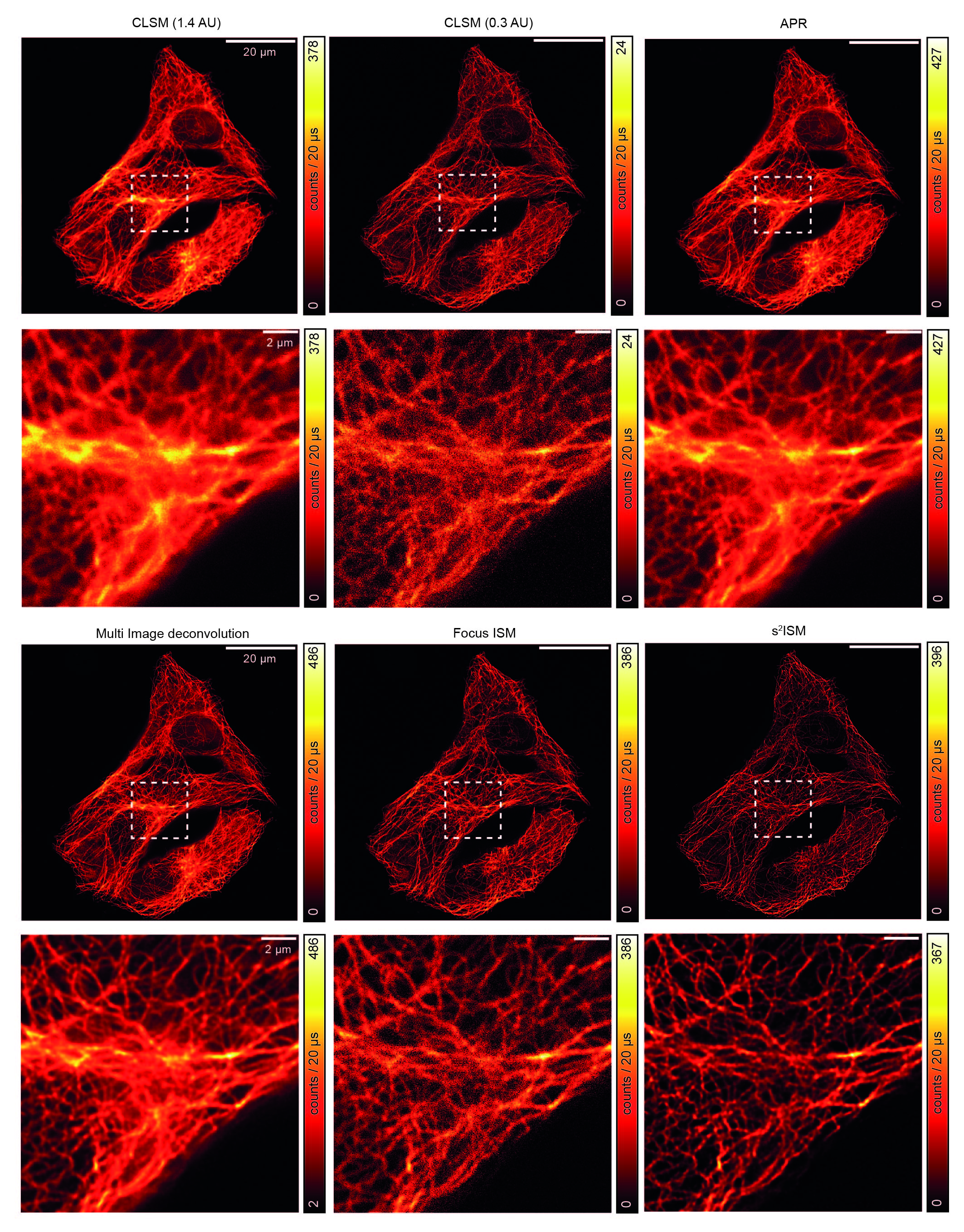}
    \caption{\textbf{Extended data: tubulin network of a group of HeLa cells.}
    Extended images from Fig. \ref{fig:4+5}a-b. Field-of-view: \qtyproduct{80 x 80}{\micro m}, image size: \qtyproduct{2000 x 2000}{} pixels, \sism and Multi-Image deconvolution iterations: 20.
    }
    \label{supfig:cell_B}
\end{figure*}

\clearpage

\begin{figure*}[h!]
    \centering
    \includegraphics[width=\textwidth]{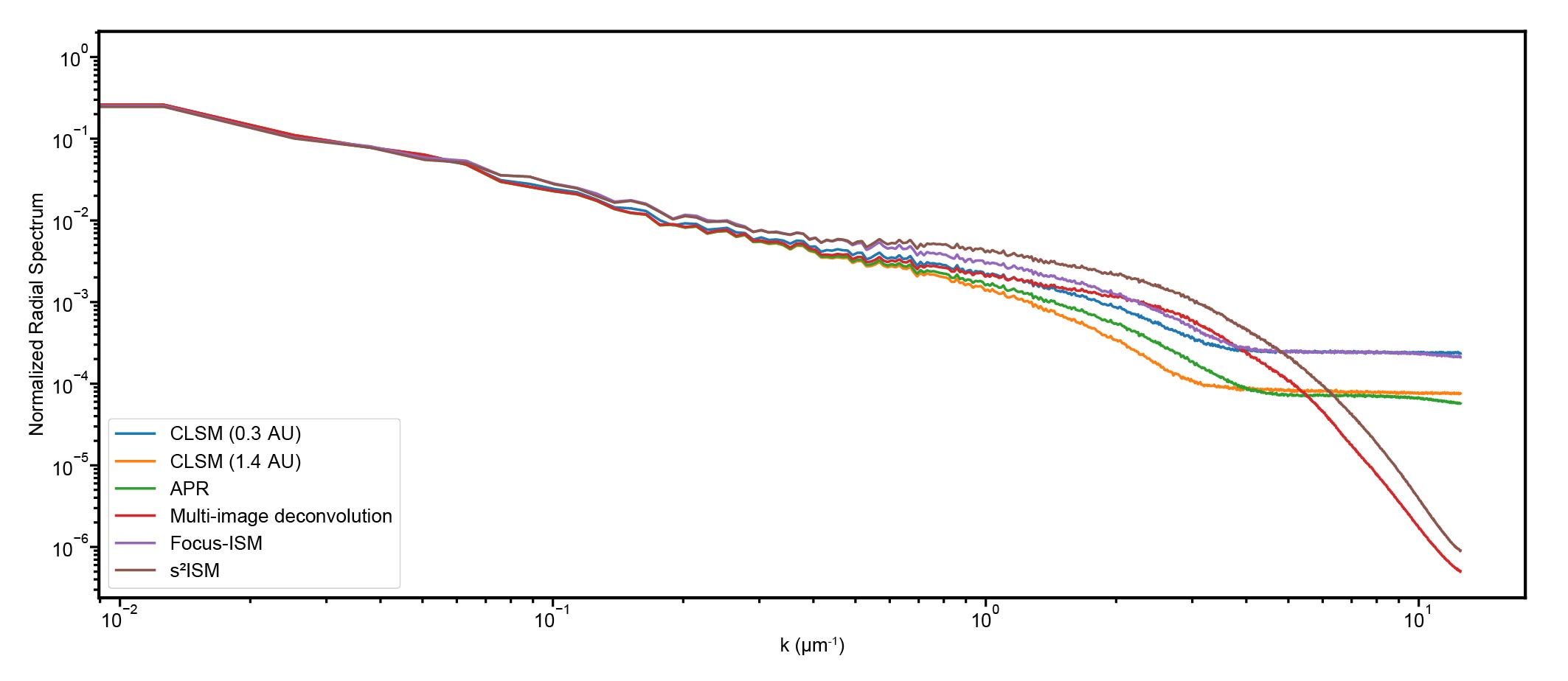}
    \caption{\textbf{Radial spectrum analysis.}
    Normalized radial spectra calculated on the images of Supp. Fig. \ref{supfig:cell_A}.
    The high-frequency plateau is the noise-level.
    A larger cut-off correlates with higher lateral resolution, while larger high-frequency values (before the cut-off) correlates with higher optical sectioning.
    Lower values after the cut-off correlates with higher signal-to-noise ratio.
    }
    \label{supfig:radial_spectra}
\end{figure*}

\clearpage

\begin{figure*}[h!]
    \centering
    \includegraphics[width=\textwidth]{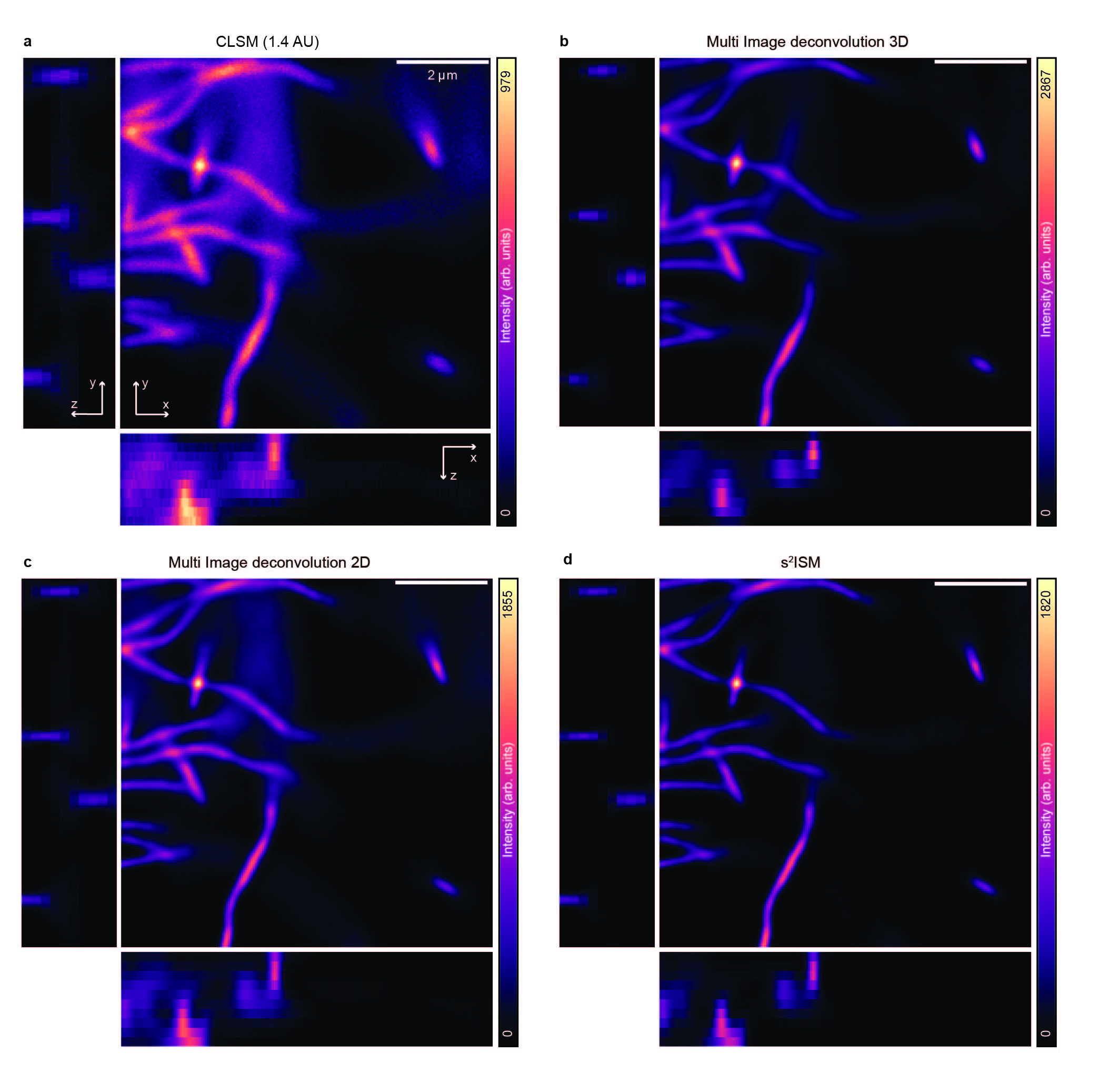}
    \caption{\textbf{Comparison of \sism and 3D multi-Image deconvolution.}
    We simulated a 3D sample of tubulin filaments -- volume size \qtyproduct{10x201x201}{} with a voxel size of \qtyproduct{200x40x40}{\nano m} $(zxy)$ -- and the correspoding 5D (3D + 2D) ISM dataset. We included shot noise in the data. Simulation parameters:  $\NA = 1.4$, $n = 1.5$, $\lambda_{\exc} = \SI{640}{\nano m}$, $\lambda_{\exc} = \SI{660}{\nano m}$.
    \textbf{a}, open-pinhole confocal image.
    \textbf{b}, 3D multi-image deconvolution (20 iterations).
    \textbf{c}, 2D multi-image deconvolution applied plane-by-plane (20 iterations per plane).
    \textbf{d}, \sism applied plane-by-plane (20 iterations per plane).
    The three slices (xy, xz, yz) are taken from the center of the stack.
    2D deconvolution is incapable of removing out-of-focus light. Instead, 3D deconvolution improves optical sectioning by reassigning the defocused light to the correct plane, granting a better SNR. However, it requires a complete volumetric dataset. \sism improves optical sectioning by removing the out-of-focus light, but requires just a planar dataset.}
    \label{supfig:3D-ISM_vs_MID-3D}
\end{figure*}

\clearpage

\clearpage

\clearpage

\begin{figure*}[h!]
    \centering
    \includegraphics[width=\textwidth]{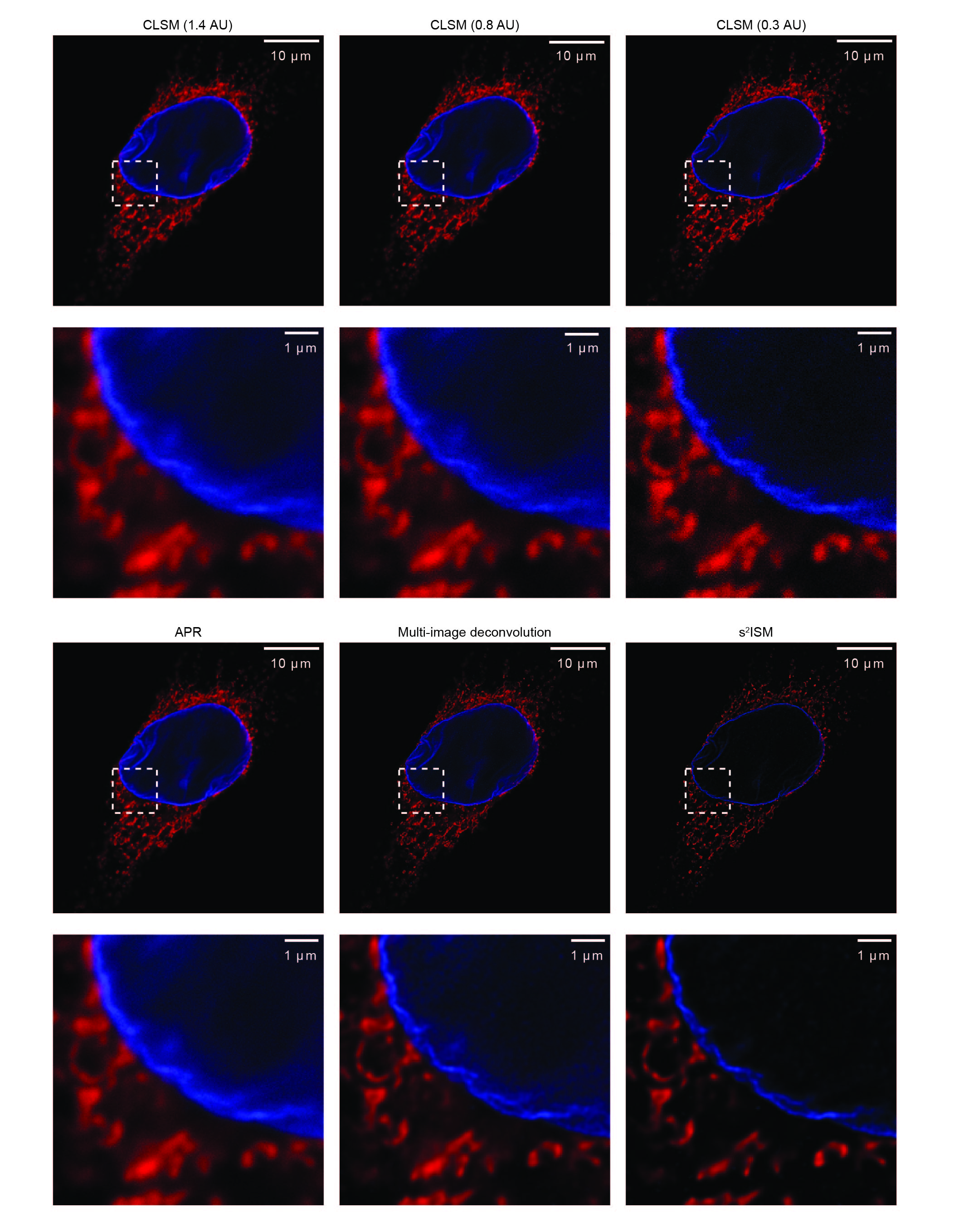}
    \caption{\textbf{Extended data: multi-color imaging of a HeLa cell.}
    Extended images from Fig. \ref{fig:6}b. Field-of-view: \qtyproduct{65 x 65}{\micro m}, image size: \qtyproduct{1625 x 1625}{} pixels,  \sism and Multi-Image deconvolution iterations: 20.}
    \label{supfig:Multicolor}
\end{figure*}

\clearpage

\begin{figure*}[!htb]
    \centering
    \includegraphics[width=\textwidth]{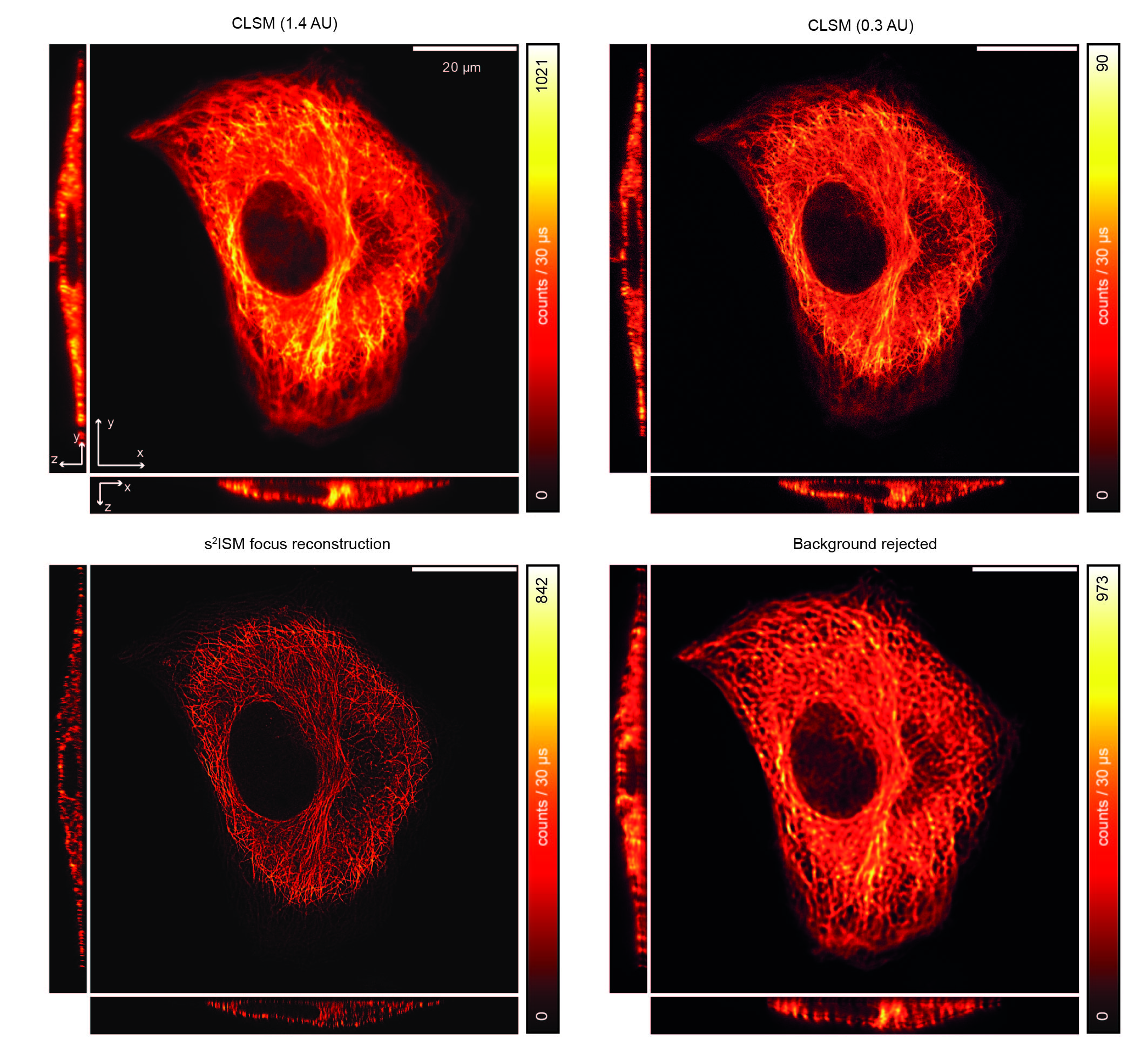}
    \caption{\textbf{Volumetric reconstruction: HeLa cell.}
    Stack of the tubulin network of a HeLa cell, obtained with linear excitation at $\lambda_{\exc} = \SI{640}{nm}$ and reconstructed plane-by-plane.
    Field-of-view: \qtyproduct{7.2 x 86 x 86}{\micro m} $(zxy)$, image size: \qtyproduct{56 x 1075 x 1075}{} pixels $(zxy)$, \sism and Multi-Image deconvolution iterations = 10 per plane. The three slices (xy, xz, yz) are taken from the center of the stack.
    }
    \label{supfig:stack_cell}
\end{figure*}

\clearpage

\begin{figure*}[h!]
    \centering
    \includegraphics[width=\textwidth]{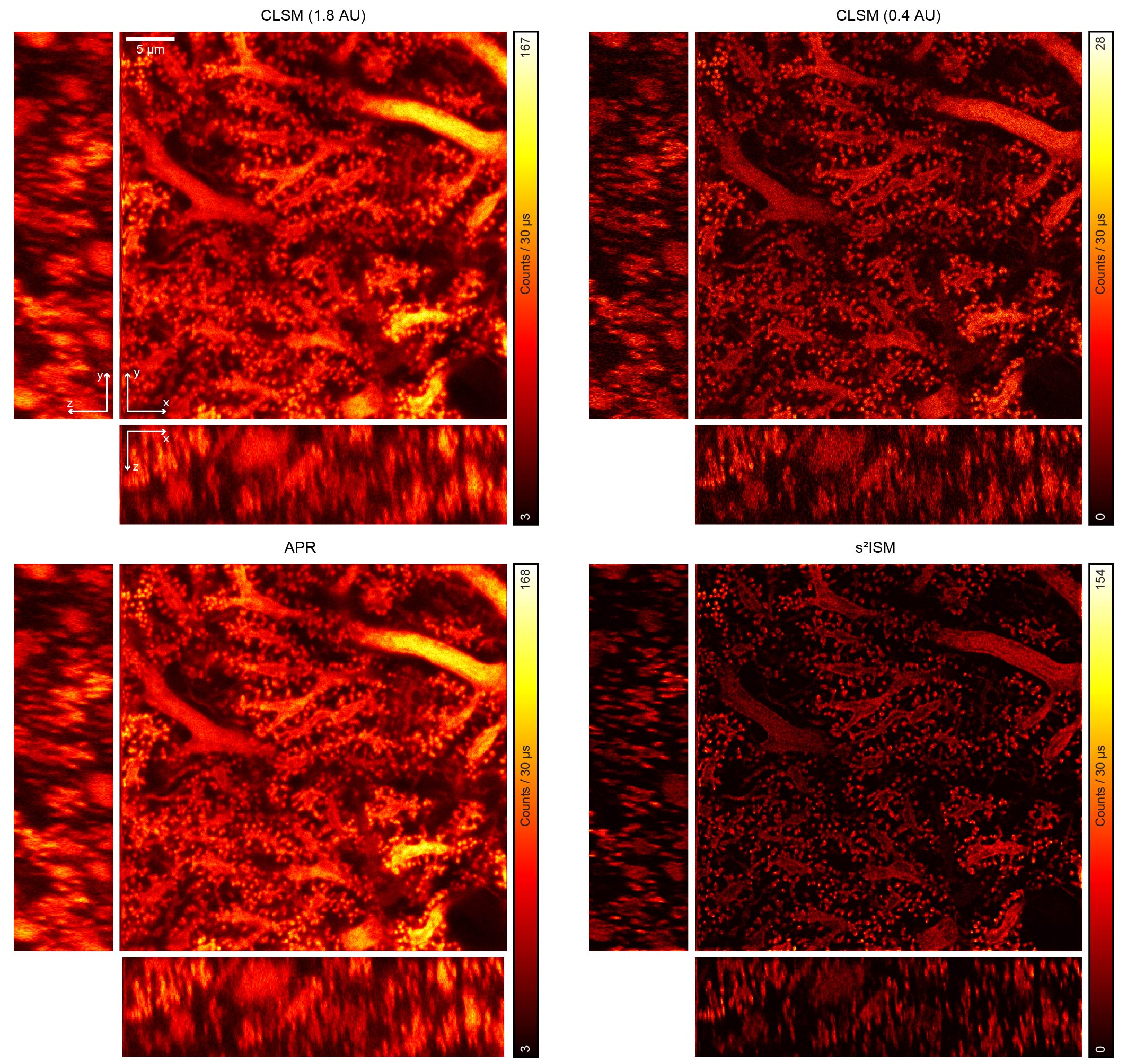}
    \caption{\textbf{Volumetric reconstruction: Purkinje cells in a cerebellum slice.}
    Stack of Purkinje cells in a cerebellum slice, obtained with linear excitation at $\lambda_{\exc} = \SI{488}{nm}$ and reconstructed plane-by-plane. Field-of-view: \qtyproduct{10.25 x 40 x 40}{\micro m} $(zxy)$, image size: \qtyproduct{50 x 1000 x 1000}{} pixels $(zxy)$, \sism and Multi-Image deconvolution iterations = 20 per plane. The three slices (xy, xz, yz) are taken from the center of the stack.
    }
    \label{supfig:Purkinje_stack}
\end{figure*}

\clearpage

\begin{figure*}[h!]
    \centering
    \includegraphics[width=\textwidth]{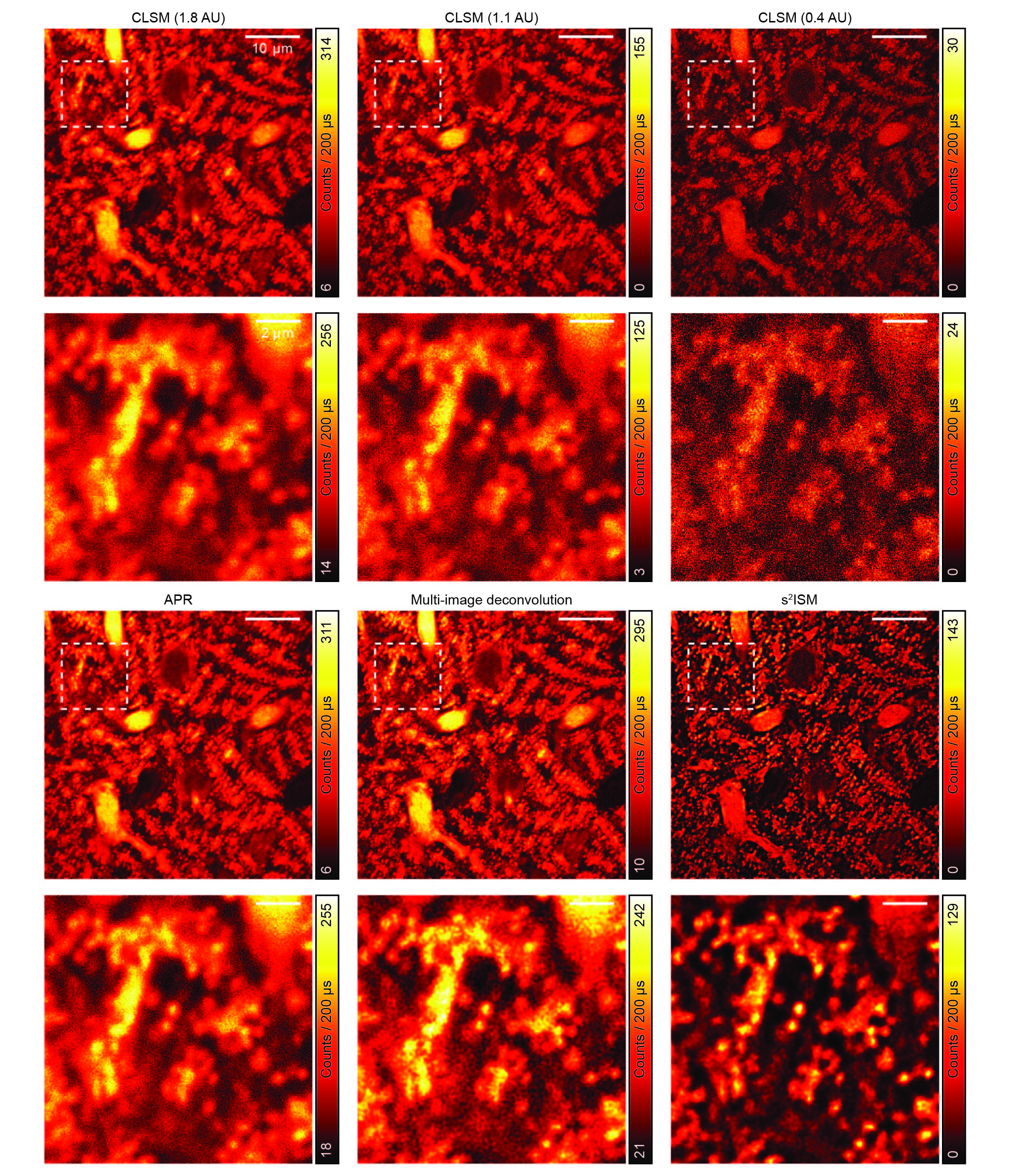}
    \caption{\textbf{Extended data: two-photon excitation imaging of a cerebellum slice.}
    Extended images from Fig. \ref{fig:6}c. Field-of-view: \qtyproduct{50 x 50}{\micro m}, image size: \qtyproduct{1250 x 1250}{} pixels, \sism and Multi-Image deconvolution iterations: 20.
    }
    \label{supfig:Purkinje}
\end{figure*}

\clearpage

\begin{figure*}[h!]
    \centering
    \includegraphics[width=\textwidth]{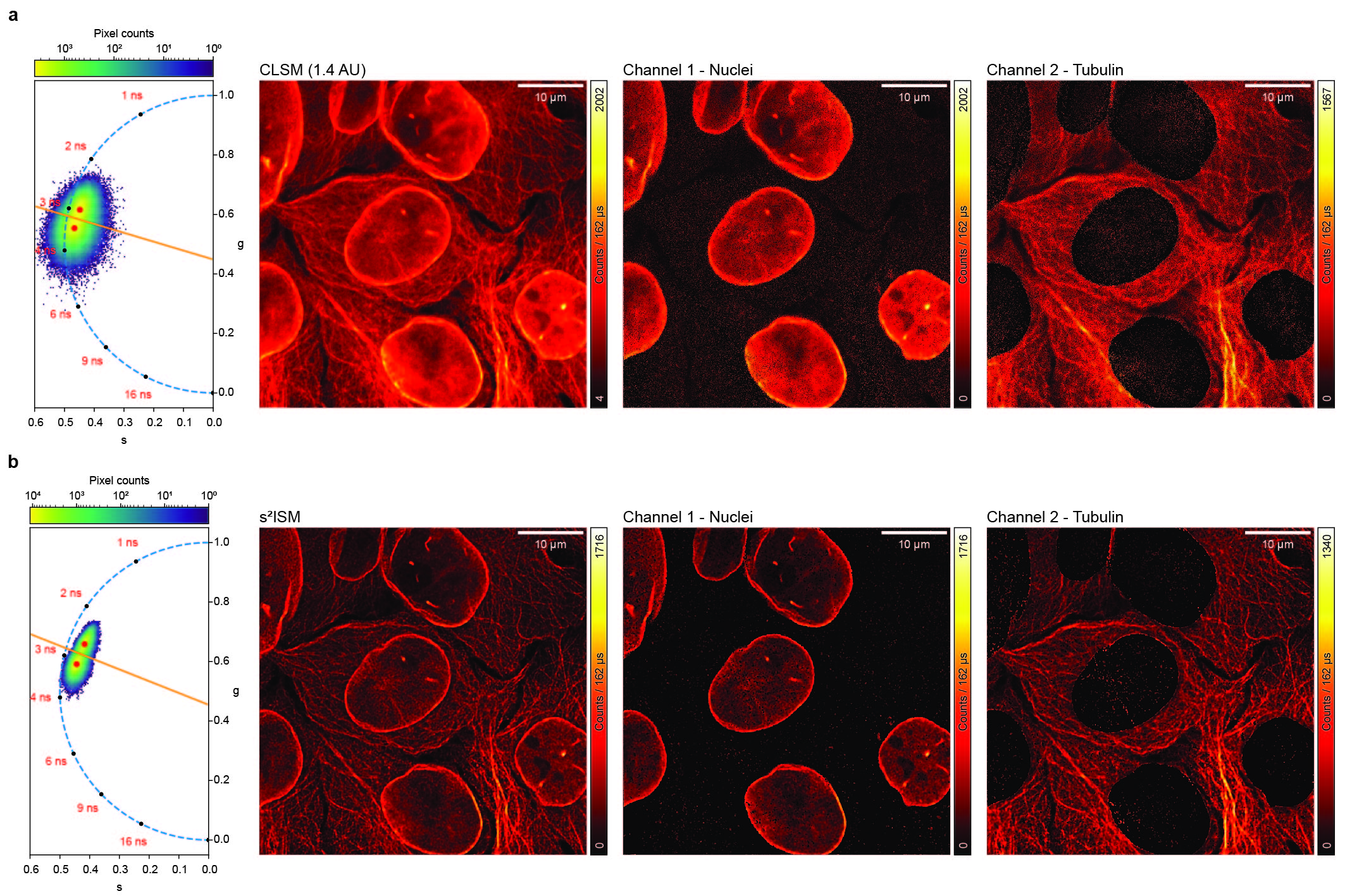}
    \caption{\textbf{Phasor segmentation.}
    The phasor cloud is fit to a Gaussian mixture with two components. Using Euclidean distance, each phasor is classified to the closest Gaussian component. The decision boundary is shown as an orange line. The results of the segmentation on the confocal and \sism images are shown in \textbf{a} and \textbf{b}, respectively. The image quality enhancement coming from \sism enables a more reliable and robust segmentation.
    }
    \label{supfig:phasor_segmentation}
\end{figure*}

\end{appendices}

\end{document}